# A guide through a family of phylogenetic dissimilarity measures among sites

**Sandrine Pavoine**

*S. Pavoine (pavoine@mnhn.fr), Centre d'Ecologie et des Sciences de la Conservation (CESCO UMR7204), Sorbonne Universités, MNHN, CNRS, UPMC, CP51, 55-61 rue Buffon, 75005, Paris, France; Mathematical Ecology Research Group, Department of Zoology, University of Oxford, Oxford OX1 3PS, UK*

*Correspondence to be sent to: Sandrine Pavoine, Muséum National d'Histoire Naturelle, UMR 7204 CESCO, 61 rue Buffon, 75005 Paris, France;*

*E-mail: pavoine@mnhn.fr*

*Abstract.* Ecological studies have now gone beyond measures of species turnover towards measures of phylogenetic and functional dissimilarity. This change of perspective has a main objective: disentangling the processes that drive species distributions from local to broad scales. A fundamental difference between phylogenetic and functional analyses is that phylogeny is intrinsically dependent on a tree-like structure whereas functional data can, most of time, only be forced to adhere a tree structure, not without some loss of information. When the branches of a phylogenetic tree have lengths, then each evolutionary unit on these branches can be considered as a basic entity on which dissimilarities among sites should be measured. Several of the recent measures of phylogenetic dissimilarities among sites thus are traditional dissimilarity indices where species are replaced by evolutionary units. The resulting indices were named PD-dissimilarity indices, in reference to early work on the phylogenetic diversity (PD) measure. Here I review and compare indices and ordination approaches that, although first developed to analyse the differences in the species compositions of sites, can be adapted to describe PD-dissimilarities among sites, thus revealing how lineages are distributed along environmental gradients, or among habitats or regions. As an illustration, I show that the amount of PD-dissimilarities among the main habitats of a disturbance gradient in Selva Lacandona of Chiapas, Mexico is strongly dependent on whether species are weighted by their abundance or not, and on the index used to measure PD-dissimilarity. Overall, the family of PD-dissimilarity indices has a critical potential for future analyses of phylogenetic diversity as it benefits from decades of research on the measure of species dissimilarity. I provide clues to help to choose among many potential indices, identifying which indices satisfy minimal basis properties, and analysing their sensitivity to abundance, size, diversity, and joint absences.

***Main text***

A change of perspective on biodiversity measures has been done mainly during the last two decades. This change operated by including, in measures of biodiversity, functional attributes (e.g. Findley 1976, Petchey and Gaston 2002), a taxonomic hierarchy (e.g. Izsak and Papp 1995, Warwick and Clarke 1995), or a phylogeny (e.g. Faith 1992). The particularity of rooted phylogenies (or taxonomies) is that they depend on a tree structure. This is one of the main arguments that can be raised for distinguishing the methodological research on phylogenetic diversity from that on functional diversity. Even if functional dendrograms can be obtained (Petchey and Gaston 2002), the tree structure is intrinsic to a phylogeny whereas that of a functional dendrogram is somehow forced.

Among the concepts that have long dominated the research for biodiversity metrics are 'turnover' and '(dis)similarity' among sites (or communities, assemblages, stations, regions, samples, relevés, quadrats, plots, etc.; e.g. Legendre and Legendre 1998). Species turnover estimates changes in the specific composition of two or more sites. Beyond species turnover, coefficients of (dis)similarity among communities can include species' abundances. In that case, two sites can then be different if they have the same species represented by different abundances. Several developments have been done to extend these concepts and their measures to phylogenetic turnover and phylogenetic (dis)similarity among sites (Lozupone and Knight 2005, Ferrier et al. 2007, Chiu et al. 2014).

Faith (1992) showed how traditional indices of diversity and dissimilarity can be transposed to phylogenetic diversity (see also e.g. Faith et al. 2009). From a methodological point of view, he replaced, in traditional coefficients, species by units of the branch lengths of a phylogenetic tree, hereafter referred to as "evolutionary units". Faith et al. (2009) named indices of "PD-dissimilarity" those indices of site-to-site dissimilarity adapted according to this translation from species to evolutionary units. Faith (1992, 2013, see also Faith and Richards 2012) showed that branch lengths can be expressed in terms of features. Evolutionary units would thus be proxies for features. With this assumption, the lengths of the branches would reflect a number of features and the sum of abundances of the species descending from a branch would reflect the abundance of the features (Chao et al. 2010, Faith and Richard 2012). Hereafter, to keep the discussion general (whatever the definition of the tree topology and branch lengths), the general concept of evolutionary unit is maintained. The translation from species to evolutionary units has inspired several developments scattered in the ecological literature. For example, instead of measuring species turnover by the proportion of species unshared by two sites, Unifrac (Lozupone and Knight 2005) measures evolutionary turnover by the proportion of evolutionary units unshared by the two sites. I review here the methods developed so far and those that could be developed using Faith's framework for measuring site-to-site PD-dissimilarities and for displaying these dissimilarities using ordination approaches. My objective is to provide an overview and a critical evaluation of this family of indices and ordination approaches through their ability to reflect evolutionary patterns among communities.

# Materials and Methods

## Principle

Consider a rooted phylogenetic tree $T$ with $S$ species as tips and $K$ branches. $A_i$ is the abundance of species $i$; $t_k$ is the set of species descending from branch $k$; $t_T$ is the whole set of species (tips of the phylogenetic tree); $b_T$ is the set of branches in the phylogenetic tree $T$; $L_k$ is the length of branch $k$; $a_k$ is the sum of abundances for all species descending from branch $k$ ($a_k = \sum_{i \in t_k} A_i$).

The principle of the change of perspective from species to evolutionary units consists in replacing species in dissimilarity coefficients by evolutionary units. A branch *k* of a phylogeny with a length equal to $L_k$ will be considered to have $L_k$ evolutionary units. Each evolutionary unit on this branch has an abundance value equal to $a_k$. The sum of abundances for all evolutionary units supported by the branch is thus $L_k a_k$.

## Site-to-site dissimilarity

Sites can be compared in a pairwise fashion according to their composition in evolutionary units exactly the same way they can be compared based on their composition in species. Consider two sites numbered 1 and 2. Following Lozupone and Knight (2005), Ferrier et al. (2007) and Nipperess et al. (2010), indices of phylogenetic similarity between two sites can be obtained thanks to three components: the number of evolutionary units present in both site 1 and site 2 (*a*), in site 1 but not in site 2 (*b*), in site 2 but not in site 1 (*c*). A fourth component, *d*, can be obtained by counting the number of evolutionary units absent from both site 1 and site 2 but present in a larger set of sites. The components *a*, *b*, *c*, and *d* can be combined in traditional coefficients of compositional turnover (see for example Gower and Legendre 1986 for a review). Formulas for three indices of PD-dissimilarity, $evoD_{Jaccard}$, $evoD_{Sørensen}$, and $evoD_{Ochiai}$, are given in Supplementary material Appendix 1 using 1-*S* with *S* respectively equal to $a/(a+b+c)$ (Jaccard 1901), $2a/(2a+b+c)$ (Sørenson 1948) and $a/\sqrt{(a+b)(a+c)}$ (Ochiai 1957). Using $2a/(2a+b+c)$, Ferrier et al. (2007) and Bryant et al. (2008) PhyloSor index is equal to 1-$evoD_{Sørenson}$. Using 1-*S* and *S* equal to $a/(a+b+c)$, Lozupone and Knight (2005) Unifrac index is $evoD_{Jaccard}$.

Following Nipperess et al. (2010), the same coefficients can be used with the abundances of the species within sites, if *a*, *b*, *c* are replaced with, respectively: the degree to which evolutionary units in sites 1 and 2 agree in their abundances ($A = \sum_{k \in b_T} L_k \min\{a_{1k}, a_{2k}\}$),the degree to which the abundances of evolutionary units in sites 1 exceed those in site 2 ($B = \sum_{k \in b_T} L_k \left(\max\{a_{1k}, a_{2k}\} - a_{2k}\right)$), the degree to which the abundances of evolutionary units in site 2 exceed those in site 1 ($C = \sum_{k \in b_T} L_k \left(\max\{a_{1k}, a_{2k}\} - a_{1k}\right)$). A fourth component (*D*) can be defined as the degree to which the maximum abundance of each evolutionary unit across a larger set of sites exceeds that in sites 1 and 2. Formulas for three indices of PD-dissimilarity using the *A*, *B*, *C* components, $evoS_{TJ}$, $evoS_{TS}$, and $evoS_{TO}$, are given in Supplementary material Appendix 1 using 1-*S* with *S* respectively equal to $A/(A+B+C)$, $2A/(2A+B+C)$ and $A/\sqrt{(A+B)(A+C)}$.

Apart from this family of dissimilarity coefficient, other coefficients of PD-dissimilarity can be derived from well-known and widely used coefficients of compositional dissimilarity. For example, adapting the $\chi^2$ (e.g. Legendre and Legendre 1998), profile (Legendre and Gallagher 2003), Hellinger (Rao 1995), and chord (e.g. Orloci 1967) distances to PD-dissimilarity leads to four interrelated indices: $evoD_{\chi^2}$, $evoD_{Profile}$, $evoD_{Hellinger}$, $evoD_{Chord}$. Adapting Bray-Curtis (Bray and Curtis 1957), Morisita-Horn (Horn 1966), scaled Canberra (Lance and Williams 1966, Legendre and Legendre 1998), and divergence (Clark 1952) indices leads to alternative interrelated PD-dissimilarity indices: $evoD_{Bray\text{-}Curtis}$, $evoD_{Morisita\text{-}Horn}$, $evoD_{ScaledCanberra}$, $evoD_{Divergence}$ (Supplementary material Appendix 1 for formulas).

Furthermore, profiles of dissimilarity indices can be developed by defining parametric indices with a parameter, here named *q*, that modifies the importance given to rare versus common evolutionary units. Chiu et al. (2014) developed a measure of $\beta$ diversity (inter-site phylogenetic diversity), which they named ${}^qD_\beta$. A main advantage of this approach is that it can be applied to a large number of sites simultaneously, which is expected to better characterize $\beta$ diversity than average pairwise dissimilarity between sites (Baselga 2013). For a comparison with the above indices, I apply it here also to the comparison of two sites only. The main characteristic of ${}^{q}D_{\beta}$ is that it is based on a between-group component of a parametric diversity measure. A little algebra shows that ${}^{q}D_{\beta}$ is an index of PD-$\beta$-diversity: it extents a compositional index of $\beta$ diversity by replacing species with evolutionary units (Supplementary material Appendix 2). I apply ${}^{q}D_{\beta}$ considering first absolute and then relative abundances for evolutionary units. Because ${}^{q}D_{\beta}$ is bounded between 1 and $m$ (where $m$ is the number of sites), I used Chiu et al. indices $1-\bar{V}_{qm} = {}^{q}D_{\beta} - 1$, $1-\bar{C}_{qm} = \left[1-\left(1/{}^{q}D_{\beta}\right)^{q-1}\right] / \left[1-\left(1/m\right)^{q-1}\right]$ and $1-\bar{U}_{qm} = \left[1-\left(1/{}^{q}D_{\beta}\right)^{1-q}\right] / \left[1-\left(1/m\right)^{1-q}\right]$ to obtain indices bounded between 0 and 1. Although not considered in Chiu et al. (2014), I also analysed ${}^{q}evoD_{Rényi}=\log_{m}({}^{q}D_{\beta})$ (here named in reference to Rényi 1961). All formulas are detailed in Supplementary material Appendix 1. It can be noticed that $evoD_{Sørensen}=1-\bar{C}_{02}$ and $evoD_{Jaccard}=1-\bar{U}_{02}$; only for q→1, ${}^{1}evoD_{Rényi} = 1-\bar{C}_{12} = 1-\bar{U}_{12}$ (see also Supplementary materiel Appendix 1).

Legendre and De Cáceres (2013) defined nine basic necessary properties for coefficients of compositional (species) dissimilarity. These properties can be adapted to the concept of PD-dissimilarity. Among the indices discussed in this paper, those that fail to satisfy at least one of Legendre and De Cáceres (2013) properties are: $evoD_{Profile}$ and $evoD_{\chi^2}$ (see Supplementary material Appendix 2 for proofs).

## Ordination

The abundances of evolutionary units within sites can also be used directly in ordination approaches that define multidimensional spaces where phylogenetic branches (supporting the evolutionary units) are positioned according to the structure of the phylogenetic tree, and sites are positioned according to their phylogenetic compositions. This corresponds to using in ordination approaches evolutionary units as the basic entities instead of species.

Using this principle, the correspondence analysis (CA) can be adapted to analyse the distributions of lineages among sites and, simultaneously, to analyse the phylogenetic composition of sites. This approach, hereafter designated as evolutionary correspondence analysis, or merely evoCA, focuses on the interdependence between sites and evolutionary units. It corresponds to applying CA to matrix ($L_k a_{jk}$) for $j$=1….$m$ and $k$=1…$K$ (Proof in Supplementary material Appendix 2). Adapting CA to evoCA, weights are given to phylogenetic branches, $L_k \sum_{j=1}^{m} a_{jk} / \sum_{j=1}^{m} \sum_{k=1}^{K} L_k a_{jk}$ for branch $k$, and to sites, $\sum_{k=1}^{K} L_k a_{jk} / \sum_{j=1}^{m} \sum_{k=1}^{K} L_k a_{jk}$ for site $j$. An alternative to evoCA is obtained by adapting the principle of the nonsymmetric correspondence analysis (NSCA, Kroonenberg and Lombardo 1999) to the analysis of the phylogenetic composition of sites. I call this approach evolutionary non-symmetric correspondence analysis, or merely evoNSCA. Consider a given evolutionary unit, evoNSCA measures how the predictability of its position in the phylogeny is improved by knowing in which site it was observed. EvoNSCA is equivalent to applying centred (non-scaled) principal component analysis (PCA) to the matrix $(a_{jk} / a_{j+})_{j,k}$ with $j$=1….$m$, $k$=1…$K$, and $a_{j+} = \sum_{k=1}^{K} L_k a_{jk}$. The weights attributed to sites are the same as in evoCA but branches are weighted by their lengths only. The distance between two sites (numbered 1 and 2) in the space of evoCA is measured by $evoD\chi^2$ and in the space of the evoNSCA analysis by $evoD_{Profile}$ (Supplementary material Appendix 2). Neither $evoD_{\chi^2}$ nor $evoD_{Profile}$ fulfils Legendre and De Cáceres (2013) basic necessary properties for coefficients of dissimilarity (Supplementary material Appendix 2). When one is interested in describing PD-dissimilarities among sites, other ordination approaches thus have to be considered.

Rao (1995) developed an alternative to CA that used Hellinger distance. I apply it here in the context of PD-dissimilarity and refer to this ordination approach as evolutionary principal component analysis based on Hellinger distance, or merely $\text{evoPCA}_{\text{Hellinger}}$. In $\text{evoPCA}_{\text{Hellinger}}$, centred (non-scaled) PCA is applied to the matrix $(\sqrt{a_{jk}/a_{j+}})_{j,k}$ with $j$=1….$m$ and $k$=1…$K$. The weights attributed to sites may be the same as in evoCA or they may be freely chosen provided they sum to 1 but the weights for branches are fixed: $L_k$ for any branch $k$. The distance between two sites in $\text{evoPCA}_{\text{Hellinger}}$ is measured by index $evoD_{Hellinger}$ (Supplementary material Appendix 1). Legendre and Gallagher (2001) have shown that the application of PCA to transformed data (as in evoNSCA and $\text{evoPCA}_{\text{Hellinger}}$) can be extended to other transformation methods. Such approaches are now referred to as transformed-based PCA or tb-PCA. Applied to evolutionary units, I use here the expression tb-evoPCA (which simply means that species are replaced with evolutionary units in traditional tb-PCA). In particular, the reasoning used to develop $\text{evoPCA}_{\text{Hellinger}}$ can be applied to develop an approach that uses the chord distance ($evoD_{Chord}$ in the context of PD-dissimilarities, see also Supplementary material Appendix 1). In this approach, named $\text{evoPCA}_{\text{Chord}}$, centred (non-scaled) PCA is applied to the matrix $(a_{jk}/\sqrt{\sum_{k=1}^{K} L_k a_{jk}^2})_{j,k}$ with $j$=1….$m$ and $k$=1…$K$; the weights attributed to sites may be freely chosen provided they sum to 1 and branches are weighted by their lengths. Compared with $\text{evoPCA}_{\text{Chord}}$, $\text{evoPCA}_{\text{Hellinger}}$ reduces the importance given to large abundances. The approaches are nevertheless linked and merge when applied to presence-absence (1/0) data, in which case $\text{evoD}_{\text{Hellinger}}$ and $\text{evoD}_{\text{Chord}}$ are both equal to $\sqrt{2evoD_{Ochiai}}$ with formula given above and in Supplementary Appendix 1 (see Legendre and Legendre 1998). In all ordination approaches, the contributions of phylogenetic branches and sites to the pattern of (inter)dependence can be calculated by multiplying, for each site and each branch, its squared normed coordinate by its weight (Lebart et al. 1984). Contributions sum to 1 per axis.

A critical point that needs to be developed here is the consequences the change of perspectives from species to evolutionary units might have on the interpretation of the results provided by the ordination approaches. CA is known to rely on unimodal responses of species to environmental gradients or heterogeneous habitats. In regions of heterogeneous environment or when sites are distributed along a long environmental gradient, ordination approaches can reveal environmental filtering; in that case, a species occur with the highest abundance in sites close to its optimum environmental conditions and its abundance should decrease with the distance from this optimum. Species that lie in the centre of the ordination diagram may have their optimum there or may have several optima and thus a multimodal response to environmental gradients (e.g. Legendre and Legendre 1998). The ordination diagram would thus be more easily interpreted in terms of environmental filtering if all species have unimodal responses to environmental gradients (see also Faith et al. 2009 for a discussion on how to use ordination approaches to analyse species competitive exclusion). Legendre and Gallagher (2001) showed that a tb-PCA relying on the Hellinger and chord distances are efficient alternatives to CA. Contrary to CA, these alternative approaches rely on dissimilarity coefficients that satisfy Legendre and De Cáceres (2013) basic necessary properties. The tb-PCA relying on the profile distance (i.e. NSCA), however, should be restricted to the analysis of homogeneous data or short gradients (Legendre and Gallagher 2001).

Applied to evolutionary units, the interpretation of the patterns provided by evoCA, evoPCA$_{Hellinger}$, evoPCA$_{Chord}$ ordination approaches will thus rely on unimodal responses of evolutionary units to environmental gradients as they traditionally relied on unimodal responses of species. Unimodal response for evolutionary units can be a realistic scenario where shared ancestry and shared environment both account for shared feature (Faith et al. 2009) and environment influences features with phylogenetic signal (Pillar and Duarte 2010). For example, Faith et al. (2009) demonstrated, in house-dust microbial communities, that evolutionary units (even those supported by non-terminal branches) displayed unimodal response patterns to environmental gradients. According to Faith et al. (2009), the phylogenetic unimodal response model where “shared environment accounts for shared features” suggest the general applicability of existing robust approaches including ordination to the analysis of PD-dissimilarity. In that context, an advantage of evoCA is that each evolutionary unit is positioned in evoCA maps at the barycentre of the species that descend from it (see proof in supplementary Appendix 1 and illustrations in supplementary Appendix 3). That way the diversity in the response, to the environmental gradient, of the species that descend from an evolutionary unit can be visualized directly on the evoCA maps.

As an alternative to these ordination approaches, the matrices of pairwise PD-dissimilarities among sites can be used in graphical approaches to visualize how phylogenetically dissimilar sites are, using the metric multidimensional scaling (also called principal coordinates analysis, PCoA) or the nonmetric multidimensional scaling methods (NMDS). As these approaches are dissimilarity-based, all information about species and evolutionary units would be lost. An important advantage of evoCA, evoNSCA and tb-evoPCA approaches thus is to allow for biplots with both evolutionary units and sites displayed; dissimilarities among sites can thus be directly explained by identifying the evolutionary units that drive those differences. NMDS is based on the minimization of a stress function that evaluates how different the fitted distances in the ordination space are from the original ecological dissimilarities. Some of its advantages are that it can cope with missing distances and that, for a fixed number of dimensions, it often provides less deformed representations of the distances than PCoA (Legendre and Legendre 1998). However, it may be time-consuming on large data sets with the risk of finding local minimum for the objective stress function rather than the true global minimum. For further comparisons among these approaches when applied to species and sites see Legendre and Legendre (1998), Borcard et al. (2011), and Legendre and Birks (2012).

## Case study

To illustrate this change of perspective from species to evolutionary units, I have considered the data collected by Medellín et al. (2000) on bats in four habitats in the Selva Lacandona of Chiapas, Mexico with Fritz et al. (2009) phylogeny pruned for retaining only the species present in Medellín et al. data set (Fig. 1). The resolution of the bat phylogeny is uncertain especially at the older nodes (Agnarsson et al. 2011). I have still chosen this illustration because of the unbalanced shape of the phylogenetic tree with long branches for a few species. This permits me to illustrate the effect on PD-dissimilarity indices of distinct species that have few relatives. The four habitats compared were distributed on a disturbance gradient from active cornfield (the most disturbed), through oldfields and cacao plantations, to rainforests (the least disturbed). Because they have yet been hardly explored, I applied to this data set PD-dissimilarity profiles using the parametric indices defined above (with parameter $q$ varying from 0 to 30) and the ordination approaches evoCA, evoNSCA, evoPCA$_{\text{Chord}}$ and evoPCA$_{\text{Hellinger}}$ (using even sites' weights and ordinary weighted mean in the latter two approaches), with both presence-absence and abundance data. The other PD-dissimilarity indices are explored in Supplementary material Appendix 4. The data set and R scripts (R Development Core Team 2014) associated with all approaches developed or discussed in this paper are available upon request to the author.

# Results

## PD-dissimilarities among sites

When rare evolutionary units were given high importance (like for presence/absence data), all parametric indices found the highest PD-dissimilarities to be between cornfields and both rainforest and oldfields; rainforest and cacao plantations were identified as the most similar. When the abundance of the evolutionary units was given more importance then the PD-dissimilarity among rainforest and cacao plantations increased to reach the highest values when the highest PD-similarity was observed between cacao plantations and oldfields (Fig. 2). The behaviour of Chiu et al. (2014) indices depended on which index was used and whether they were applied to relative or absolute abundances. Because ${}^{q}evoD_{Rényi}$ and $1-\bar{V}_{q2}$ provided very similar results, I displayed only ${}^{q}evoD_{Rényi}$ in Figure 2. With relative abundances, indices ${}^{q}evoD_{Rényi}$ and $1-\bar{V}_{q2}$ rapidly decreased with increasing *q* and remained close to 0 (Fig. 2). In contrast, when absolute abundances were used, these indices first decreased and then increased with increasing *q* (Fig. 2). The values of ${}^{q}evoD_{Rényi}$ and $1-\bar{V}_{q2}$ actually did not decrease monotonously with *q* even when relative abundances were used (exact values can be obtained thanks to the Supplementary material Appendixes 5-8). Index $1-\bar{U}_{q2}$ decreased in both cases with increasing *q*, although $1-\bar{C}_{q2}$ first decreased and then increased (Fig. 2). Using absolute abundances led $1-\bar{C}_{q2}$ to converge to the maximum dissimilarity (exceeding 0.9) approximately when *q* exceeded 21, except when comparing cacao plantations with oldfields. This strongly contrasted with $1-\bar{U}_{q2}$ that rapidly reached close-to-zero values for all site-to-site comparisons, which in contrast indicated that all habitats were phylogenetically similar. Those differences among the indices can be explained because they quantify different aspects of the connections between the phylogenetic tree and the abundance of the species (Chiu et al. 2014). The indices thus complement each other. In addition, the two indices $1-\bar{U}_{q2}$, $1-\bar{C}_{q2}$, together with ${}^{q}evoD_{Rényi}$ and $1-\bar{V}_{q2}$, rank pairs of habitats similarly in order of their PD-dissimilarity (Fig. 3). For example, for the highest values of *q*, they agreed that the highest PD-dissimilarity was between rainforest and cacao plantations with absolute abundances and between rainforest and cornfields with relative abundances. As highlighted above, the main advantage of these indices is that they can quantify multiple-site dissimilarity, which is different from the average pairwise dissimilarity between sites (Fig. 2, red lines).

## Ordination approaches

As evoCA provided close results as $\text{evoPCA}_{\text{Hellinger}}$ and evoNSCA very similar results as $\text{evoPCA}_{\text{Chord}}$, I present below only the results obtained with $\text{evoPCA}_{\text{Chord}}$ and $\text{evoPCA}_{\text{Hellinger}}$ (but see Appendix 3 for evoCA and evoNSCA). $\text{EvoPCA}_{\text{Hellinger}}$ and $\text{evoPCA}_{\text{Chord}}$ are similar when presence/absence data are used. The axes of these analyses represented 39%, 37%, and 24% of the phylogenetic dissimilarities among habitats. The rainforest, cacao plantations, oldfields, cornfields were almost equidistant so that the three axes were necessary to display the dissimilarities among them (Fig. 4). The fact that *Thyroptera tricolor*, which descends from a long branch in the phylogenetic tree (Fig. 1), was observed only in the rainforest and in the cacao plantations contributed to make these habitats phylogenetically different from oldfields and cornfields. The terminal branch supporting *Micronycteris brachyotis* also, but to a lesser extent, contributed to the distinctiveness of the rainforest, whereas those supporting *M. megalotis*, *Myotis keaysi*, and to a lesser extent *Tonatia evotis* also contributed to the distinctiveness of the cacao plantations. The distinctiveness of the oldfields was mostly driven by the long terminal branch that supports *Mormoops megalophylla*. The cornfields were mostly characterized by the absence of a large number of species (including the descendants from node #30, Fig. 1); but it was also characterized by the clade descending from node #17 (Fig. 1) only found in this habitat.

When abundance data were used, cornfields were characterized by the high relative abundance of *S. lilium*, oldfields by higher relative abundances of species supported by node numbered #19 on Figure 1 (including genus *Carollia*) and more specifically by species *C. brevicauda* (Fig. 5), and the rainforest either by the high relative abundance of *Pteronotus parnellii* (according to $\text{evoPCA}_{\text{Hellinger}}$, Fig. 5a), or by the high relative abundances of species supported by node numbered #21 (all species from subfamily Stenodermatinae except *S. lilium*), #24 (species from genus *Artibeus* within subfamily Stenodermatinae) and #26 (species *Artibeus lituratus* and *A. jamaicensis*) (according to $\text{evoPCA}_{\text{Chord}}$, Fig. 5b). When considering abundance data, the terminal branch from which *S. lilium* descends had the highest contribution to the phylogenetic differences among habitats.

# Discussion

PD-dissimilarity indices are the set of indices obtained by replacing species with evolutionary units in the traditional compositional dissimilarity indices. As a consequence, all the results developed on PD-dissimilarity indices also apply to species dissimilarity indices as the latter can simply be obtained considering a star phylogeny with unit branch lengths. In this special case, each species contributes an independent evolutionary unit. I first explore below the results yielded by PD-dissimilarity indices when comparing bat communities along the gradient of disturbance. Then I provide a guide on the pros and cons of these indices.

## Bat PD-dissimilarity along the disturbance gradient

Parametric indices of PD-dissimilarity revealed that cornfields were particularly distinct due to an absence of rare lineages, whereas the rainforest was distinct by differences in the abundance of widespread lineages. Ordination approaches revealed a set of evolutionary units (identified by the branches that sustain them and by the species that descend from them) that contribute to the PD-dissimilarities among habitats. With presence/absence data, the PD-dissimilarities among habitats were mostly driven by rare species, some of which are specialist known to be sensitive to disturbance including *Thyroptera tricolor* (Chaverri and Kunz 2011), and the Phyllostomines (species descending from node #7 in Fig. 1, Fenton et al. 1992). When higher weights were given to the most abundant lineages, most of the species that permitted to distinguish the compositions of the four habitat types were descendants of node #18 in Fig. 1 that include the Carolliinae (descendents of node #19, mostly characterizing oldfields) and the Stenodermatinae (descendents of node #20; with the relative abundance of *Sturnina lilium* compared to the other species characterizing cornfields in opposition to the rainforest). Although observed in the four habitats, *S. lilium*, a generalist species (Medellín et al. 2000), was very abundant in cornfields compared to the other habitat types. Its importance in cornfields was also increased by the fact that other species were missing in this habitat type. In contrast, Carolliinae and other Stenodermatinae species are more specialized. For example *Carollia brevicauda* abundant in cacao plantations and oldfields is an understory species that feeds on successional trees and tall shrubs. *Artibeus lituratus* the most abundant species in rainforest is a canopy tree specialist feeding on figs and fruits (Medellín et al. 2000). These oppositions among related species may reflect the very diverse traits and ecological strategies exhibited by bat species even among those that are related (e.g. Stevens et al. 2003). The low contribution of vampire bats to the analysis of this study area is related to the fact that disturbance was not associated with cattle on which vampire bats feed (Medellín et al. 2000).

For low values of $q$, the phylogenetic patterns revealed by the parametric indices generally agreed. However when $q$ increased they depended on the index used. Variations in the indices related to Chiu et al. (2014) framework on Hill numbers (${}^{q}evoD_{Rényi}$, $1-\bar{V}_{q2}$ , $1-\bar{C}_{q2}$ and $1-\bar{U}_{q2}$) depended on whether relative or absolute abundances of the evolutionary units in habitats were used. With absolute abundances ${}^{q}evoD_{Rényi}$ (and $1-\bar{V}_{q2}$) first decreased and then increased with increasing $q$. This would mean that the habitats mostly differed in the absolute abundance of dominant evolutionary units (i.e. typically clades, genera, descending from node #18, subfamily Stenodermatinae) and in the presence of rare evolutionary units (like those on the branch that supports *Thyroptera tricolor*). However in contrast, with $1-\bar{U}_{q2}$ (with both relative and absolute abundances), ${}^{q}evoD_{Rényi}$ and $1-\bar{V}_{q2}$ (with relative abundances), differences among habitats mostly vanished with increasing $q$, indicating that the commonest evolutionary units are shared by all habitats.

Although Chiu et al. (2014) indices ranked pairs of habitats similarly in order of their PD-dissimilarity, $1-\bar{C}_{q2}$ and $1-\bar{U}_{q2}$ provided opposed results in terms of how high they estimated the PD-dissimilarity to be: $1-\bar{U}_{q2}$ continuously decreased with $q$ whatever the habitats compared; $1-\bar{C}_{q2}$, instead, increased towards maximum dissimilarity with absolute abundances for all pairs of habitats except when comparing cacao plantations and oldfields. Considering low values for $q$ as recommended by Chiu et al. (2014) would not have permitted to detect the complete opposition between $1-\bar{C}_{q2}$ and $1-\bar{U}_{q2}$. Chiu et al. (2014) indices complement each other and provide different viewpoints reflecting the effective proportion of unshared lineages in each site ($1-\bar{C}_{q2}$) vs. in the pooled sites ($1-\bar{U}_{q2}$). When $q$ increases, parametric indices are more and more impacted by the most abundant evolutionary units. Given the shape of the phylogenetic tree and the rarity, in this case study, of the three phylogenetically distinct species, *Thyroptera tricolor*, *Bauerus dubiaquercus*, and *Myotis keaysi*, the most abundant evolutionary units always are those supported by the branch that connects node #1 to node #2 in Figure 1. These evolutionary units are shared by all habitats which explains why $1-\bar{U}_{q2}$ tends to zero when *q* increases. The fact that $1-\bar{C}_{q2}$ increases towards unity with absolute abundances means that the effective proportion of unshared lineages in each site reaches a maximum. Given that the most abundant evolutionary units in each site are shared by the other sites, $1-\bar{C}_{q2}$, applied to absolute abundances, might here thus be impacted by differences among habitats in the total abundance in evolutionary units, which are partly impacted by sampling bias. Indeed, sampling effort in oldfields was twice as much as in the other habitats because oldfields were first divided into young abandoned fields (8-12 years since abandonment) and old abandoned fields (>12 years) (Medellín et al. 2000). The lowest difference in total abundance was between oldfields and cacao plantations.

Overall, the bat case study illustrates that the transposition of species dissimilarity to PD-dissimilarity is indeed possible and meaningful as it correctly acknowledges that species are not equivalent having experienced different histories. However it also demonstrates that the choice of an index can determine the conclusions of a study. The different indices thus depict different aspects, or facets, of PD-dissimilarity.

## Developing PD-dissimilarity: guidelines through the indices

I have only provided examples of species dissimilarity indices that can be used to develop PD-dissimilarity indices. There are myriads of indices developed in the literature and all can be adapted to the PD-dissimilarity framework (see Supplementary material Appendix 2). Ecologists are thus faced with the enormous amount of available indices. Finding a single (consensual) summary statistic to measure biological dissimilarity (biodissimilarity) is appealing for most ecologists because it would make comparisons among studies possible (e.g. Hubálek 2000). Indeed cross-study comparison with different indices could be misleading (Bloom 1981, Koleff et al. 2003). Unfortunately, finding a consensual index is not that simple, not to say that this is utopic. The choice of the statistic is dependent on the objective of the study and on the kind of data available (Wolda 1981, Jackson et al. 1989). For example, biogeographic and conservation studies, dealing with broad spatial scales, will often be more interested in presence/absence data, as their broad scale renders measures of abundance data prohibitively expensive, whereas local analyses of phylogenetic community ecology will be often more interested in abundance data. Apart from the main constraint driven by data availability, the choice of a measure is often largely subjective and based on tradition, or on a posteriori criteria such as the interpretability of the results (Jackson et al. 1989).

The results of an analysis using an index do not necessarily reflect all the information contained in the ecological data matrices. This stresses the importance of choosing an appropriate index as this choice determines the issue of the analysis (Hubálek 1982, Legendre and Legendre 1998) and of providing explicit justifications for the use of a given index or of a set of indices (Bloom 1981). Once an index has been chosen, care should be taken in interpreting and communicating the results of the index. For example, while a variety of dissimilarity indices vary between the same limits, typically 0 and 1, they do not give comparable values for the same amount of actual (dis)similarity. This renders automatic interpretation, like very low, low, moderate, high, and very high (dis)similarity based on a 0.2 interval impossible (Bloom 1981). The indices have thus to be fully understood to ensure correct interpretation.

To avoid the subjective choice of an index, statisticians and numerical ecologists have tested the performance of the indices of compositional (species) dissimilarity using objective criteria, such as the dependence of dissimilarity indices on sample size. However, the many theoretical and empirical comparative tests published so far have led to a variety of contradictory recommendations. Many such studies have stuck to the conclusion that the main determinant for this choice is the hypothesis to be tested but the region and taxa analysed can also influence this choice (Wolda 1981). For example, Hubálek (1982) run theoretical and empirical tests for presence/absence-based similarity indices. He drew a list of necessary and optional criteria that such indices should fulfil. Despite that he also concluded that for various kinds of studies different similarity indices could be optimal, that for certain purposes other admissible coefficients might work better, and that, under some circumstances, even indices incompatible with the identified, required criteria can yield very good results. For those that wanted to be guided towards a single powerful index, this conclusion is obviously disappointing. However it reflects a reality: biodissimilarity requires a diversity of indices for all its aspects to be efficiently analysed.

Inconsistencies among indices are inevitable whenever one attempts to reduce a multidimensional concept into a single number (Patil and Tallie 1979). Different coefficients emphasize different aspects of community data. Comparing and contrasting their results can thus yield important ecological insights on the actual patterns of dissimilarity among sites (Anderson et al. 2011). At the same time, this plurality of dissimilarity indices might be perceived as frightening so that regular attempts are made to provide clues for differentiating among the indices and eventually classify them (e.g. Legendre and Legendre 1998, Legendre and De Cáceres 2013). For example, according to Legendre and De Cáceres (2013) basic necessary properties for dissimilarity indices, the use of $evoD_{Profile}$ and $evoD_{\chi^2}$ should not be recommended.

As underlined above, a key criterion for choosing a PD-dissimilarity index is its sensitivity to species' abundances. PD-dissimilarity patterns among bat communities in the Selva

Lacandona of Chiapas, Mexico, depended on whether presence/absence or abundance data were used. PD-dissimilarities among habitats with presence/absence data might be driven by the occurrence of rare species that are highly sensitive to disturbance (as Medellin et al. 2000 suggested) or they might result from random sampling effects. Indices using presence/absence data cannot be estimated correctly (in terms of bias and variance) when sampling is incomplete and the probabilities of detection of missing species are unknown (e.g. Chao et al. 2006). Overall the more an index relies on rare species the more it is likely to be impacted by sampling artefacts. Inversely, in case of skewed distributions of species abundances, most of abundance-based indices would be impacted by a few very abundant species only. Using an index that self normalizes data, might permit to reveal more subtle ecological patterns (e.g. logarithmic or square-root transformations of abundances, Legendre and Legendre 1998). For example, in $evoD_{ScaledCanberra}$ and $evoD_{Divergence}$, a difference between abundant evolutionary units contributes less to the PD-dissimilarity than the same difference between rarer evolutionary units. On the contrary, $evoD_{Bray\text{-}Curtis}$ weights evenly differences among rare and abundant evolutionary units; and $evoD_{Morisita\text{-}Horn}$ gives more weight to differences among abundant evolutionary units (Supplementary material Appendix 1 for formulas). The choice for an index based on its sensitivity to rare vs. abundant evolutionary units depends on the objective of the study. Alternatively, parametric indices can be used to control the relative importance given to rare vs. abundant evolutionary units.

As for compositional dissimilarity indices, the degree of dependence of PD-dissimilarity indices on the diversity or on the size of the compared communities may also be a key criterion of choice (Wolda 1981, Jackson et al. 1989). Among the PD-dissimilarity indices introduced above, several indices normalize data by communities' size (=sum of abundances of the evolutionary units; e.g. $evoD_{Hellinger}$). In contrast, $evoD_{Chord}$ normalizes the vector containing the abundances of evolutionary units in a site by its norm (square root of the sum of squared abundances), which measures the degree of skewness in the abundances of evolutionary units and is thus inversely related to an aspect of evolutionary diversity (evodiversity). Another criterion for choosing a PD-dissimilarity index is whether the joint absence of an evolutionary unit in two sites should be considered as a sign of evosimilarity among the sites. Legendre and Legendre (1998) warned against the use of joint absences in compositional (species) dissimilarity indices. In particular $evoD_{\chi^2}$ and $evoD_{Profile}$, and their associated ordination approaches, evoCA and evoNSCA, respectively, are sensitive to joint absences. The presence of a lineage in a site, and especially its abundance, might reveal how close the species from the lineage are from optimal conditions. In contrast, its absence might be due to many different processes and does not necessarily reflect environmental similarities or geographic proximity among sites. On the other hand, joint absences may be relevant information to include when focusing on phenomena that can cause the absence of a lineage from a site, such as disturbance (as in the bat case study), or predation (Anderson et al. 2011). In experimental studies, if several plots are initiated with the same composition and then submitted to different environmental conditions, the joint absence of some species after a period of time might be informative on the congruent impact of different environmental conditions on species, and lineages.

## Conclusions

Measures of PD-dissimilarity generalize species dissimilarity to include phylogenetic information. As such, they have the potential to complement measures of functional dissimilarity to decipher historical and biogeographic processes from ecological and stochastic processes. Profiles of PD-dissimilarity, where the relative importance given to rare opposed to abundant evolutionary units is controlled, have been understudied in the ecological literature and could be the focus of new research. The insight they could provide in comparison to using a single formula has still to be investigated. The divergence in the conclusions of comparative studies of dissimilarity indices shows that more research is needed to compare and classify indices and eventually to help ecologists choose the right indices for each of their analyses. To avoid the arbitrary influence of a few programs that render the use of a few indices freely available, the collaborative development of software should continue. To avoid subjective choices, further research studies are needed to analyse mathematical properties and biological interest of the indices. Then it would be easier to accept that a variety of indices is needed to answer a variety of ecological and evolutionary questions about biodiversity.

## References

Agnarsson, I. et al. 2011. A time-calibrated species-level phylogeny of bats (Chiroptera, Mammalia). – PloS Currents 3: RRN1212.

Anderson, M. J. et al. 2011. Navigating the multiple meanings of β diversity: a roadmap for the practicing ecologist. – Ecol. Lett. 14: 19–28.

Baselga, A. 2013. Multiple site dissimilarity quantifies compositional heterogeneity among several sites, while average pairwise dissimilarity may be misleading. Ecography 36: 124–128.

Bloom, S. A. 1981. Similarity indices in community studies: potential pitfalls. – Mar. Ecol. Progr. Ser. 5: 125–128.

Borcard, D. et al. 2011. Numerical ecology with R. Springer.

Bray, R. J. and Curtis, J. T. 1957. An ordination of the upland forest communities of southern

Wisconsin. – Ecol. Monogr. 27: 325–349.

Bryant, J. A. et al. 2008. Microbes on mountainsides: contrasting elevational patterns of bacterial and plant diversity. – Proc. Natl. Acad. Sci. USA 105: 11505–11511.

Chao, A. et al. 2006. Abundance-based similarity and their estimation when there are unseen species in samples. – Biometrics 62: 361–371.

Chao, A. et al. 2010. Phylogenetic diversity measures based on Hill numbers. – Philos. Trans. R. Soc. B-Biol. Sci. 365: 3599–3609.

Chaverri, G. and Kunz, T. H. 2011. Response of a bat specialist to the loss of a critical resource. – PloS ONE 6: e28821.

Chiu, C.-H. et al. 2014. Phylogenetic beta diversity, similarity, and differentiation measures based on Hill numbers. – Ecol. Monogr. 84: 21–44.

Clark, P. J. 1952. An extension of the coefficient of divergence for use with multiple characters. – Copeia 1952: 61–64.

Faith, D. P. 1992. Conservation evaluation and phylogenetic diversity. – Biol. Conserv. 61: 1–10.

Faith, D. P. 2013. Biodiversity and evolutionary history: useful extensions of the PD phylogenetic diversity assessment framework. – Ann. N. Y. Acad. Sci. 1289: 69–89.

Faith, D. P. and Richards, Z. T. 2012. Climate change impacts on the tree of life: changes in phylogenetic diversity illustrated for Acropora corals. – Biology 1: 906–932.

Faith, D. P. et al. 2009. The cladistic basis for the phylogenetic diversity (PD) measure links evolutionary features to environmental gradients and supports broad applications of microbial ecology's "phylogenetic beta diversity" Framework. – Int. J. Mol. Sci. 10: 4723–4741.

Fenton, M. B. et al. 1992. Phyllostomid bats (Chiroptera: Phyllostomidae) as indicators of habitat disruption in the Neotropics. – Biotropica 24: 440–446.

Ferrier, S. et al. 2007. Using generalized dissimilarity modelling to analyse and predict patterns of beta diversity in regional biodiversity assessment. – Divers. Distrib. 13: 252–264.

Findley, J. S. 1976. The structure of bat communities. – Am. Nat. 110:129–139.

Fritz, S. A. et al. 2009. Geographic variation in predictors of mammalian extinction risk: big is bad, but only in the tropics. – Ecol. Lett. 12: 538–549.

Gower, J. C. and Legendre, P. 1986. Metric and Euclidean properties of dissimilarity coefficients. – J. Classif. 3: 5–48.

Horn, H. S. 1966. Measurement of "overlap" in comparative ecological studies. – Am. Nat. 100: 419–424.

Hubálek, Z. 1982. Coefficients of association and similarity, based on binary (presence-absence) data: an evaluation. – Biol. Rev. 57: 669–689.

Hubálek, Z. 2000. Measures of species diversity in ecology: an evaluation. – Folia Zool. 9: 241–260.

Izsák, J. and Papp, L. 1995. Application of the quadratic entropy indices for diversity studies of drosophilid assemblages. – Environ. Ecol. Stat. 2: 213–224.

Jaccard, P. 1901. Étude comparative de la distribution florale dans une portion des Alpes et des Jura. – Bulletin de la Société Vaudoise des Sciences Naturelles 37: 547–579.

Jackson, D. A. et al. 1989. Similarity coefficients: measures of co-occurrence and association or simply measures of occurrence? – Am. Nat. 133: 436–453.

Koleff, P. et al. 2003. Measuring beta diversity for presence-absence data. – J. Anim. Ecol. 72: 367–382.

Kroonenberg, P. M. and Lombardo, R. 1999. Nonsymmetric correspondence analysis: a tool for analysing contingency tables with a dependence structure. – Multivar. Behav. Res. 34: 367–396.

Lance, G. N. and Williams, W. T. 1966. Computer programs for classification. – Proc. ANCCAC Conference, paper 12/3.

Lebart, L. et al. 1984. Multivariate descriptive analysis: correspondence and related techniques for large matrices. – John Wiley and Sons.

Legendre, P. and Birks, H. J. B. 2012. From classical to canonical ordination. Chapter 8. – In: Birks et al. (eds) Tracking environmental change using lake sediments, volume 5: data handling and numerical techniques. – Springer, pp. 201-248.

Legendre, P. and De Cáceres, M. 2013. Beta diversity as the variance of community data: dissimilarity coefficients and partitioning. – Ecol. Lett. 16: 951–963.

Legendre, P. and Gallagher, E. D. 2001. Ecologically meaningful transformations for ordination of species data. – Oecologia 129: 271–280.

Legendre, P. and Legendre, L. 1998. Numerical ecology. 2nd English ed. – Elsevier Science BV.

Lozupone, C. A. and Knight, R. 2005. UniFrac: a new phylogenetic method for comparing microbial communities. – Appl. Environ. Microb. 71: 8228–8235.

Medellín, R. et al. 2000. Bat diversity and abundance as indicators of disturbance in Neotropical rainforest. – Conserv. Biol. 14: 1666–1675.

Nipperess, D. A. et al. 2010. Resemblance in phylogenetic diversity among ecological assemblages. – J. Veg. Sci. 21: 809–820.

Ochiai, A. 1957. Zoogeographic studies on the soleoid fishes found in Japan and its neighbouring regions. – Bull. Japan Soc. Sci. Fisheries 22: 526–530.

Orloci, L. 1967. An agglomerative method for classification of plant communities. – J. Ecol. 55: 193–206.

Patil, G. P. and Taillie, C. 1979. An overview of diversity. – In: Grassle, J. F. et al. (eds), Ecological diversity in theory and practice. – International Cooperative Publishing House,

pp. 3-27.

Petchey, O. L. and Gaston, K. 2002. Functional diversity (FD), species richness and community composition. – Ecol. Lett. 5: 402–411.

Pillar, V. D. and Duarte, L. D. S. 2010. A framework for metacommunity analysis of phylogenetic structure. – Ecology Letters 13: 587–596.

R Development Core Team. 2014. R: a Language and environment for statistical computing. – R Foundation for Statistical Computing.

Rao, C. R. 1995. A review of canonical coordinates and an alternative to correspondence analysis using Hellinger distance. – Qüestiió 19: 23–63.

Rényi, A. 1961. On measures of entropy and information. – In: Neyman, J. (ed.), Berkeley symposium on mathematical statistics and probability, pp. 547-561.

Sørensen, T. 1948. A method of establishing groups of equal amplitude in plant sociology based on similarity of species content. – K. dan Vidensk Selsk Biol. Skr. 5: 1–34.

Stevens, R. D. et al. 2003. Patterns of functional diversity across an extensive environmental gradient: vertebrate consumers, hidden treatments and latitudinal trends. – Ecol. Lett. 6: 1099–1108.

Warwick, R. M. and Clarke, K. R. 1995. New 'biodiversity' measures reveal a decrease in taxonomic distinctness with increasing stress. – Mar. Ecol. Prog. Ser. 129: 301–305.

Wolda, H. 1981. Similarity indices, sample size and diversity. – Oecologia 50: 296–302.

**Figures**

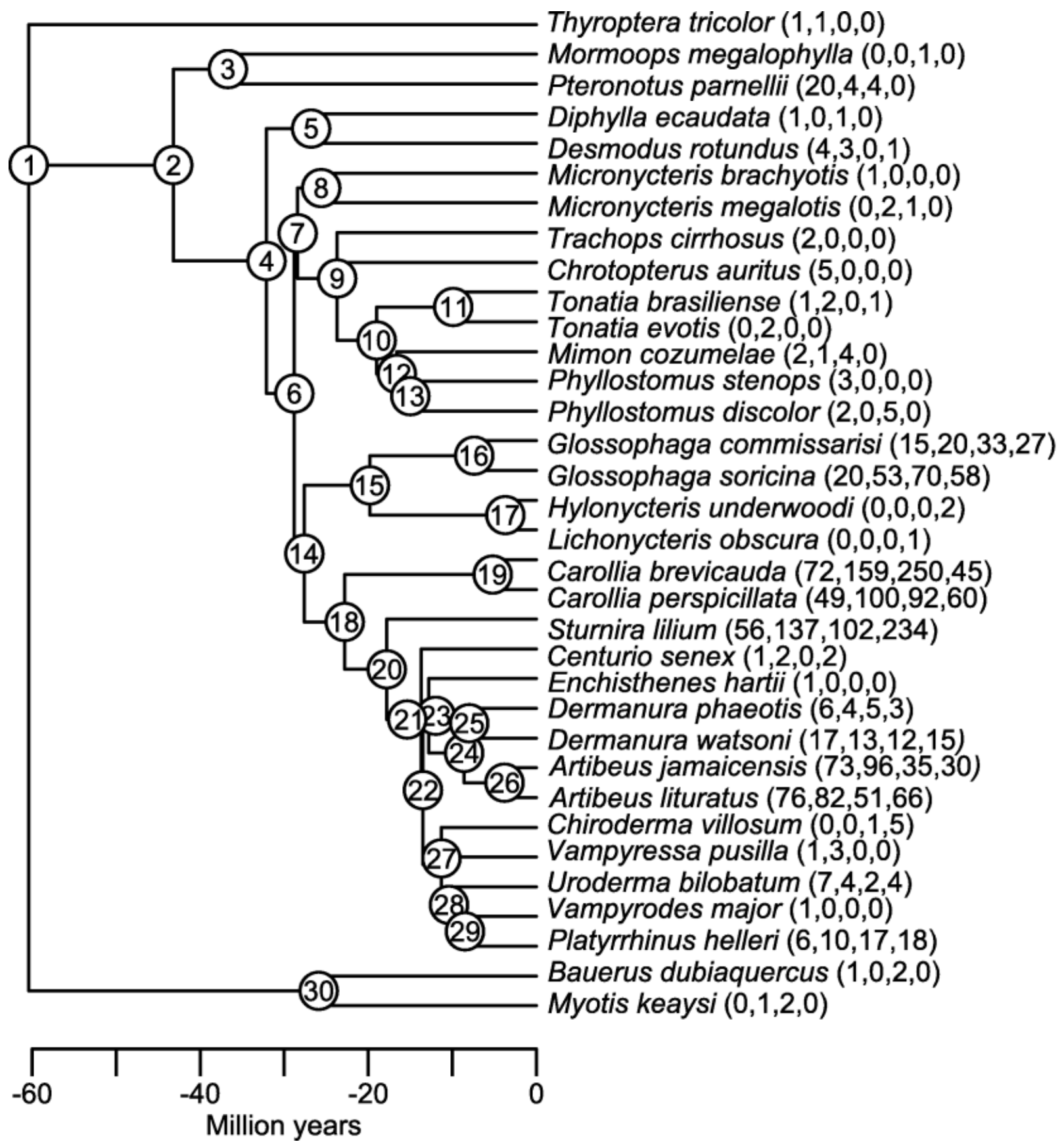

Figure 1. Dated phylogenetic tree for bat species observed in the Selva Lacandona of Chiapas, Mexico by Medellín et al. (2000). The values provided in brackets are the numbers of captures of bat species in the rainforest, cacao plantations, oldfields and cornfields in that order.

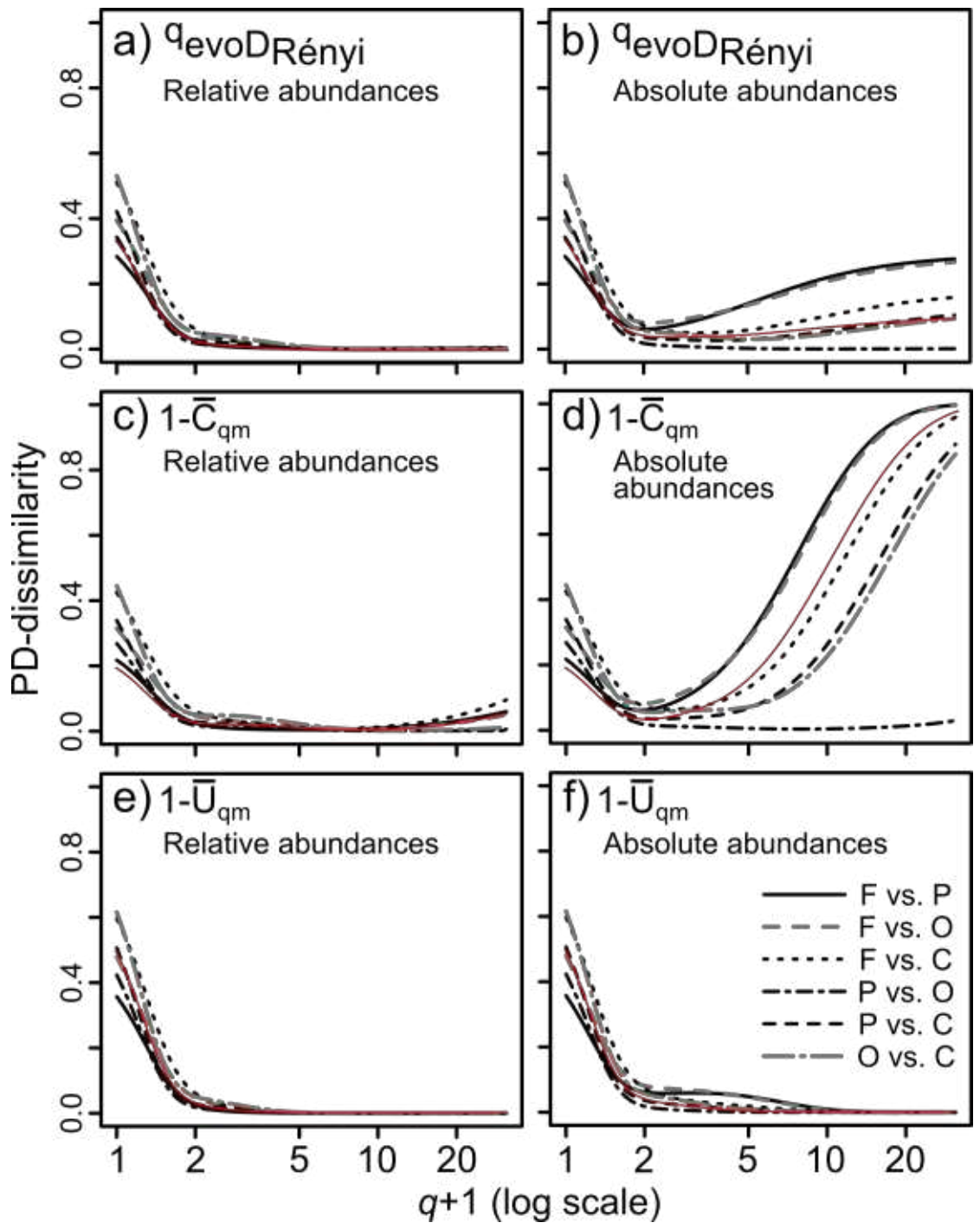

Figure 2. Profiles of pairwise PD-dissimilarities among habitats: a) ${}^{q}evoD_{Rényi}$ with relative abundances for the evolutionary units, b) ${}^{q}evoD_{Rényi}$ with absolute abundances for the evolutionary units, c) $1-\bar{C}_{q2}$ with relative abundances, d) $1-\bar{C}_{q2}$ with absolute abundances, e) $1-\bar{U}_{q2}$ with relative abundances, f) $1-\bar{U}_{q2}$ with absolute abundances (see Supplementary material Appendix 1 for index equations). Codes for habitats are: F=rainforest; P=cacao plantations; O=oldfields; C=cornfields.

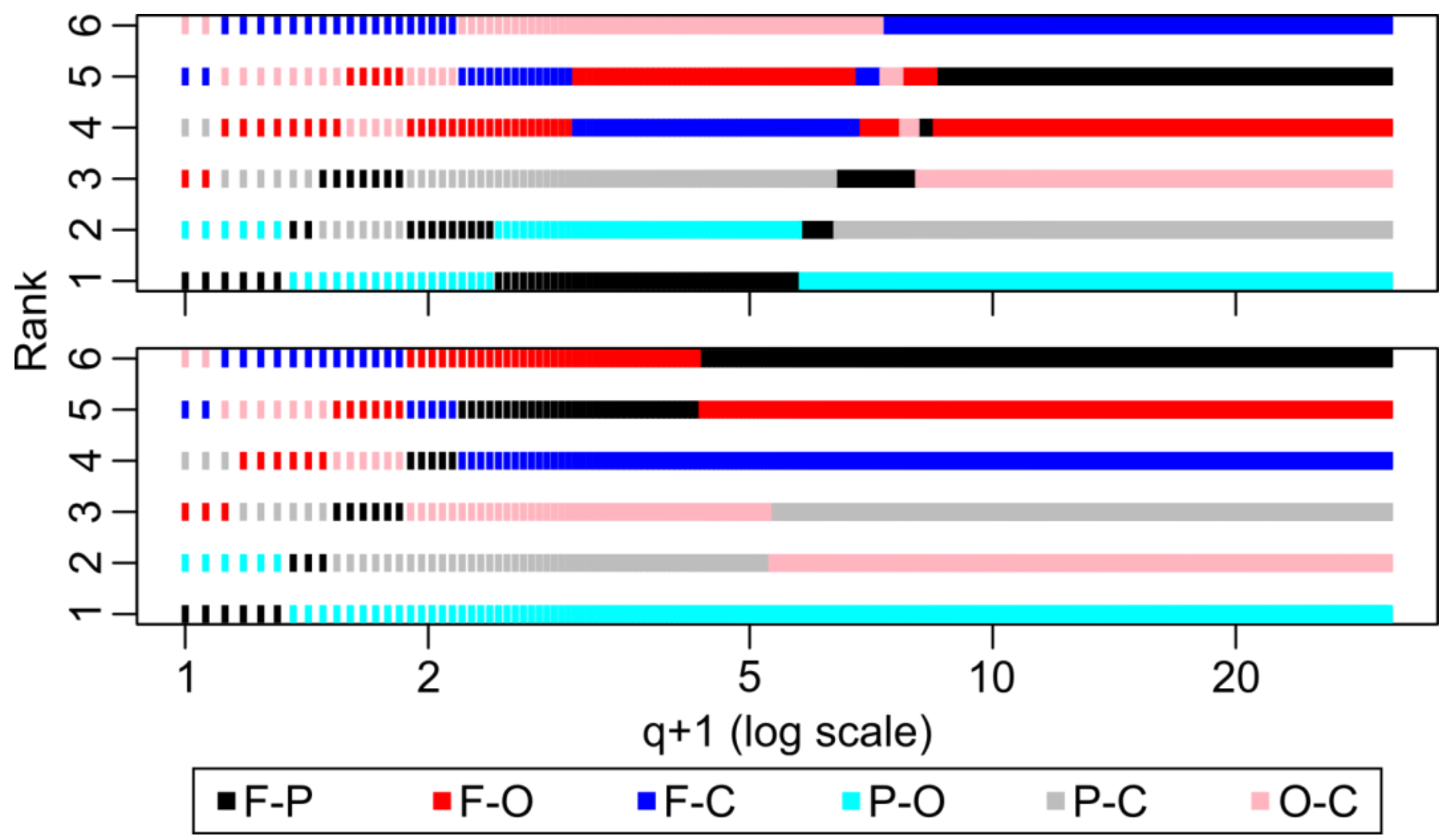

Figure 3. Ranks attributed to pairwise dissimilarities between habitats by indices ${}^{q}evoD_{Rényi}$, $1-\bar{V}_{q2}$, $1-\bar{C}_{q2}$, and $1-\bar{U}_{q2}$, calculated per value of $q$: with relative abundances (top panel); with absolute abundances (bottom panel). Codes for habitats are: F=rainforest; P=cacao plantations; O=oldfields; C=cornfields. F-P for example indicates that rainforest is compared to cacao plantations. Ranks are given in increasing order, with 1 meaning smallest observed dissimilarity and 6 highest dissimilarity.

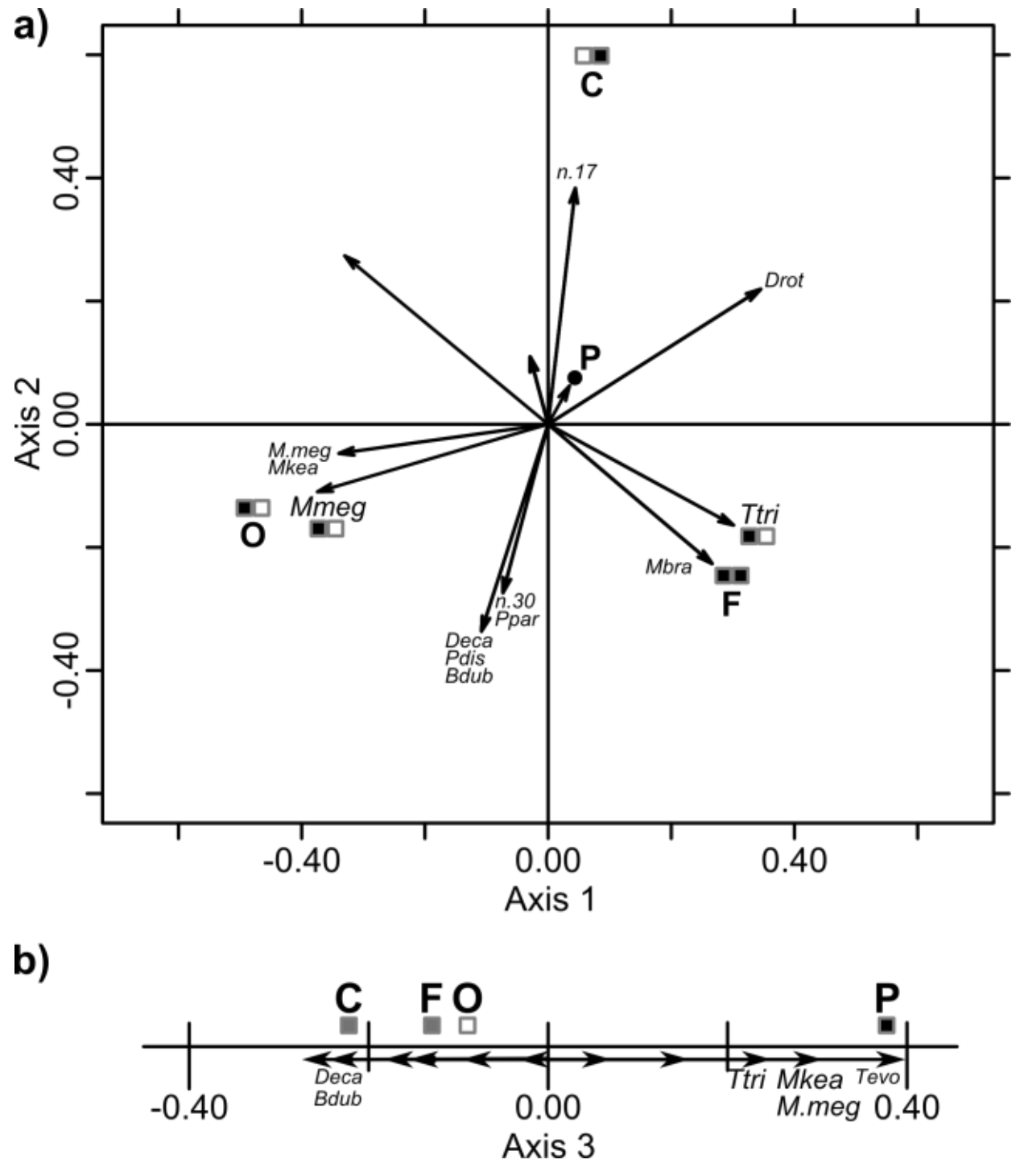

Figure 4. Factorial maps obtained with both $evoPCA_{Hellinger}$ and $evoPCA_{Chord}$ with the data expressed as presence/absence of the evolutionary units in each habitat: a) first and second axes; b) third axis. The three axes expressed 39%, 37%, and 24%, respectively, of the averaged squared phylogenetic distances among communities. Codes for habitats are: F=rainforest; P=cacao plantations; O=oldfields; C=cornfields. Branches with high contributions to the axes are indicated by using the name of their direct descendant node (or tip, that is to say species, for terminal branches). Codes for species are: Bdub (*Bauerus dubiaquercus*), Deca (*Diphylla ecaudata*), Drot (*Desmodus rotundus*), Mbra (*Micronycteris brachyotis*) and M.meg (*M. megalotis*), Mkea (*Myotis keaysi*), Mmeg (*Mormoops megalophylla*), Pdis (*Phyllostomus discolor*), Ppar (*Pteronotus parnelii*), Tevo (*Tonatia evotis*), Ttri (*Thyroptera tricolor*). Codes for internal nodes start with letter *n* and are followed by node numbers as indicated in Figure 1. Branches, so named, whose contributions are higher than 5% (but lower than 10%) to, at least, an axis are indicated with the smallest font. Branches whose contribution to, at least, an axis is higher than 10% are indicated with the largest font and their contributions are roughly evaluated by the colour of boxes: (in (a) left box for the first axis, right box for the second axis) white colour for a contribution lower than 10%, grey for a contribution higher than 10% but lower than 20%, and black for a contribution higher than 20%. The influence of each branch in determining the positions of the habitats is indicated by an arrow (whose direction and size are determined in proportion to the coordinates of the branches).

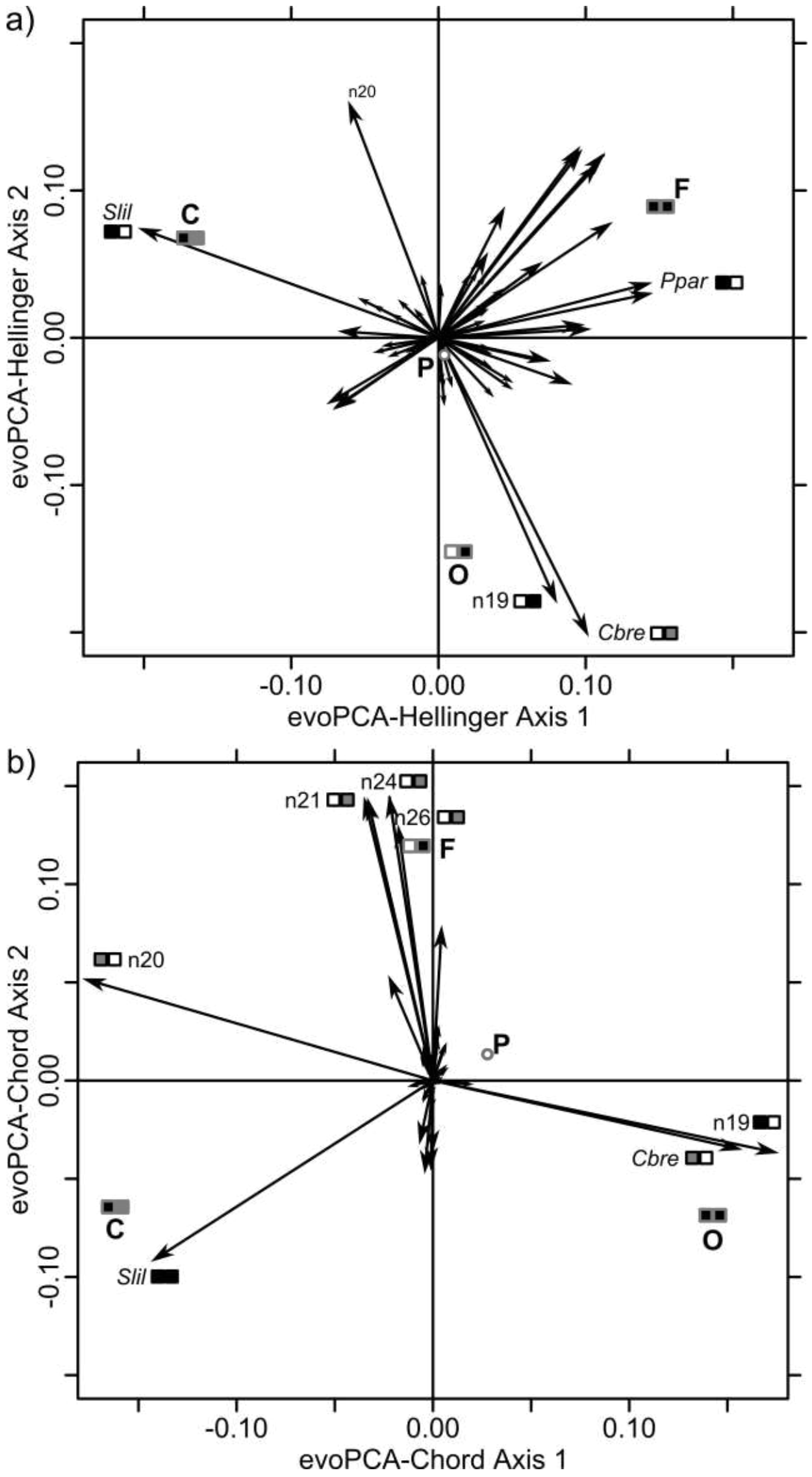

Figure 5. Factorial maps obtained with a) evoPCA$_{\text{Hellinger}}$, and b) evoPCA$_{\text{Chord}}$ with abundance data. The first two axes of each analysis were displayed. The first two axes of evoPCA$_{\text{Hellinger}}$ expressed 56% and 37%, respectively, of the averaged squared phylogenetic distances among habitats as measured by index $evoD_{Hellinger}$. The first two axes of evoPCA$_{\text{Chord}}$ expressed 66% and 33%, respectively, of the averaged squared phylogenetic distances among habitats as measured by index $evoD_{Chord}$. Codes for species are: Cbre (*Carollia brevicauda*), Ppar (*Pteronotus parnelii*), Slil (*Sturnira lilium*). Codes for habitats, branches, internal nodes, and for the contributions of habitats and branches to the analyses are given in the legend of Figure 3.

## Appendix 1. Formulas for all PD-dissimilarity indices used in this paper

The indices of PD-dissimilarity described in the main text are summarized in Table 1-1. I specify in this table whether the indices have Euclidean properties. The metric multidimensional scaling (mMDS) method best performs when the matrices of PD-dissimilarities are Euclidean, a property which ensures a perfect correspondence between the canonical distances among points in mMDS and PD-dissimilarities. Non-Euclidean distances can be analysed using non-metric multidimensional scaling (nMDS).

Note that, with presence/absence data, $1-\bar{V}_{02} = 1-\bar{C}_{02} = evoD_{Sørensen}$ , ${}^{0}evoD_{Rényi} = \log_2(evoD_{Sørensen} + 1)$, and $1-\bar{U}_{02} = evoD_{Jaccard}$. Different parametric indices of PD-dissimilarity might merge for some value of $q$ even if they have very different values for other values of $q$: for example, when sites are weighted by their relative abundance in evolutionary units, $evoD_{Morisita\text{-}Horn} = 1-\bar{C}_{22}$ and, with both relative and absolute abundances, $1-\bar{C}_{12} = 1-\bar{U}_{12} = {}^{1}evoD_{Rényi}$ .

TABLE 1-1. Examples of coefficients of site-to-site PD-dissimilarity.

| **References for coefficients of compositional (dis)similarity** | **Transformation used** | **Resulting formula of PD-dissimilarity** | **Euclidean** | **Squared Euclidean** |
|---|---|---|---|---|
| | | ***Incidence (presence/absence) data*** | | |
| $S_{Jaccard}$ (Jaccard 1901) | $(1-S_{Jaccard})$ | $evoD_{Jaccard}=1-\frac{\sum_{k\in b_T}L_k\delta_{1k}\delta_{2k}}{\sum_{k\in b_T}L_k\delta_{1k}+\sum_{k\in b_T}L_k\delta_{2k}-\sum_{k\in b_T}L_k\delta_{1k}\delta_{2k}}$ | No | Yes |
| $S_{Sørensen}$ (Sørensen 1948) | $(1-S_{Sørensen})$ | $evoD_{Sørensen}=1-\frac{2\sum_{k\in b_T}L_k\delta_{1k}\delta_{2k}}{\sum_{k\in b_T}L_k\delta_{1k}+\sum_{k\in b_T}L_k\delta_{2k}}$ | No | Yes |
| $S_{Ochiai}$ (Ochiai 1957) | $(1-S_{Ochiai})$ | $evoD_{Ochiai}=1-\frac{\sum_{k\in b_T}L_k\delta_{1k}\delta_{2k}}{\sqrt{\sum_{k\in b_T}L_k\delta_{1k}}\sqrt{\sum_{k\in b_T}L_k\delta_{2k}}}$ | No | Yes |
| | | ***Abundance data*** | | |
| | | *Example of indices derived from Tamás et al. (2001) parameters* | | |
| $S_{TJ}$: Tamás et al. (2001) parameters applied to $S_{Jaccard}$ | $(1-S_{TJ})$ | $evoD_{TJ}=\frac{\sum_{k\in b_T}L_k\left\|a_{1k}-a_{2k}\right\|}{\sum_{k\in b_T}L_k\max\{a_{1k},a_{2k}\}}$ | No | Yes |
| $S_{TS}$: Tamás et al. (2001) parameters | $(1-S_{TS})$ | $evoD_{TS}=evoD_{Bray-Curtis}=\frac{\sum_{k\in b_T}L_k\left\|a_{1k}-a_{2k}\right\|}{\sum_{k\in b_T}L_k\left(a_{1k}+a_{2k}\right)}$ | No | Yes |
| applied to $S_{Sørensen}$ | | | | |
| $S_{TO}$: Tamás et al. (2001) parameters applied to $S_{Ochiai}$ | $(1-S_{TO})$ | $evoD_{TO}=1-\frac{\sum_{k\in b_T} L_k \min\{a_{1k},a_{2k}\}}{\sqrt{\sum_{k\in b_T} L_k a_{1k}}\sqrt{\sum_{k\in b_T} L_k a_{2k}}}$ | No | Yes |
| *Example of indices derived from Minkowski's distance; raw abundances* | | | | |
| Minkowski distance | None | ${}^{q}evoD_{Minkowski}=\left[\sum_{k\in b_T} L_k \lvert a_{1k}-a_{2k}\rvert^{q}\right]^{1/q}$, $q>0$ | No | No |
| Manhattan distance | None | $evoD_{Manhattan}=\sum_{k\in b_T} L_k \lvert a_{1k}-a_{2k}\rvert$ | No | Yes |
| Euclidean distance | None | $evoD_{Euclidean}=\sqrt{\sum_{k\in b_T} L_k (a_{1k}-a_{2k})^2}$ | Yes | Yes |
| *Example of indices derived from Euclidean distance after data transformations*[b] | | | | |
| $D_{\text{Chord}}$: Chord distance (Orloci 1967) | None | $evoD_{Chord}=\sqrt{\sum_{k\in b_T} L_k\left(\frac{a_{1k}}{\sqrt{\sum_{k\in b_T} L_k a_{1k}^2}}-\frac{a_{2k}}{\sqrt{\sum_{k\in b_T} L_k a_{2k}^2}}\right)^2}$ | Yes | Yes |
| Hellinger distance (Rao 1995) | None | $evoD_{Hellinger}=\sqrt{\sum_{k\in b_T} L_k\left(\sqrt{\frac{a_{1k}}{\sum_{k\in b_T} L_k a_{1k}}}-\sqrt{\frac{a_{2k}}{\sum_{k\in b_T} L_k a_{2k}}}\right)^2}$ | Yes | Yes |
| Profiles distance (Legendre & Gallagher 2001) | None | $evoD_{Profile}=\sqrt{\sum_{k\in b_T} L_k\left(\frac{a_{1k}}{\sum_{k\in b_T} L_k a_{1k}}-\frac{a_{2k}}{\sum_{k\in b_T} L_k a_{2k}}\right)^2}$ | Yes | Yes |
| $\chi^2$ distance | None | $evoD_{\chi^2}=\sqrt{\sum_{k\in b_T} L_k \frac{\sum_{k\in b_T} L_k \sum_j a_{jk}}{\sum_j a_{jk}}\left(\frac{a_{1k}}{\sum_{k\in b_T} L_k a_{1k}}-\frac{a_{2k}}{\sum_{k\in b_T} L_k a_{2k}}\right)^2}$ | Yes | Yes |
| | | *Coefficient derived from Hill numbers* | | |
| ${}^qD_\beta$ (Chiu et al. 2014) applied with absolute abundances (see Appendix B) $q>0$, $q\neq 1$ | ${}^qD_\beta$-1 | $1-\bar{V}_{q2,abs}=2\left\{\frac{\sum_{k\in b_T} L_k\left(a_{1k}+a_{2k}\right)^q}{\sum_{k\in b_T} L_k\left[\left(a_{1k}\right)^q+\left(a_{2k}\right)^q\right]}\right\}^{\frac{1}{1-q}}-1,\ q\geq 0,\ q\neq 1$ | No | No |
| $1-\bar{V}_{02}$ with absolute | | $1-\bar{V}_{02,abs}=evoD_{Sørensen}$ | No | Yes |
| abundances | | | | |
| $1-\bar{V}_{12}$ with absolute abundances | | $1-\bar{V}_{12,abs}=2\exp\left[\begin{array}{l}\frac{1}{\sum_{k\in b_T}L_k\left(a_{1k}+a_{2k}\right)}\sum_{k\in b_T}L_k a_{1k}\ln\left(\frac{a_{1k}}{a_{1k}+a_{2k}}\right)\\ \frac{1}{\sum_{k\in b_T}L_k\left(a_{1k}+a_{2k}\right)}\sum_{k\in b_T}L_k a_{2k}\ln\left(\frac{a_{2k}}{a_{1k}+a_{2k}}\right)\end{array}\right]-1$ | No | No |
| ${}^qD_\beta$ (Chiu et al. 2014) applied with absolute abundances (see Supp. Mat. Appendix 2) $q>0$, $q\neq1$ | $\log_2({}^qD_\beta)$ | ${}^q evoD_{Rényi,abs}=1+\frac{1}{1-q}\log_2\left\{\frac{\sum_{k\in b_T}L_k\left(a_{1k}+a_{2k}\right)^q}{\sum_{k\in b_T}L_k\left[\left(a_{1k}\right)^q+\left(a_{2k}\right)^q\right]}\right\},\ q\geq0,\ q\neq1$ | No | No |
| Version of ${}^q evoD_{Rényi,abs}$ with $q$=0 | | ${}^0 evoD_{Rényi,abs}=\log_2\left\{evoD_{Sørensen}+1\right\}$ | No | No |

| | | | |
|---|---|---|---|
| ${}^{q}evoD_{Rényi,abs}$ with $q\to1$ | $${}^{1}evoD_{Rényi,abs}=1+\frac{1}{\sum_{k\in b_T}L_k\left(a_{1k}+a_{2k}\right)}\sum_{k\in b_T}L_k a_{1k}\log_2\left(\frac{a_{1k}}{a_{1k}+a_{2k}}\right)+\frac{1}{\sum_{k\in b_T}L_k\left(a_{1k}+a_{2k}\right)}\sum_{k\in b_T}L_k a_{2k}\log_2\left(\frac{a_{2k}}{a_{1k}+a_{2k}}\right)$$ | No | Yes* |
| $1-\bar{C}_{q2}$ (Chiu et al. 2014) applied with absolute abundances | $$1-\bar{C}_{q2,abs}=\frac{1}{1-2^{1-q}}\left\{1-2^{1-q}\frac{\sum_{k\in b_T}L_k\left(a_{1k}+a_{2k}\right)^q}{\sum_{k\in b_T}L_k\left[\left(a_{1k}\right)^q+\left(a_{2k}\right)^q\right]}\right\}$$ | No | No |
| $1-\bar{C}_{02,abs}$ (with $q$=0) | $$1-\bar{C}_{02,abs}={}^{0}evoD_{Hill,abs}=evoD_{Sørensen}$$ | No | Yes |
| Version of $1-\bar{C}_{q2,abs}$ with $q\to1$ | $$1-\bar{C}_{12,abs}={}^{1}evoD_{Rényi,abs}$$ | No | Yes* |
| $1-\bar{U}_{q2}$ (Chiu et al. 2014) applied with absolute abundances | $$1-\bar{U}_{q2,abs}=\frac{1}{1-2^{q-1}}\left\{1-2^{q-1}\frac{\sum_{k\in b_T}L_k\left[\left(a_{1k}\right)^q+\left(a_{2k}\right)^q\right]}{\sum_{k\in b_T}L_k\left(a_{1k}+a_{2k}\right)^q}\right\}$$ | No | No |

| | | | | |
|---|---|---|---|---|
| $1-\bar{U}_{02,abs}$ (with $q$=0) | | $1-\bar{U}_{02,abs}=evoD_{Jaccard}$ | No | Yes |
| Version of $1-\bar{U}_{q2,abs}$ with $q\rightarrow 1$ | | $1-\bar{U}_{12,abs}=1-\bar{C}_{12,abs}$ | No | Yes* |
| ${}^{q}D_{\beta}$ (Chiu et al. 2014) applied with relative abundances (see Appendix B) $q>0$, $q\neq 1$ | ${}^{q}D_{\beta}$-1 | $1-\bar{V}_{q2,rel}=2\left\{\frac{\sum_{k\in b_T}L_k\left(\frac{a_{1k}}{\sum_{k\in b_T}L_k a_{1k}}+\frac{a_{2k}}{\sum_{k\in b_T}L_k a_{2k}}\right)^{q}}{\sum_{k\in b_T}L_k\left[\left(\frac{a_{1k}}{\sum_{k\in b_T}L_k a_{1k}}\right)^{q}+\left(\frac{a_{2k}}{\sum_{k\in b_T}L_k a_{2k}}\right)^{q}\right]}\right\}^{\frac{1}{1-q}}-1$ | No | No |
| $1-\bar{V}_{02}$ with relative abundances | | $1-\bar{V}_{02,rel}=evoD_{Sørensen}$ | No | Yes |
| $1-\bar{V}_{12}$ with relative abundances | | $1-\bar{V}_{12,rel}=2\exp\left[\frac{1}{2}\sum_{k\in b_T}L_k\left(\frac{a_{1k}}{\sum_{k\in b_T}L_k a_{1k}}\right)\ln\left(\frac{\frac{a_{1k}}{\sum_{k\in b_T}L_k a_{1k}}}{\frac{a_{1k}}{\sum_{k\in b_T}L_k a_{1k}}+\frac{a_{2k}}{\sum_{k\in b_T}L_k a_{2k}}}\right)+\frac{1}{2}\sum_{k\in b_T}L_k\left(\frac{a_{2k}}{\sum_{k\in b_T}L_k a_{2k}}\right)\ln\left(\frac{\frac{a_{2k}}{\sum_{k\in b_T}L_k a_{2k}}}{\frac{a_{1k}}{\sum_{k\in b_T}L_k a_{1k}}+\frac{a_{2k}}{\sum_{k\in b_T}L_k a_{2k}}}\right)\right]-1$ | No | No |
| $^qD_\beta$ (Chiu et al. 2014) applied with relative abundances (see Appendix B) $q>0$, $q\neq1$ | $\log_2(^qD_\beta)$ | $^q evoD_{Rényi,rel}=1+\frac{1}{1-q}\log_2\left\{\frac{\sum_{k\in b_T}L_k\left(\frac{a_{1k}}{\sum_{k\in b_T}L_k a_{1k}}+\frac{a_{2k}}{\sum_{k\in b_T}L_k a_{2k}}\right)^q}{\sum_{k\in b_T}L_k\left[\left(\frac{a_{1k}}{\sum_{k\in b_T}L_k a_{1k}}\right)^q+\left(\frac{a_{2k}}{\sum_{k\in b_T}L_k a_{2k}}\right)^q\right]}\right\},\ q\geq0,\ q\neq1$ | No | No |
| Version of $^q evoD_{Rényi,rel}$ with $q=0$ | | $^0 evoD_{Rényi,rel}=\log_2\left\{evoD_{Sørensen}+1\right\}$ | No | No |

| | | | |
|---|---|---|---|
| ${}^{q}evoD_{Rényi,rel}$ with $q \to 1$ | $${}^{1}evoD_{Rényi,rel} = 1 + \frac{1}{2}\sum_{k \in b_T} L_k \left( \frac{a_{1k}}{\sum_{k \in b_T} L_k a_{1k}} \right) \log_2 \left( \frac{\frac{a_{1k}}{\sum_{k \in b_T} L_k a_{1k}}}{\frac{a_{1k}}{\sum_{k \in b_T} L_k a_{1k}} + \frac{a_{2k}}{\sum_{k \in b_T} L_k a_{2k}}} \right) + \frac{1}{2}\sum_{k \in b_T} L_k \left( \frac{a_{2k}}{\sum_{k \in b_T} L_k a_{2k}} \right) \log_2 \left( \frac{\frac{a_{2k}}{\sum_{k \in b_T} L_k a_{2k}}}{\frac{a_{1k}}{\sum_{k \in b_T} L_k a_{1k}} + \frac{a_{2k}}{\sum_{k \in b_T} L_k a_{2k}}} \right)$$ | No | Yes* |
| $1-\bar{C}_{q2}$ (Chiu et al. 2014) applied with relative abundances | $$1-\bar{C}_{q2,rel} = \frac{1}{1-2^{1-q}} \left\{ 1 - 2^{1-q} \frac{\sum_{k \in b_T} L_k \left( \frac{a_{1k}}{\sum_{k \in b_T} L_k a_{1k}} + \frac{a_{2k}}{\sum_{k \in b_T} L_k a_{2k}} \right)^q}{\sum_{k \in b_T} L_k \left[ \left( \frac{a_{1k}}{\sum_{k \in b_T} L_k a_{1k}} \right)^q + \left( \frac{a_{2k}}{\sum_{k \in b_T} L_k a_{2k}} \right)^q \right]} \right\}$$ | No | No |
| $1-\bar{C}_{02,rel}$ (with $q$=0) | $$1-\bar{C}_{02,rel} = {}^{0}evoD_{Hill,rel} = evoD_{Sørensen}$$ | No | Yes |
| Version of $1-\bar{C}_{q2,rel}$ with $q \to 1$ | $$1-\bar{C}_{12,rel} = {}^{1}evoD_{Rényi,rel}$$ | No | Yes* |

| | | | | |
|---|---|---|---|---|
| $1-\bar{U}_{q2}$ (Chiu et al. 2014) applied with absolute abundances | | $1-\bar{U}_{q2,rel}=\frac{1}{1-2^{q-1}}\left\{1-2^{q-1}\frac{\sum_{k\in b_T}L_k\left[\left(\frac{a_{1k}}{\sum_{k\in b_T}L_k a_{1k}}\right)^q+\left(\frac{a_{2k}}{\sum_{k\in b_T}L_k a_{2k}}\right)^q\right]}{\sum_{k\in b_T}L_k\left(\frac{a_{1k}}{\sum_{k\in b_T}L_k a_{1k}}+\frac{a_{2k}}{\sum_{k\in b_T}L_k a_{2k}}\right)^q}\right\}$ | No | No |
| $1-\bar{U}_{02,rel}$ (with $q$=0) | | $1-\bar{U}_{02,rel}=evoD_{Jaccard}$ | No | Yes |
| Version of $1-\bar{U}_{q2,rel}$ with $q\to1$ | | $1-\bar{U}_{12,rel}=1-\bar{C}_{12,rel}$ | No | Yes* |
| | | *Other families of PD-dissimilarity indices* | | |
| | | $\frac{\sum_{k\in b_T}L_k\left\|a_{1k}-a_{2k}\right\|^q}{\sum_{k\in b_T}L_k\left(a_{1k}^q+a_{2k}^q\right)}$, $q$>0 | No | No |
| $D_{Bray\text{-}Curtis}$: Bray and Curtis (1957) | None | $evoD_{Bray-Curtis}=\frac{\sum_{k\in b_T}L_k\left\|a_{1k}-a_{2k}\right\|}{\sum_{k\in b_T}L_k\left(a_{1k}+a_{2k}\right)}$ | No | Yes |
| $S_{MH}$: Morisita-Horn index (Morisita 1959; Horn 1966) | $1-S_{MH}$ | $evoD_{Morisita-Horn} = \frac{\sum_{k \in b_T} L_k (a_{1k} - a_{2k})^2}{\sum_{k \in b_T} L_k (a_{1k}^2 + a_{2k}^2)}$ | No | Yes |
| | | $\left[ \frac{1}{\sum_{k \in b_T} L_k} \sum_{k \in b_T} L_k \left[ \frac{\lvert a_{1k} - a_{2k} \rvert}{(a_{1k} + a_{2k})} \right]^q \right]^{1/q}$, $q$>0 | No | No |
| Scaled Canberra metric (Lance and Williams 1966) | None | $evoD_{ScaledCanberra} = \frac{1}{\sum_{k \in b_T} L_k} \sum_{k \in b_T} L_k \frac{\lvert a_{1k} - a_{2k} \rvert}{(a_{1k} + a_{2k})}$ | No | Yes |
| Coefficient of divergence (Clark 1952) | None | $evoD_{Divergence} = \sqrt{\frac{1}{\sum_{k \in b_T} L_k} \sum_{k \in b_T} L_k \left( \frac{a_{1k} - a_{2k}}{a_{1k} + a_{2k}} \right)^2}$ | Yes | Yes |

*Notes:* The Euclidean properties for parametric indices might depend on the parameter *q*. Squared Euclidean means that the square root of the dissimilarity coefficient has Euclidean properties. The application of the Bray-Curtis formula to measure phylogenetic dissimilarity was first introduced by Nipperess et al. (2010). In Canberra metric and the coefficient of divergence, the summation excludes *k* for which $a_{1k}$=$a_{2k}$=0. For the division by sum of units ($\sum_{k \in b_T} L_k$ ; scaling factor) in Canberra metric see Legendre and De Cáceres (2013) or Legendre and Legendre (1998) and references therein. Notations: Indices 1 and 2

refer to the two compared sites; *j* stands for any site in a larger set of sites; $b_T$ is the set of branches in a phylogenetic tree *T*; $L_k$ is the length of branch *k* in the phylogenetic tree; $\delta_{jk}$ denote the presence ($\delta_{jk}$=1) or absence ($\delta_{jk}$=0) of branch *k* in site *j*; $a_{jk}$ is the sum of the abundances, in site *j*, of all species descending from branch *k*. In the column "Euclidean", "Yes" means that the distance always has Euclidean properties; "No" means that the Euclidean properties depend on the data set used. Similarly, in the column "Squared Euclidean", "Yes" means that the square root of the distance always has Euclidean properties; "No" means that the Squared Euclidean properties depend on the data set used. In both columns "Euclidean" and "Squared Euclidean".

* Applications of the Jensen-Shannon distance

## References

Bray, R. J. and Curtis, J. T. 1957. An ordination of the upland forest communities of southern Wisconsin. – Ecol. Monogr. 27: 325–349.

Chiu, C.-H. et al. 2014. Phylogenetic beta diversity, similarity, and differentiation measures based on Hill numbers. – Ecol. Monogr. 84:21–44.

Clark P. J. 1952. An extension of the coefficient of divergence for use with multiple characters. – Copeia 1952:61–64.

Horn, H. S. 1966. Measurement of "overlap" in comparative ecological studies. – Am. Nat. 100: 419–424.

Jaccard, P. 1901. Étude comparative de la distribution florale dans une portion des Alpes et des Jura. – Bulletin de la Société Vaudoise des Sciences Naturelles 37: 547–579.

Lance, G. N. and Williams, W. T. 1966. Computer programs for classification. – Proc. ANCCAC Conference, paper 12/3.

Legendre, P. and De Cáceres, M. 2013. Beta diversity as the variance of community data: dissimilarity coefficients and partitioning. – Ecol. Lett. 16: 951–963.

Legendre, P. and Gallagher, E.D. 2001. Ecologically meaningful transformations for ordination of species data. – Oecologia 129: 271–280.

Legendre, P. and Legendre, L. 1998. Numerical ecology. 2nd English ed. – Elsevier Science BV.

Morisita, M. 1959. Measuring of interspecific association and similarity between communities. – Memoirs of the Faculty of Science, Kyushu University, Series E (Biology) 3: 65-80.

Nipperess, D. A. et al. 2010. Resemblance in phylogenetic diversity among ecological assemblages. – J. Veg. Sci. 21: 809–820.

Ochiai, A. 1957. Zoogeographic studies on the soleoid fishes found in Japan and its neighbouring regions. – Bulletin of the Japanese Society of Scientific Fisheries 22: 526–530.

Orloci, L. 1967. An agglomerative method for classification of plant communities. – J. Ecol. 55: 193–206.

Rao, C.R. 1995. A review of canonical coordinates and an alternative to correspondence analysis using Hellinger distance. – Qüestiió 19: 23–63.

Sørensen, T. 1948. A method of establishing groups of equal amplitude in plant sociology based on similarity of species content. – Kongelige Danske Videnskabernes Selskabs Biologiske Skrifter 5: 1–34.

Tamás, J et al. 2001. An extension of presence/absence coefficients to abundance data: a new look at absence. J. Veg. Sci. 12: 401–410.

# Appendix 2. Proofs – Mathematical developments on the dissimilarity indices and ordination approaches

As in Chiu et al. (2014), the height of the phylogenetic tree included in all analyses can be adjusted and the tree can be ultrametric or not (see Fig. 2-1).

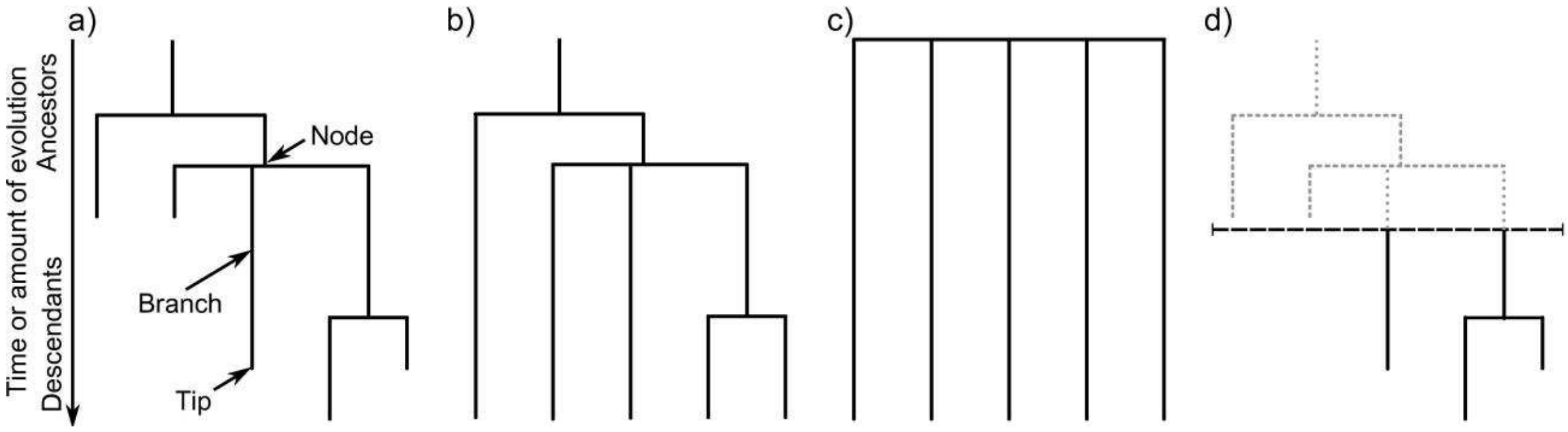

Figure 2-1. Example of phylogenetic trees which PD-dissimilarity indices can be applied to: (a) a non-ultrametric tree, (b) an ultrametric tree, (c) a star-shaped tree, and (d) a portion of tree. A tree is ultrametric if the sum of branch lengths from any tip to the root is constant. Chiu et al. (2014) underlined in their paper that evolutionary diversity is measured considering a period of evolution from present to a given time in the past. In theory, the PD-dissimilarity methodology can also be applied to any portion of tree (time-scaled or not).

## 1. Index ${}^{q}D_{\beta}$

Using the decomposition of Hill numbers developed by Chiu et al. (2014), the following formulas can be derived for γ and α evolutionary diversity:

$$
{}^{q}\gamma_{evoHill} = \begin{cases} \left[ \sum_{k \in b_T} L_k \left( \sum_{j=1}^{m} w_j \dfrac{a_{jk}}{\sum_{k \in b_T} L_k a_{jk}} \right)^{q} \right]^{\frac{1}{(1-q)}} & q > 0, q \neq 1 \\ \exp\left[ -\sum_{k \in b_T} L_k \left( \sum_{j=1}^{m} w_j \dfrac{a_{jk}}{\sum_{k \in b_T} L_k a_{jk}} \right) \log\left( \sum_{j=1}^{m} w_j \dfrac{a_{jk}}{\sum_{k \in b_T} L_k a_{jk}} \right) \right] & q = 1 \end{cases}
$$

$$
{}^{q}\alpha_{evoHill} = \begin{cases} \dfrac{1}{m} \left[ \sum_{k \in b_T} L_k \sum_{j=1}^{m} (w_j)^{q} \left( \dfrac{a_{jk}}{\sum_{k \in b_T} L_k a_{jk}} \right)^{q} \right]^{\frac{1}{(1-q)}} & q > 0, q \neq 1 \\ \dfrac{1}{m} \exp\left[ -\sum_{k \in b_T} L_k \sum_{i=1}^{m} \left( w_j \dfrac{a_{jk}}{\sum_{k \in b_T} L_k a_{jk}} \right) \log\left( w_j \dfrac{a_{jk}}{\sum_{k \in b_T} L_k a_{jk}} \right) \right] & q = 1 \end{cases}
$$

The weights attributed to sites can be freely chosen (using for example total abundance, biomass within each site or the area of the sites) provided that they sum to 1. Among the many possibilities, one could define the weight of site $j$ as

$$w_j = \frac{\sum_{k \in b_T} L_k a_{jk}}{\sum_{k \in b_T} L_k \sum_{j=1}^{m} a_{jk}}$$

i.e. as the ratio of the total abundance in evolutionary units within site $j$ ($\sum_{k \in b_T} L_k a_{jk}$) to the total abundance in all evolutionary units over the whole study area ($\sum_{k \in b_T} L_k \sum_{j=1}^{m} a_{jk}$) [Note that for ultrametric trees, $\sum_{k \in b_T} L_k a_{jk} / \sum_{k \in b_T} L_k \sum_{j=1}^{m} a_{jk} = \sum_{i \in t_T} A_{ij} / \sum_{i \in t_T} \sum_{j=1}^{m} A_{ij}$, where $A_{ij}$ is the abundance of species $i$ in site $j$]. With this particular definition of site weights, the coefficients ${}^q\gamma_{evoHill}$ and ${}^q\alpha_{evoHill}$ are equal to Chiu et al. (2014) ${}^qPD_\gamma$ and ${}^qPD_\alpha$ indices of phylogenetic $\gamma$ and $\alpha$ diversity, respectively. For any $q \geq 0$, a $\beta$ diversity index can be defined as ${}^q\beta_{evoHill} = {}^q\gamma_{evoHill} / {}^q\alpha_{evoHill} = \boldsymbol{{}^qD_\beta}$ (Chiu et al. 2014).

Moreover, choosing

$$w_j = \frac{\sum_{k \in b_T} L_k a_{jk}}{\sum_{k \in b_T} L_k \sum_{j=1}^{m} a_{jk}}$$

leads to $\boldsymbol{{}^qD_\beta}$ expressed as absolute abundances and choosing $w_j=1/m$ leads to $\boldsymbol{{}^qD_\beta}$ expressed as relative abundances (see also Supplementary material Appendix 1; see main text for notations).

## 2. Indices ${}^q\alpha_{evoHill}$, ${}^q\beta_{evoHill}$, and ${}^q\gamma_{evoHill}$ compared to Chiu et al. (2014) ${}^qPD_\alpha$, ${}^qPD_\beta$, and ${}^qPD_\gamma$

Chiu et al. (2014) first developed a new decomposition for Hill numbers considering species abundances (without phylogenetic data). Let $A_{ij}$ be the abundance of species $i$ in site $j$. The sum of abundances of all species, $A_{+j} = \sum_{i=1}^{S} A_{ij}$, is the size of the assemblage in site $j$. The sum of abundances across all sites is $A_{++} = \sum_{i=1}^{S} \sum_{j=1}^{m} A_{ij}$. Let $P_{ij}$ be the proportion of species $i$ in site $j$: $P_{ij} = A_{ij} / A_{+j}$. Finally the weight attributed to site $j$ is $W_j = A_{+j} / A_{++}$ (relative assemblage size). Then, for $q$>0,

$$ {}^{q}D_{\gamma} = \begin{cases} \left\{ \sum_{i=1}^{S} \left( \sum_{j=1}^{m} W_j P_{ij} \right)^{q} \right\}^{\frac{1}{1-q}} & q \neq 1 \\ \exp\left\{ -\sum_{i=1}^{S} \left( \sum_{j=1}^{m} W_j P_{ij} \right) \ln\left( \sum_{j=1}^{m} W_j P_{ij} \right) \right\} & q = 1 \end{cases} $$

$$ {}^{q}D_{\alpha} = \begin{cases} \frac{1}{m} \left\{ \sum_{i=1}^{S} \sum_{j=1}^{m} \left( W_j P_{ij} \right)^{q} \right\}^{\frac{1}{1-q}} & q \neq 1 \\ \exp\left\{ -\sum_{i=1}^{S} \sum_{j=1}^{m} W_j P_{ij} \log\left( W_j P_{ij} \right) - \ln\left( m \right) \right\} & q = 1 \end{cases} $$

and ${}^{q}D_{\beta} = {}^{q}D_{\gamma} / {}^{q}D_{\alpha}$.

I now replace species with evolutionary units. As in the main text, $b_T$ is the set of branches in a phylogenetic tree $T$; $L_k$ is the length of branch $k$ in the phylogenetic tree; $a_{jk}$ is the sum of the abundances, in site $j$, of all species descending from branch $k$; $m$ is the number of sites; $w_j$ is a positive weight attributed to site $j$ ($\sum_{j=1}^{m} w_j = 1$); $t_T$ is the set of species (tips) in the whole tree; $t_k$ is the set of species that descend from branch $k$.

The abundance of an evolutionary unit supported by branch $k$ in site $j$ is $a_{jk} = \sum_{i \in t_k} A_{ij}$ . The sum of abundance of all evolutionary units is $a_{j+} = \sum_{k \in b_T} L_k a_{jk}$ . The sum of the abundances of all evolutionary units across all sites is $a_{++} = \sum_{k \in b_T} \sum_{j=1}^{m} L_k a_{jk}$ . Let $p_{ij}$ be the proportion, in site $j$, of an evolutionary unit present on branch $k$: $p_{jk} = a_{jk} / a_{j+}$ .

The weight attributed to site $j$ is $w_j = a_{+j} / a_{++}$. Replacing species by evolutionary units in Chiu et al. (2014) decomposition of Hill numbers thus leads to

$$ {}^{q}\gamma_{evoHill} = \begin{cases} \left\{ \sum_{k \in b_T} L_k \left( \sum_{j=1}^{m} w_j p_{jk} \right)^{q} \right\}^{\frac{1}{1-q}} & q \neq 1 \\ \exp\left\{ -\sum_{k \in b_T} L_k \left( \sum_{j=1}^{m} w_j p_{jk} \right) \ln\left( \sum_{j=1}^{m} w_j p_{jk} \right) \right\} & q = 1 \end{cases} $$

$$
{}^{q}\alpha_{evoHill} = \begin{cases} \dfrac{1}{m}\left\{ \sum_{k \in b_T} L_k \sum_{j=1}^{m} \left(w_j p_{jk}\right)^q \right\}^{\frac{1}{1-q}} & q \neq 1 \\ \exp\left\{ -\sum_{k \in b_T} L_k \sum_{j=1}^{m} w_j p_{jk} \ln\left(w_j p_{jk}\right) - \ln(m) \right\} & q = 1 \end{cases}
$$

and ${}^{q}\beta_{evoHill}= {}^{q}\gamma_{evoHill} / {}^{q}\alpha_{evoHill}$.

Now, let $\pi_{jk} = a_{jk} / A_{j+}$. Chiu et al. (2014) developed the following equations for $\alpha$, $\beta$, $\gamma$ “branch diversity”:

$$
{}^{q}PD_{\gamma} = \begin{cases} \left\{ \sum_{k \in b_T} L_k \left( \dfrac{\sum_{j=1}^{m} W_j \pi_{jk}}{\sum_{j=1}^{m} W_j \sum_{k \in b_T} L_k \pi_{jk}} \right)^q \right\}^{\frac{1}{1-q}} & q \neq 1 \\ \left( \sum_{j=1}^{m} W_j \sum_{k \in b_T} L_k \pi_{jk} \right) \exp\left\{ -\sum_{k \in b_T} \dfrac{L_k}{\sum_{j=1}^{m} W_j \sum_{k \in b_T} L_k \pi_{jk}} \left( \sum_{j=1}^{m} W_j \pi_{jk} \right) \ln\left( \sum_{j=1}^{m} W_j \pi_{jk} \right) \right\} & q = 1 \end{cases}
$$

$$
{}^{q}PD_{\alpha} = \begin{cases} \dfrac{1}{m}\left\{ \sum_{k \in b_T} L_k \sum_{j=1}^{m} \left( \dfrac{W_j \pi_{jk}}{\sum_{j=1}^{m} W_j \sum_{k \in b_T} L_k \pi_{jk}} \right)^q \right\}^{\frac{1}{1-q}} & q \neq 1 \\ \exp\left\{ -\sum_{k \in b_T} L_k \sum_{j=1}^{m} \dfrac{W_j \pi_{jk}}{\sum_{j=1}^{m} W_j \sum_{k \in b_T} L_k \pi_{jk}} \ln\left( \dfrac{W_j \pi_{jk}}{\sum_{j=1}^{m} W_j \sum_{k \in b_T} L_k \pi_{jk}} \right) - \ln(m) \right\} & q = 1 \end{cases}
$$

and ${}^{q}PD_{\beta}={}^{q}PD_{\gamma}/{}^{q}PD_{\alpha}$.

In the equations of ${}^{q}PD_{\alpha}$, ${}^{q}PD_{\beta}$, and ${}^{q}PD_{\gamma}$, sites are said to be weighted by their relative size: $W_j = A_{+j} / A_{++}$ is the ratio of the number of individuals (sum of species’ abundances) in $j$ to the number of individuals in all pooled sites. However a simplification of the equations of ${}^{q}PD_{\alpha}$, ${}^{q}PD_{\beta}$, and ${}^{q}PD_{\gamma}$ shows that ${}^{q}PD_{\alpha}= {}^{q}\alpha_{evoHill}$, ${}^{q}PD_{\beta}= {}^{q}\beta_{evoHill}$, and ${}^{q}PD_{\gamma}= {}^{q}\gamma_{evoHill}$. ${}^{q}PD_{\alpha}$, ${}^{q}PD_{\beta}$, and ${}^{q}PD_{\gamma}$ can also be viewed as evodiversity indices where sites are weighted by their relative abundance in evolutionary units instead of species:

$$\frac{W_j\pi_{jk}}{\sum_{j=1}^{m}W_j\sum_{k\in b_T}L_k\pi_{jk}}=\frac{\left(A_{+j}/A_{++}\right)\left(a_{jk}/A_{j+}\right)}{\sum_{j=1}^{m}\left(A_{+j}/A_{++}\right)\sum_{k\in b_T}L_k\left(a_{jk}/A_{j+}\right)}$$

$$=\frac{\left(A_{+j}\right)\left(a_{jk}/A_{j+}\right)}{\sum_{j=1}^{m}\left(A_{+j}\right)\sum_{k\in b_T}L_k\left(a_{jk}/A_{j+}\right)}$$

$$=\frac{a_{jk}}{\sum_{j=1}^{m}\sum_{k\in b_T}L_k a_{jk}}$$

$$=w_j p_{jk}$$

This shows all equalities except that

$$\left(\sum_{j=1}^{m}W_j\sum_{k\in b_T}L_k\pi_{jk}\right)\exp\left\{-\sum_{k\in b_T}\frac{L_k}{\sum_{j=1}^{m}W_j\sum_{k\in b_T}L_k\pi_{jk}}\left(\sum_{j=1}^{m}W_j\pi_{jk}\right)\ln\left(\sum_{j=1}^{m}W_j\pi_{jk}\right)\right\}$$

$$=\exp\left\{-\sum_{k\in b_T}L_k\left(\sum_{j=1}^{m}w_j p_{jk}\right)\ln\left(\sum_{j=1}^{m}w_j p_{jk}\right)\right\}$$

This equality can be shown by the following steps:

Given that

$$\sum_{k\in b_T}L_k\left(\sum_{j=1}^{m}w_j p_{jk}\right)=\sum_{k\in b_T}L_k\sum_{j=1}^{m}w_j\frac{a_{jk}}{\sum_{k\in b_T}L_k a_{jk}}=1$$

and that

$$\frac{W_j\pi_{jk}}{\sum_{j=1}^{m}W_j\sum_{k\in b_T}L_k\pi_{jk}}=w_j p_{jk}$$

$$\left(\sum_{j=1}^{m}W_j\sum_{k\in b_T}L_k\pi_{jk}\right)\exp\left\{-\sum_{k\in b_T}\frac{L_k}{\sum_{j=1}^{m}W_j\sum_{k\in b_T}L_k\pi_{jk}}\left(\sum_{j=1}^{m}W_j\pi_{jk}\right)\ln\left(\sum_{j=1}^{m}W_j\pi_{jk}\right)\right\}$$

$$=\left(\sum_{j=1}^{m}W_j\sum_{k\in b_T}L_k\pi_{jk}\right)\exp\left\{-\sum_{k\in b_T}L_k\left(\sum_{j=1}^{m}w_j p_{jk}\right)\ln\left[\left(\sum_{j=1}^{m}w_j p_{jk}\right)\left(\sum_{j=1}^{m}W_j\sum_{k\in b_T}L_k\pi_{jk}\right)\right]\right\}$$

$$=\exp\left\{\begin{array}{l}-\sum_{k\in b_T}L_k\left(\sum_{j=1}^{m}w_j p_{jk}\right)\ln\left(\sum_{j=1}^{m}w_j p_{jk}\right)-\sum_{k\in b_T}L_k\left(\sum_{j=1}^{m}w_j p_{jk}\right)\ln\left[\left(\sum_{j=1}^{m}W_j\sum_{k\in b_T}L_k\pi_{jk}\right)\right]\\+\ln\left(\sum_{j=1}^{m}W_j\sum_{k\in b_T}L_k\pi_{jk}\right)\end{array}\right\}$$

$$=\exp\left\{-\sum_{k\in b_T}L_k\left(\sum_{j=1}^{m}w_j p_{jk}\right)\ln\left(\sum_{j=1}^{m}w_j p_{jk}\right)\right\}$$

## 3. Details on the ordination approach evoCA

Notations (reminder): *j* stands for any site; *k* stands for any branch in a phylogenetic tree *T*; $L_k$ is the length of branch *k* in the phylogenetic tree; $a_{jk}$ is the sum of the abundances, in site *j*, of all species descending from branch *k*.

EvoCA defines axes (with coordinates for branches and sites) that best describe the relationships between sites and evolutionary units. EvoCA analyses the interdependence between the two variables "site" and "evolutionary unit". In absence of phylogenetic structure among sites, there should be no connection between the two variables. In that case, knowing that an evolutionary unit was found in a site gives us no information on its phylogenetic position. Knowing that this evolutionary unit belongs to a given branch on the phylogenetic tree gives us no more information on the site(s) in which it could have been observed.

Let $\mathbf{Z}=(z_{jk})$ a square matrix with $j=1\ldots m$ and $k=1\ldots K$ (where *m* is the number of sites and *K* is the number of branches on the phylogenetic tree *T*) and

$$z_{jk} = \frac{a_{jk} - a_{j+}a_{+k} / a_{++}}{a_{j+}a_{+k} / a_{++}}$$

where $a_{j+} = \sum_{k=1}^{K} L_k a_{jk}$, $a_{+k} = \sum_{j=1}^{m} a_{jk}$, and $a_{++} = \sum_{j=1}^{m}\sum_{k=1}^{K} L_k a_{jk}$. Let $\mathbf{D}_J$ be a diagonal matrix with $a_{j+} / a_{++}$ the $j^{th}$ diagonal entry, $j=1, \ldots, m$. Let $\mathbf{D}_K$ be a diagonal matrix with $L_k a_{+k} / a_{++}$ the $k^{th}$ diagonal entry, $k=1, \ldots, K$. The generalized singular value decomposition (GSVD) of the triplet ($\mathbf{Z}$, $\mathbf{D}_J$, $\mathbf{D}_K$) consists in determining $\mathbf{U}$ ($K \times v$ matrix of eigenvectors) and $\mathbf{\Lambda}$ ($v \times v$ diagonal matrix of eigenvalues) so that

$$\mathbf{D}_K^{1/2}\mathbf{Z}^t\mathbf{D}_J\mathbf{Z}\mathbf{D}_K^{1/2} = \mathbf{U}\mathbf{\Lambda}\mathbf{U}^t$$

It can be shown (see next paragraph) that this approach simply corresponds to applying the correspondence analysis to matrix ($L_k a_{jk}$) for $j=1\ldots m$ and $k=1\ldots K$. The GSVD defines two systems of coordinates: a $m \times v$ matrix $\mathbf{Z}\mathbf{D}_K^{1/2}\mathbf{U}$ for the coordinates of the communities and a $K \times v$ matrix $\mathbf{D}_K^{-1/2}\mathbf{U}\mathbf{\Lambda}^{1/2}$ for the coordinates of the phylogenetic branches. The matrix of variance-covariance among the coordinates of the communities of each axis and that among the coordinates of the phylogenetic branches of each axis are equal to the diagonal matrix $\mathbf{\Lambda}$, which implies that axes are orthogonal. Normed coordinates with unit variance can be obtained as $\mathbf{Z}\mathbf{D}_K^{1/2}\mathbf{U}\mathbf{\Lambda}^{-1/2}$ for the communities and $\mathbf{D}_K^{-1/2}\mathbf{U}$ for the phylogenetic branches. It can be shown (proof below) that the raw coordinate of any phylogenetic branch $k$ is at the weighted barycentre (i.e. mean) of the sites (positioned by their normed coordinates) which its descendant species occur in. Weighting of the sites is by $a_{jk}/a_{+k}$ for all $j = 1, \ldots, m$. It can also be shown (proof below) that the raw coordinate of any site $j$ is at the weighted barycentre of the phylogenetic branches positioned by their normed coordinates. For a site $j$, weighting of the phylogenetic branches is by $L_k a_{jk}/a_{j+}$ for all $k = 1, \ldots, K$. Consider any branch $k$, and $d$=1, …, $D_k$ its descendant species (tips). A species is positioned in the space of evoCA at the same position as the terminal branch that connects it to the rest of the phylogenetic tree. It can be shown (proof below) that the coordinate of branch $k$ is at the weighted barycentre of the coordinates of its descendant species. Weighting is by $A_{d+}/a_{+k}$ for all $d$=1, …, $D_k$ with $A_{d+}$ the sum of the abundances of species $d$ over all sites.

● **EvoCA is equivalent to CA applied to matrix** Consider $y_{jk}= L_k a_{jk}$ and the corresponding matrix $\mathbf{Y}=[L_j a_{ij}]$ where $j$ in 1…$m$ and $k$ in 1…$K$. The marginal sums of $\mathbf{Y}$ are: $y_{j+} = \sum_{k=1}^{K} L_k a_{jk}$ , $y_{+k} = \sum_{j=1}^{m} L_k a_{jk}$ , and $y_{++} = \sum_{j=1}^{m}\sum_{k=1}^{K} L_k a_{jk}$ . EvoCA is equivalent to the correspondence analysis (CA) of $\mathbf{Y}$. Indeed the CA of $\mathbf{Y}$ depends on the generalized singular value decomposition (GSVD) of the triplet ($\mathbf{Z}$, $\mathbf{D}_J$, $\mathbf{D}_K$) where

$$\mathbf{Z} = \left[ \frac{y_{jk} - y_{j+} y_{+k} / y_{++}}{y_{j+} y_{+k} / y_{++}} \right] \text{ ; } \mathbf{D}_J = diag\left\{ y_{j+} / y_{++} \right\} \text{; and } \mathbf{D}_K = diag\left\{ y_{+k} / y_{++} \right\}$$

where $j$ varies in 1…$m$ and $k$ in 1…$K$.

$\mathbf{D}_J = diag\left\{ \sum_{k=1}^{K} L_k a_{jk} \Big/ \sum_{j=1}^{m}\sum_{k=1}^{K} L_k a_{jk} \right\}$ here is equal to the $\mathbf{D}_J$ defined for evoCA.

$\mathbf{D}_K = diag\left\{ L_k \sum_{j=1}^{m} a_{jk} \Big/ \sum_{j=1}^{m}\sum_{k=1}^{K} L_k a_{jk} \right\}$ is equal to the $\mathbf{D}_K$ defined for evoCA.

$$\mathbf{Z} = \left[ \frac{L_k a_{jk} - \dfrac{\left(\sum_{k=1}^{K} L_k a_{jk}\right)\left(L_k \sum_{j=1}^{m} a_{jk}\right)}{\sum_{j=1}^{m}\sum_{k=1}^{K} L_k a_{jk}}}{\dfrac{\left(\sum_{k=1}^{K} L_k a_{jk}\right)\left(L_k \sum_{j=1}^{m} a_{jk}\right)}{\sum_{j=1}^{m}\sum_{k=1}^{K} L_k a_{jk}}} \right]$$

Simplifying the equation of $\mathbf{Z}$ by $L_k$, leads to

$$\mathbf{Z} = \left[ \frac{a_{jk} - \frac{\left(\sum_{k=1}^{K} L_k a_{jk}\right)\left(\sum_{j=1}^{m} a_{jk}\right)}{\sum_{j=1}^{m}\sum_{k=1}^{K} L_k a_{jk}}}{\frac{\left(\sum_{k=1}^{K} L_k a_{jk}\right)\left(\sum_{j=1}^{m} a_{jk}\right)}{\sum_{j=1}^{m}\sum_{k=1}^{K} L_k a_{jk}}} \right]$$

which is the equation used to define **Z** for evoCA. □

**● The raw coordinate of any phylogenetic branch *k* is at the weighted barycentre (i.e. mean) of the sites (positioned by their normed coordinates) which its descendant species occur in:**

For any branch *k*, the weighting of the sites is by $a_{jk}/a_{+k}$ for all $j = 1, \ldots, m$ and with $a_{+k} = \sum_{j=1}^{m} a_{jk}$. This is a well-known property of the correspondence analysis (e.g. Legendre and Legendre 1998). This property is thus proved by the fact that evoCA is CA applied to matrix **Y** introduced above. In the CA of matrix **Y**, any branch *k* is at the weighted barycentre of the sites, with the weight of any site *j* equal to $y_{jk}/y_{+k} = L_k a_{jk} / \sum_{j=1}^{m} L_k a_{jk}$ which simplifies to $a_{jk}/a_{+k}$. □

**● The raw coordinate of any site *j* is at the weighted barycentre (i.e. mean) of the phylogenetic branches positioned by their normed coordinates:**

For a site *j*, weighting of the phylogenetic branches is by $L_k a_{jk}/a_{j+}$ for all $k = 1, \ldots, K$ and with $a_{j+} = \sum_{k=1}^{K} L_k a_{jk}$. Similarly, this is a well-known property of the correspondence analysis (e.g. Legendre and Legendre 1998). This property is thus proved by the fact that evoCA is CA applied to matrix **Y** introduced above. In the CA of matrix **Y**, any site *j* is at the weighted barycentre of the branches, with the weight of any branch *k* equal to $y_{jk}/y_{j+} = L_k a_{jk} / \sum_{k=1}^{K} L_k a_{jk}$.

□

**● The coordinate of branch *k* is at the weighted barycentre (*i.e.* mean) of the coordinates of its descendant species:**

Let $c_{jl}$ be the normed coordinate of site *j* on axis *l*. As shown above the coordinate of branch *k* on axis *l* is $c_{kl} = \sum_{j=1}^{m} a_{jk} c_{jl} / a_{+k}$. Let $d=1, \ldots, D_k$ be the species that descend from branch *k*. Let $A_{dj}$ the abundance of species *d* in site *j*. The terminal branch from which only species *d* descends thus has coordinate $c_{dl} = \sum_{j=1}^{m} A_{dj} c_{jl} / A_{d+}$ on axis *l*, with $A_{d+}$ the sum of the abundances of species *d* over all sites. $a_{jk}$ is the sum of the abundances of species that descend from branch $k$: $a_{jk} = \sum_{d=1}^{D_k} A_{dj}$. The coordinate of branch $k$ on axis $l$ is thus $\sum_{j=1}^{m}\sum_{d=1}^{D_k} A_{dj} c_{jl} / a_{+k}$. This can be rewritten as

$$\sum_{d=1}^{D_k}\sum_{j=1}^{m} A_{dj} c_{jl} / a_{+k} = \sum_{d=1}^{D_k} \frac{A_{d+}}{a_{+k}} \sum_{j=1}^{m} \frac{A_{dj} c_{jl}}{A_{d+}} = \sum_{d=1}^{D_k} \frac{A_{d+}}{a_{+k}} c_{dl}$$

□

# 4. Basic necessary properties for PD-dissimilarity coefficients

Notations (reminder): Indices 1 and 2 refer to the two compared sites; $j$ stands for any site; $b_T$ is the set of branches in a phylogenetic tree $T$; $L_k$ is the length of branch $k$ in the phylogenetic tree; $a_{jk}$ is the sum of abundance for all species descending from branch $k$.

I checked whether the coefficients of PD-dissimilarity given in Supplementary material Appendix 1 satisfy Legendre and De Cáceres (2013) nine basic necessary properties (P1-9). Although most demonstrations are closely related with each other, I have chosen to detail all demonstrations here. The results are summarized in Table 2-1 below.

As the species dissimilarity version of PD-dissimilarity coefficients are particular cases where species have independent evolution (star phylogeny with unit branch lengths), the coefficients that were shown by Legendre and De Cáceres (2013) to fail to satisfy the basic necessary properties do not satisfy them when phylogenetic information is added. As a result, $evoD_{Minkowski}$ (that includes $evoD_{Manhattan}$ and $evoD_{Euclidean}$), $evoD_{Profile}$, and $evoD_{\chi^2}$ do not satisfy the basic necessary properties.

● P1 (Minimum of zero and positiveness) and P2 (symmetry) were satisfied by all coefficients.

● Property P3 assumes that increasing the difference in the abundance of a species (here an evolutionary unit) among the two sites, increases the dissimilarity between the two sites. As recommended by Legendre and De Cáceres (2013), property P3 was verified using OCCAS method (Hadju 1981) and replacing species with evolutionary units, thus considering two evolutionary units only (Table 3 in Hadju 1981). For parametric indices, I considered parameter *q*, or *r*, varying from 0.001 to 30. I verified P3 only for the indices that were not considered in Legendre and De Cáceres (2013). For all other indices I used the results obtained by Legendre and De Cáceres (2013).

With the approach described above, P3 was found to be correct for all coefficients with however a special warning for index $evoD_{Divergence}$. Indeed, the coefficient of divergence is unchanged by modifying the abundance of a species present in only one site. As Legendre and De Cáceres (2013) considered that the coefficient of Divergence satisfy P3, I thus considered a weaker version of P3: when a species (or an evolutionary unit) is present in both sites, increasing the differences in abundance for the species (or the evolutionary unit) among the two sites increases the dissimilarity among the sites. When a species is present in one site only, increasing the abundance for the species (or the evolutionary unit) in the site where it is present does not decrease the dissimilarity among the sites.

● Property P4 considers that a compositional dissimilarity coefficient should be unaffected when a species is added in both sites with an abundance equal to zero but that it should decrease if it is added with the same, positive abundance in both sites. Here I considered an evolutionary unit instead of a species.

Indices that use Nipperess et al. (2010) components *A*, *B*, C satisfy P4. Indeed, when an evolutionary unit is added with the same, positive abundance *Ab* in both sites. Nipperess et al. component *A* is increased ( $A=\sum_{j=1}^{K} L_j \min\{a_{j1},a_{j2}\}+Ab$ ). Components *B*, *C* are unchanged. For the same reason, indices that use Nipperess et al. (2010) components *a*, *b*, *c* satisfy P4.

Index $evoD_{Chord}$ satisfies P4. Let *T* be a phylogenetic tree and 1 and 2 denote two sites. $evoD_{Chord}$(1,2|T) can be written as $evoD_{Chord}(1,2\,|\,T)=\sqrt{\sum_{k\in b_T} L_k\left(b_{1k}-b_{2k}\right)^2}$ where $\sum_{k\in b_T} L_k b_{1k}^2=1$.

Adding the new evolutionary unit leads to a new phylogenetic tree named *Te*. Adding it with an equal weight in both sites 1 and 2 can be seen as adding a positive value *A* as follows:

$$evoD_{Chord}\left(1,2\,|\,Te\right)=\sqrt{\sum_{k\in b_T}L_k\left(\frac{b_{1k}}{\sqrt{\sum_{k\in b_T}L_kb_{1k}^2+A^2}}-\frac{b_{2k}}{\sqrt{\sum_{k\in b_T}L_kb_{2k}^2+A^2}}\right)^2+\left(\frac{A}{\sqrt{\sum_{k\in b_T}L_kb_{1k}^2+A^2}}-\frac{A}{\sqrt{\sum_{k\in b_T}L_kb_{2k}^2+A^2}}\right)^2}$$

$$=\sqrt{\sum_{k\in b_T}L_k\left(\frac{b_{1k}}{\sqrt{1+A^2}}-\frac{b_{2k}}{\sqrt{1+A^2}}\right)^2}$$

$$=\frac{1}{\sqrt{1+A^2}}\sqrt{\sum_{k\in b_T}L_k\left(b_{1k}-b_{2k}\right)^2}$$

Because $0<1/\sqrt{1+A^2}<1$,

$$\frac{1}{\sqrt{1+A^2}}\sqrt{\sum_{k\in b_T}L_k\left(b_{1k}-b_{2k}\right)^2}\le\sqrt{\sum_{k\in b_T}L_k\left(b_{1k}-b_{2k}\right)^2}$$

Index *evoD*$_{Hellinger}$ satisfies P4. Indeed, *evoD*$_{Hellinger}$ can be written as $evoD_{Hellinger}\left(1,2\,|\,T\right)=\sqrt{\sum_{k\in b_T}L_k\left(\sqrt{b_{1k}}-\sqrt{b_{2k}}\right)^2}$ where $\sum_{k\in b_T}L_kb_{1k}=1$. Adding the new evolutionary unit with an equal relative abundance in both sites can be viewed as adding a positive value *A* such that *evoD*$_{Hellinger}$ equals

$$evoD_{Hellinger}\left(1,2\,|\,Te\right)=\sqrt{\sum_{k\in b_T}L_k\left(\sqrt{\frac{b_{1k}}{\sum_{k\in b_T}L_kb_{1k}+A}}-\sqrt{\frac{b_{2k}}{\sum_{k\in b_T}L_kb_{2k}+A}}\right)^2+\left(\sqrt{\frac{A}{\sum_{k\in b_T}L_kb_{1k}+A}}-\sqrt{\frac{A}{\sum_{k\in b_T}L_kb_{2k}+A}}\right)^2}$$

$$=\sqrt{\sum_{k\in b_T}L_k\left(\sqrt{\frac{b_{1k}}{1+A}}-\sqrt{\frac{b_{2k}}{1+A}}\right)^2}$$

$$=\frac{1}{\sqrt{1+A}}\sqrt{\sum_{k\in b_T}L_k\left(\sqrt{b_{1k}}-\sqrt{b_{2k}}\right)^2}$$

Because $0<1/\sqrt{1+A}<1$,

$$\frac{1}{\sqrt{1+A}}\sqrt{\sum_{k\in b_T}L_k\left(\sqrt{b_{1k}}-\sqrt{b_{2k}}\right)^2}\le\sqrt{\sum_{k\in b_T}L_k\left(\sqrt{b_{1k}}-\sqrt{b_{2k}}\right)^2}$$

Index *evoD*$_{Profile}$ satisfies P4. Indeed, *evoD*$_{Profile}$ can be written as $evoD_{Profile}\left(1,2\,|\,T\right)=\sqrt{\sum_{k\in b_T}L_k\left(b_{1k}-b_{2k}\right)^2}$ where $\sum_{k\in b_T}L_kb_{1k}=1$. Adding the new evolutionary unit with an equal relative abundance in both sites can be viewed as adding a positive value *A* such that *evoD*$_{Profile}$ equals

$$evoD_{Profile}(1,2|Te)=\sqrt{\sum_{k\in b_T}L_k\left(\frac{b_{1k}}{\sum_{k\in b_T}L_k b_{1k}+A}-\frac{b_{2k}}{\sum_{k\in b_T}L_k b_{2k}+A}\right)^2+\left(\frac{A}{\sum_{k\in b_T}L_k b_{1k}+A}-\frac{A}{\sum_{k\in b_T}L_k b_{2k}+A}\right)^2}$$

$$=\sqrt{\sum_{k\in b_T}L_k\left(\frac{b_{1k}}{1+A}-\frac{b_{2k}}{1+A}\right)^2}$$

$$=\frac{1}{(1+A)}\sqrt{\sum_{k\in b_T}L_k\left(b_{1k}-b_{2k}\right)^2}$$

Because $0<1/(1+A)<1$,

$$\frac{1}{(1+A)}\sqrt{\sum_{k\in b_T}L_k\left(b_{1k}-b_{2k}\right)^2}\leq\sqrt{\sum_{k\in b_T}L_k\left(b_{1k}-b_{2k}\right)^2}$$

Index $evoD_{\chi^2}$ and the $\chi^2$ distance are sensitive to both species relative and absolute abundances. Adding a new evolutionary unit to both sites with equal absolute abundances do not necessarily decrease the distance between them as shown by the following counter-example:

Consider four communities C1, C2, C3, C4 and five species S1, S2, S3, S4, S5. Consider that the five species have independent evolution (star phylogeny with unit branch lengths): i.e. each species contributes an independent evolutionary unit. The distributions of species abundance in the communities are

C1 : S1=5, S2=5, S3=5, S4=5, S5=0

C2 : S1=4, S2=4, S3=4, S4=4, S5=0

C3 : S1=5, S2=5, S3=5, S4=5, S5=0.1

C4 : S1=4, S2=4, S3=4, S4=4, S5=0.1

Compared to C1, C3 contains a new species (S5) and thus a new evolutionary unit with abundance 0.1; compared to C2, C4 also contains a new species (S5) and thus a new evolutionary unit with abundance 0.1.

${}^2evoD_{\chi^2}$(C1,C2)≈ 0

${}^2evoD_{\chi^2}$(C3,C4)≈ 0.02352

I will thus consider adding the new evolutionary unit so that the resulting RELATIVE abundance is similar in both sites.

$$evoD_{\chi^2}\left(1,2\,|\,T\right)=\sqrt{\sum\nolimits_{k\in b_T} L_k \frac{\sum\nolimits_{k\in b_T} L_k \sum\nolimits_j a_{jk}}{\sum\nolimits_j a_{jk}}\left(\frac{a_{1k}}{\sum\nolimits_{k\in b_T} L_k a_{1k}}-\frac{a_{2k}}{\sum\nolimits_{k\in b_T} L_k a_{2k}}\right)^2}$$

adding the new evolutionary unit with the same relative abundance in both sites can be viewed as adding a positive value $A$ so that

$$evoD_{\chi^2}\left(1,2\,|\,Te\right)$$

$$=\sqrt{\begin{aligned}&\sum\nolimits_{k\in b_T} L_k \frac{\sum\nolimits_{k\in b_T} L_k \sum\nolimits_j a_{jk}+A\frac{\sum\nolimits_{k\in b_T} L_k a_{1k}+\sum\nolimits_{k\in b_T} L_k a_{2k}}{\sum\nolimits_{k\in b_T} L_k a_{1k}}}{\sum\nolimits_j a_{jk}}\left(\frac{a_{1k}}{\sum\nolimits_{k\in b_T} L_k a_{1k}+A}-\frac{a_{2k}}{\sum\nolimits_{k\in b_T} L_k a_{2k}+A\frac{\sum\nolimits_{k\in b_T} L_k a_{2k}}{\sum\nolimits_{k\in b_T} L_k a_{1k}}}\right)^2\\&+\frac{\sum\nolimits_{k\in b_T} L_k \sum\nolimits_j a_{jk}+A\frac{\sum\nolimits_{k\in b_T} L_k a_{1k}+\sum\nolimits_{k\in b_T} L_k a_{2k}}{\sum\nolimits_{k\in b_T} L_k a_{1k}}}{A\frac{\sum\nolimits_{k\in b_T} L_k a_{1k}+\sum\nolimits_{k\in b_T} L_k a_{2k}}{\sum\nolimits_{k\in b_T} L_k a_{1k}}}\left(\frac{A}{\sum\nolimits_{k\in b_T} L_k a_{1k}+A}-\frac{A\frac{\sum\nolimits_{k\in b_T} L_k a_{2k}}{\sum\nolimits_{k\in b_T} L_k a_{1k}}}{\sum\nolimits_{k\in b_T} L_k a_{2k}+A\frac{\sum\nolimits_{k\in b_T} L_k a_{2k}}{\sum\nolimits_{k\in b_T} L_k a_{1k}}}\right)^2\end{aligned}}$$

$$=\sqrt{\sum\nolimits_{k\in b_T} L_k \frac{\sum\nolimits_{k\in b_T} L_k \sum\nolimits_j a_{jk}+A\frac{\sum\nolimits_{k\in b_T} L_k a_{1k}+\sum\nolimits_{k\in b_T} L_k a_{2k}}{\sum\nolimits_{k\in b_T} L_k a_{1k}}}{\sum\nolimits_j a_{jk}}\left(\frac{a_{1k}}{\sum\nolimits_{k\in b_T} L_k a_{1k}+A}-\frac{a_{2k}}{\sum\nolimits_{k\in b_T} L_k a_{2k}+A\frac{\sum\nolimits_{k\in b_T} L_k a_{2k}}{\sum\nolimits_{k\in b_T} L_k a_{1k}}}\right)^2}$$

$$=\sqrt{\left[\sum\nolimits_{k\in b_T} L_k \sum\nolimits_j a_{jk}+A\frac{\sum\nolimits_{k\in b_T} L_k a_{1k}+\sum\nolimits_{k\in b_T} L_k a_{2k}}{\sum\nolimits_{k\in b_T} L_k a_{1k}}\right]\left[\sum\nolimits_{k\in b_T} L_k \frac{1}{\sum\nolimits_j a_{jk}}\left(\frac{a_{1k}}{\sum\nolimits_{k\in b_T} L_k a_{1k}+A}-\frac{a_{2k}\frac{\sum\nolimits_{k\in b_T} L_k a_{1k}}{\sum\nolimits_{k\in b_T} L_k a_{2k}}}{\sum\nolimits_{k\in b_T} L_k a_{1k}+A}\right)^2\right]}$$

Consider the second term:

$$\sum\nolimits_{k\in b_T} L_k \frac{1}{\sum\nolimits_j a_{jk}}\left(\frac{a_{1k}}{\sum\nolimits_{k\in b_T} L_k a_{1k}+A}-\frac{a_{2k}\frac{\sum\nolimits_{k\in b_T} L_k a_{1k}}{\sum\nolimits_{k\in b_T} L_k a_{2k}}}{\sum\nolimits_{k\in b_T} L_k a_{1k}+A}\right)^2$$

It is equal to

$$\left[\sum_{k\in b_T} L_k \frac{1}{\sum_j a_{jk}} \left( \frac{a_{1k}}{\sum_{k\in b_T} L_k a_{1k}} \frac{\sum_{k\in b_T} L_k a_{1k}}{\sum_{k\in b_T} L_k a_{1k} + A} - \frac{a_{2k}}{\sum_{k\in b_T} L_k a_{2k}} \frac{\sum_{k\in b_T} L_k a_{1k}}{\sum_{k\in b_T} L_k a_{1k} + A} \right)^2 \right]$$

$$= \left( \frac{\sum_{k\in b_T} L_k a_{1k}}{\sum_{k\in b_T} L_k a_{1k} + A} \right)^2 \left[ \sum_{k\in b_T} L_k \frac{1}{\sum_j a_{jk}} \left( \frac{a_{1k}}{\sum_{k\in b_T} L_k a_{1k}} - \frac{a_{2k}}{\sum_{k\in b_T} L_k a_{2k}} \right)^2 \right]$$

evoD$\chi^2$(1;2|Te) can thus be rewritten as:

$$\sqrt{\left[ \sum_{k\in b_T} L_k \sum_j a_{jk} + A \frac{\sum_{k\in b_T} L_k a_{1k} + \sum_{k\in b_T} L_k a_{2k}}{\sum_{k\in b_T} L_k a_{1k}} \right] \left( \frac{\sum_{k\in b_T} L_k a_{1k}}{\sum_{k\in b_T} L_k a_{1k} + A} \right)^2 \left[ \sum_{k\in b_T} L_k \frac{1}{\sum_j a_{jk}} \left( \frac{a_{1k}}{\sum_{k\in b_T} L_k a_{1k}} - \frac{a_{2k}}{\sum_{k\in b_T} L_k a_{2k}} \right)^2 \right]}$$

The last term

$$\sum_{k\in b_T} L_k \frac{1}{\sum_j a_{jk}} \left( \frac{a_{1k}}{\sum_{k\in b_T} L_k a_{1k}} - \frac{a_{2k}}{\sum_{k\in b_T} L_k a_{2k}} \right)^2$$

is nonnegative and independent of *A*.

The variations of evoD$\chi^2$(1;2|Te) as a function of *A* can thus be analysed thanks to

$$h(A) = \left[ \sum_{k\in b_T} L_k \sum_j a_{jk} + A \frac{\sum_{k\in b_T} L_k a_{1k} + \sum_{k\in b_T} L_k a_{2k}}{\sum_{k\in b_T} L_k a_{1k}} \right] \left( \frac{\sum_{k\in b_T} L_k a_{1k}}{\sum_{k\in b_T} L_k a_{1k} + A} \right)^2$$

$$\frac{dh}{dA} = \frac{\sum_{k\in b_T} L_k a_{1k} + \sum_{k\in b_T} L_k a_{2k}}{\sum_{k\in b_T} L_k a_{1k}} \left( \frac{\sum_{k\in b_T} L_k a_{1k}}{\sum_{k\in b_T} L_k a_{1k} + A} \right)^2$$

$$-2 \left[ \sum_{k\in b_T} L_k \sum_j a_{jk} + A \frac{\sum_{k\in b_T} L_k a_{1k} + \sum_{k\in b_T} L_k a_{2k}}{\sum_{k\in b_T} L_k a_{1k}} \right] \left( \sum_{k\in b_T} L_k a_{1k} \right)^2 \left( \frac{1}{\sum_{k\in b_T} L_k a_{1k} + A} \right)^3$$

$$= \frac{\sum_{k\in b_T} L_k a_{1k} + \sum_{k\in b_T} L_k a_{2k}}{\sum_{k\in b_T} L_k a_{1k}} \left( \frac{\sum_{k\in b_T} L_k a_{1k}}{\sum_{k\in b_T} L_k a_{1k} + A} \right)^2 \left[ 1 - 2 \frac{\sum_{k\in b_T} L_k a_{1k}}{\sum_{k\in b_T} L_k a_{1k} + A} \frac{\sum_{k\in b_T} L_k \sum_j a_{jk}}{\sum_{k\in b_T} L_k a_{1k} + \sum_{k\in b_T} L_k a_{2k}} - 2 \frac{A}{\sum_{k\in b_T} L_k a_{1k} + A} \right]$$

*dh/dA* is negative iff

$$1-2\frac{\sum_{k\in b_T} L_k a_{1k}}{\sum_{k\in b_T} L_k a_{1k}+A}\frac{\sum_{k\in b_T} L_k \sum_j a_{jk}}{\sum_{k\in b_T} L_k a_{1k}+\sum_{k\in b_T} L_k a_{2k}}-2\frac{A}{\sum_{k\in b_T} L_k a_{1k}+A}<0$$

$$\Leftrightarrow 2\frac{\sum_{k\in b_T} L_k a_{1k}}{\sum_{k\in b_T} L_k a_{1k}+A}\frac{\sum_{k\in b_T} L_k \sum_j a_{jk}}{\sum_{k\in b_T} L_k a_{1k}+\sum_{k\in b_T} L_k a_{2k}}+2\frac{A}{\sum_{k\in b_T} L_k a_{1k}+A}\frac{\sum_{k\in b_T} L_k a_{1k}+\sum_{k\in b_T} L_k a_{2k}}{\sum_{k\in b_T} L_k a_{1k}+\sum_{k\in b_T} L_k a_{2k}}>1$$

$$\Leftrightarrow \left(\sum_{k\in b_T} L_k a_{1k}\right)\left(\sum_{k\in b_T} L_k \sum_j a_{jk}\right)+A\left(\sum_{k\in b_T} L_k a_{1k}+\sum_{k\in b_T} L_k a_{2k}\right)>\frac{1}{2}\left(\sum_{k\in b_T} L_k a_{1k}+A\right)\left(\sum_{k\in b_T} L_k a_{1k}+\sum_{k\in b_T} L_k a_{2k}\right)$$

$$\Leftrightarrow \left(\sum_{k\in b_T} L_k a_{1k}\right)\left(\sum_{k\in b_T} L_k \sum_j a_{jk}\right)+\frac{1}{2}A\left(\sum_{k\in b_T} L_k a_{1k}+\sum_{k\in b_T} L_k a_{2k}\right)>\frac{1}{2}\left(\sum_{k\in b_T} L_k a_{1k}\right)\left(\sum_{k\in b_T} L_k a_{1k}+\sum_{k\in b_T} L_k a_{2k}\right)$$

The last inequality is always true given that

$$\left(\sum_{k\in b_T} L_k a_{1k}\right)\left(\sum_{k\in b_T} L_k \sum_j a_{jk}\right)>\frac{1}{2}\left(\sum_{k\in b_T} L_k a_{1k}\right)\left(\sum_{k\in b_T} L_k a_{1k}+\sum_{k\in b_T} L_k a_{2k}\right)$$

*evoDχ²* thus satisfy P4 (considering RELATIVE abundances as for e.g. $evoD_{Profile}$).

For the indices derived from the decomposition of Hill numbers, it is sufficient to analyze $1-\bar{V}_{q2}$ as the logarithm function used in ${}^{q}evoD_{Rényi}$, is monotonously increasing. Here I consider that a new evolutionary unit is added in both sites with an abundance value equal to *A* (relative or absolute abundance depending on whether the coefficient of PD-dissimilarity considers relative or absolute abundances).

With absolute abundances, the behavior of $1-\bar{V}_{q2,abs}$ ($q\neq1$) can be studied thanks to the coefficient

$$\left\{\frac{\sum_{k\in b_T} L_k\left(a_{1k}+a_{2k}\right)^q}{\sum_{k\in b_T} L_k\left[\left(a_{1k}\right)^q+\left(a_{2k}\right)^q\right]}\right\}^{\frac{1}{1-q}}$$

Adding the new evolutionary unit can be viewed as adding a value of abundance *A* as follows:

$$\left\{\frac{\sum_{k\in b_T} L_k\left(a_{1k}+a_{2k}\right)^q+2^q A^q}{\sum_{k\in b_T} L_k\left[\left(a_{1k}\right)^q+\left(a_{2k}\right)^q\right]+2A^q}\right\}^{\frac{1}{1-q}}$$

Consider $u=L_k\left(a_{1k}+a_{2k}\right)^q$, $x=2^q A^q$, $v=\sum_{k\in b_T} L_k\left[\left(a_{1k}\right)^q+\left(a_{2k}\right)^q\right]$ and $y=2A^q$.

$$\left\{\frac{u}{v}\right\}^{\frac{1}{1-q}}\geq\left\{\frac{u+x}{v+y}\right\}^{\frac{1}{1-q}}\Leftrightarrow\begin{cases}\frac{u}{v}\geq\frac{u+x}{v+y} & q<1\\ \frac{u}{v}\leq\frac{u+x}{v+y} & q>1\end{cases}$$

$$\Leftrightarrow\begin{cases}\dfrac{v+y}{v}\geq\dfrac{u+x}{u} & q<1\\ \dfrac{v+y}{v}\leq\dfrac{u+x}{u} & q>1\end{cases}\Leftrightarrow\begin{cases}\dfrac{y}{v}\geq\dfrac{x}{u} & q<1\\ \dfrac{y}{v}\leq\dfrac{x}{u} & q>1\end{cases}\Leftrightarrow\begin{cases}\dfrac{u}{v}\geq\dfrac{x}{y} & q<1\\ \dfrac{u}{v}\leq\dfrac{x}{y} & q>1\end{cases}\Leftrightarrow\begin{cases}\dfrac{u}{v}\geq\left(\dfrac{1}{2}\right)^{1-q} & q<1\\ \dfrac{u}{v}\leq\left(\dfrac{1}{2}\right)^{1-q} & q>1\end{cases}\Leftrightarrow\begin{cases}2\left(\dfrac{u}{v}\right)^{\frac{1}{1-q}}\geq 1 & q<1\\ 2\left(\dfrac{u}{v}\right)^{\frac{1}{1-q}}\geq 1 & q>1\end{cases}$$

The inequalities are always true (Chiu et al. 2014) so that indices $1-\bar{V}_{q2,abs}$ and ${}^{q}evoD_{Rényi,abs}$, $q\neq1$, satisfy P4.

When $q\rightarrow1$, the limit of $1-\bar{V}_{q2}$ is

$$1-\bar{V}_{12,abs}(1,2\,|\,T)=2\exp\left[-\frac{1}{\sum_{k\in b_T}L_k(a_{1k}+a_{2k})}\sum_{k\in b_T}L_k a_{1k}\ln\left(1+\frac{a_{2k}}{a_{1k}}\right)-\frac{1}{\sum_{k\in b_T}L_k(a_{1k}+a_{2k})}\sum_{k\in b_T}L_k a_{2k}\ln\left(1+\frac{a_{1k}}{a_{2k}}\right)\right]-1$$

Adding the new evolutionary unit can be viewed as adding a value A so that

$$1-\bar{V}_{12,abs}(1,2\,|\,Te)=2\exp\left\{\frac{1}{\sum_{k\in b_T}L_k(a_{1k}+a_{2k})+2A}\left[\begin{array}{l}-\sum_{k\in b_T}L_k a_{1k}\ln\left(1+\dfrac{a_{2k}}{a_{1k}}\right)-\sum_{k\in b_T}L_k a_{2k}\ln\left(1+\dfrac{a_{1k}}{a_{2k}}\right)\\ -A\ln\left(1+\dfrac{A}{A}\right)-A\ln\left(1+\dfrac{A}{A}\right)\end{array}\right]\right\}-1$$

$$=2\exp\left\{\frac{1}{\sum_{k\in b_T}L_k(a_{1k}+a_{2k})+2A}\left[-\sum_{k\in b_T}L_k a_{1k}\ln\left(1+\frac{a_{2k}}{a_{1k}}\right)-\sum_{k\in b_T}L_k a_{2k}\ln\left(1+\frac{a_{1k}}{a_{2k}}\right)-2A\ln(2)\right]\right\}-1$$

The value $A$ decreases $1-\bar{V}_{12}$ iff

$$\frac{1}{\sum_{k\in b_T}L_k(a_{1k}+a_{2k})+2A}\left[-\sum_{k\in b_T}L_k a_{1k}\ln\left(1+\frac{a_{2k}}{a_{1k}}\right)-\sum_{k\in b_T}L_k a_{2k}\ln\left(1+\frac{a_{1k}}{a_{2k}}\right)-2A\ln(2)\right]$$
$$\leq\frac{1}{\sum_{k\in b_T}L_k(a_{1k}+a_{2k})}\left[-\sum_{k\in b_T}L_k a_{1k}\ln\left(1+\frac{a_{2k}}{a_{1k}}\right)-\sum_{k\in b_T}L_k a_{2k}\ln\left(1+\frac{a_{1k}}{a_{2k}}\right)\right]$$

The inequality is always true because $A>0$ and $\ln(2)>0$.

$1-\bar{V}_{12,abs}$ and ${}^{1}evoD_{Rényi,abs}$ satisfy P4.

Now consider relative abundances,

$$1-\bar{V}_{q2,rel}(1,2\,|\,T)=2\left\{\frac{\sum_{k\in b_T}L_k\left(\dfrac{a_{1k}}{\sum_{k\in b_T}L_k a_{1k}}+\dfrac{a_{2k}}{\sum_{k\in b_T}L_k a_{2k}}\right)^q}{\sum_{k\in b_T}L_k\left[\left(\dfrac{a_{1k}}{\sum_{k\in b_T}L_k a_{1k}}\right)^q+\left(\dfrac{a_{2k}}{\sum_{k\in b_T}L_k a_{2k}}\right)^q\right]}\right\}^{\frac{1}{1-q}}-1,\ q>0,\ q\neq1$$

Let $b_{jk} = a_{jk} \Big/ \sum_{k \in b_T} L_k a_{jk}$ .

$$1-\bar{V}_{q2,rel} = 2\left\{ \frac{\sum_{k \in b_T} L_k \left(b_{1k} + b_{2k}\right)^q}{\sum_{k \in b_T} L_k \left[\left(b_{1k}\right)^q + \left(b_{2k}\right)^q\right]} \right\}^{\frac{1}{1-q}} -1,\ q \geq 0,\ q \neq 1$$

Adding the new evolutionary unit can be viewed as adding a positive value A such that

$$1-\bar{V}_{q2,rel}\left(1,2\,|\,Te\right) = 2\left\{ \frac{\sum_{k \in b_T} L_k \left(\frac{b_{1k}}{1+A} + \frac{b_{2k}}{1+A}\right)^q + \left(\frac{2A}{1+A}\right)^q}{\sum_{k \in b_T} L_k \left[\left(\frac{b_{1k}}{1+A}\right)^q + \left(\frac{b_{2k}}{1+A}\right)^q\right] + 2\left(\frac{A}{1+A}\right)^q} \right\}^{\frac{1}{1-q}} -1,\ q \geq 0,\ q \neq 1$$

$$= 2\left\{ \frac{\sum_{k \in b_T} L_k \left(b_{1k} + b_{2k}\right)^q + 2^q A^q}{\sum_{k \in b_T} L_k \left[\left(b_{1k}\right)^q + \left(b_{2k}\right)^q\right] + 2A^q} \right\}^{\frac{1}{1-q}} -1$$

The demonstration can thus follow the same reasoning as for $1-\bar{V}_{q2,abs}$ with absolute abundances. With relative abundances, $1-\bar{V}_{q2,rel}$ and ${}^q evoD_{Rényi,rel}$, $q \neq 1$, thus satisfy P4.

With relative abundances and when $q \to 1$, the limit of $1-\bar{V}_{q2,rel}$ is

$$1-\bar{V}_{q2,rel}\left(1,2\,|\,T\right) = 2\exp\left[ -\frac{1}{2}\sum_{k \in b_T} L_k \left(\frac{a_{1k}}{\sum_{k \in b_T} L_k a_{1k}}\right) \ln\left(1 + \frac{\frac{a_{2k}}{\sum_{k \in b_T} L_k a_{2k}}}{\frac{a_{1k}}{\sum_{k \in b_T} L_k a_{1k}}}\right) - \frac{1}{2}\sum_{k \in b_T} L_k \left(\frac{a_{2k}}{\sum_{k \in b_T} L_k a_{2k}}\right) \ln\left(1 + \frac{\frac{a_{1k}}{\sum_{k \in b_T} L_k a_{1k}}}{\frac{a_{2k}}{\sum_{k \in b_T} L_k a_{2k}}}\right) \right] -1$$

This can be simplified as

$$1-\bar{V}_{12,rel}\left(1,2\,|\,T\right) = 2\exp\left[ -\frac{1}{2}\sum_{k \in b_T} L_k b_{1k} \log\left(1 + \frac{b_{2k}}{b_{1k}}\right) - \frac{1}{2}\sum_{k \in b_T} L_k b_{2k} \log\left(1 + \frac{b_{1k}}{b_{2k}}\right) \right] -1$$

Adding the new evolutionary unit can be viewed as adding a positive value $A$ so that

$$1-\bar{V}_{12,rel}\left(1,2\,|\,Te\right) = 2\exp\left[ -\frac{1}{2}\sum_{k \in b_T} L_k \frac{b_{1k}}{1+A} \ln\left(1 + \frac{b_{2k}}{b_{1k}}\right) - \frac{1}{2}\sum_{k \in b_T} L_k \frac{b_{2k}}{1+A} \ln\left(1 + \frac{b_{1k}}{b_{2k}}\right) - \frac{A}{1+A}\ln\left(2\right) \right] -1$$

$$= 2\exp\left\{ \left(\frac{1}{1+A}\right)\left[ -\frac{1}{2}\sum_{k \in b_T} L_k b_{1k} \ln\left(1 + \frac{b_{2k}}{b_{1k}}\right) - \frac{1}{2}\sum_{k \in b_T} L_k b_{2k} \ln\left(1 + \frac{b_{1k}}{b_{2k}}\right) - A\ln\left(2\right) \right] \right\} -1$$

The value $A$ decreases $1-\bar{V}_{12,rel}$ iff

$$\left(\frac{1}{1+A}\right)\left[-\frac{1}{2}\sum_{k\in b_T} L_k b_{1k}\ln\left(1+\frac{b_{2k}}{b_{1k}}\right)-\frac{1}{2}\sum_{k\in b_T} L_k b_{2k}\ln\left(1+\frac{b_{1k}}{b_{2k}}\right)-A\ln(2)\right]$$
$$\leq -\frac{1}{2}\sum_{k\in b_T} L_k b_{1k}\ln\left(1+\frac{b_{2k}}{b_{1k}}\right)-\frac{1}{2}\sum_{k\in b_T} L_k b_{2k}\ln\left(1+\frac{b_{1k}}{b_{2k}}\right)$$

The inequality is always true because $A$>0 and ln(2)>0. With relative abundances, $1-\bar{V}_{12,rel}$ and ${}^{1}evoD_{Rényi,rel}$ satisfy P4.

With absolute abundances,

$$1-\bar{C}_{q2,abs}(1,2\,|\,T)=\frac{1}{1-2^{1-q}}\left\{1-2^{1-q}\frac{\sum_{k\in b_T} L_k\left(a_{1k}+a_{2k}\right)^q}{\sum_{k\in b_T} L_k\left[\left(a_{1k}\right)^q+\left(a_{2k}\right)^q\right]}\right\}$$

Adding the new evolutionary unit,

$$1-\bar{C}_{q2,abs}(1,2\,|\,Te)=\frac{1}{1-2^{1-q}}\left\{1-2^{1-q}\frac{\sum_{k\in b_T} L_k\left(a_{1k}+a_{2k}\right)^q+2^qA^q}{\sum_{k\in b_T} L_k\left[\left(a_{1k}\right)^q+\left(a_{2k}\right)^q\right]+2A^q}\right\}$$

Consider $u=L_k\left(a_{1k}+a_{2k}\right)^q$, $x=2^qA^q$, $v=\sum_{k\in b_T} L_k\left[\left(a_{1k}\right)^q+\left(a_{2k}\right)^q\right]$ and $y=2A^q$.

$$\frac{1}{1-2^{1-q}}\left\{1-2^{1-q}\frac{u}{v}\right\}\geq\frac{1}{1-2^{1-q}}\left\{1-2^{1-q}\frac{u+x}{v+y}\right\}\Leftrightarrow\begin{cases}1-2^{1-q}\dfrac{u}{v}\leq 1-2^{1-q}\dfrac{u+x}{v+y} & q<1\\ 1-2^{1-q}\dfrac{u}{v}\geq 1-2^{1-q}\dfrac{u+x}{v+y} & q>1\end{cases}$$

$$\Leftrightarrow\begin{cases}2^{1-q}\dfrac{u}{v}\geq 2^{1-q}\dfrac{u+x}{v+y} & q<1\\ 2^{1-q}\dfrac{u}{v}\leq 2^{1-q}\dfrac{u+x}{v+y} & q>1\end{cases}\Leftrightarrow\begin{cases}\dfrac{u}{v}\geq\dfrac{u+x}{v+y} & q<1\\ \dfrac{u}{v}\leq\dfrac{u+x}{v+y} & q>1\end{cases}$$

The demonstration can then follow that of $1-\bar{V}_{q2,abs}$ or

$$\Leftrightarrow\begin{cases}\dfrac{v+y}{v}\geq\dfrac{u+x}{u} & q<1\\ \dfrac{v+y}{v}\leq\dfrac{u+x}{u} & q>1\end{cases}\Leftrightarrow\begin{cases}\dfrac{y}{v}\geq\dfrac{x}{u} & q<1\\ \dfrac{y}{v}\leq\dfrac{x}{u} & q>1\end{cases}\Leftrightarrow\begin{cases}\dfrac{u}{v}\geq\dfrac{x}{y} & q<1\\ \dfrac{u}{v}\leq\dfrac{x}{y} & q>1\end{cases}\Leftrightarrow\begin{cases}\dfrac{u}{v}\geq\left(\dfrac{1}{2}\right)^{1-q} & q<1\\ \dfrac{u}{v}\leq\left(\dfrac{1}{2}\right)^{1-q} & q>1\end{cases}\Leftrightarrow\begin{cases}2^{1-q}\dfrac{u}{v}\geq 1 & q<1\\ 2^{1-q}\dfrac{u}{v}\leq 1 & q>1\end{cases}$$

$$\Leftrightarrow\begin{cases}-2^{1-q}\dfrac{u}{v}\leq -1 & q<1\\ -2^{1-q}\dfrac{u}{v}\geq -1 & q>1\end{cases}\Leftrightarrow\begin{cases}1-2^{1-q}\dfrac{u}{v}\leq 0 & q<1\\ 1-2^{1-q}\dfrac{u}{v}\geq 0 & q>1\end{cases}\Leftrightarrow\begin{cases}\dfrac{1}{1-2^{1-q}}\left[1-2^{1-q}\dfrac{u}{v}\right]\geq 0 & q<1\\ \dfrac{1}{1-2^{1-q}}\left[1-2^{1-q}\dfrac{u}{v}\right]\geq 0 & q>1\end{cases}$$

When $q\to 1$, $1-\bar{C}_{12,abs}={}^{1}evoD_{Rényi,abs}$ (see above).

With relative abundances,

$$1-\bar{C}_{q2,rel}(1,2|T)=\frac{1}{1-2^{1-q}}\left\{1-2^{1-q}\frac{\sum_{k\in b_T}L_k\left(\frac{a_{1k}}{\sum_{k\in b_T}L_k a_{1k}}+\frac{a_{2k}}{\sum_{k\in b_T}L_k a_{2k}}\right)^q}{\sum_{k\in b_T}L_k\left[\left(\frac{a_{1k}}{\sum_{k\in b_T}L_k a_{1k}}\right)^q+\left(\frac{a_{2k}}{\sum_{k\in b_T}L_k a_{2k}}\right)^q\right]}\right\}$$

Considering $b_{jk}=a_{jk}/\sum_{k\in b_T}L_k a_{jk}$ for any site $j$ and branch $k$

$$1-\bar{C}_{q2,rel}(1,2|T)=\frac{1}{1-2^{1-q}}\left\{1-2^{1-q}\frac{\sum_{k\in b_T}L_k\left(b_{1k}+b_{2k}\right)^q}{\sum_{k\in b_T}L_k\left[\left(b_{1k}\right)^q+\left(b_{2k}\right)^q\right]}\right\}$$

Adding the new evolutionary unit,

$$1-\bar{C}_{q2,rel}\left(1,2|Te\right)=\frac{1}{1-2^{1-q}}\left\{1-2^{1-q}\frac{\sum_{k\in b_T}L_k\left(\frac{b_{1k}}{1+A}+\frac{b_{2k}}{1+A}\right)^q+\left(\frac{2A}{1+A}\right)^q}{\sum_{k\in b_T}L_k\left[\left(\frac{b_{1k}}{1+A}\right)^q+\left(\frac{b_{2k}}{1+A}\right)^q\right]+2\left(\frac{A}{1+A}\right)^q}\right\}$$

$$=\frac{1}{1-2^{1-q}}\left\{1-2^{1-q}\frac{\sum_{k\in b_T}L_k\left(b_{1k}+b_{2k}\right)^q+2^qA^q}{\sum_{k\in b_T}L_k\left[\left(b_{1k}\right)^q+\left(b_{2k}\right)^q\right]+2A^q}\right\}$$

The demonstration can thus follow the same reasoning as for $1-\bar{C}_{q2,abs}$.

When $q\rightarrow 1$, $1-\bar{C}_{12,rel}={}^{1}evoD_{Rényi,rel}$ (see above).

With absolute abundances,

$$1-\bar{U}_{q2,abs}(1,2|T)=\frac{1}{1-2^{q-1}}\left\{1-2^{q-1}\frac{\sum_{k\in b_T}L_k\left[\left(a_{1k}\right)^q+\left(a_{2k}\right)^q\right]}{\sum_{k\in b_T}L_k\left(a_{1k}+a_{2k}\right)^q}\right\}$$

Adding the new evolutionary unit,

$$1-\bar{U}_{q2,abs}\left(1,2|Te\right)=\frac{1}{1-2^{q-1}}\left\{1-2^{q-1}\frac{\sum_{k\in b_T}L_k\left[\left(a_{1k}\right)^q+\left(a_{2k}\right)^q\right]+2A^q}{\sum_{k\in b_T}L_k\left(a_{1k}+a_{2k}\right)^q+2^qA^q}\right\}$$

Consider $u=L_k\left(a_{1k}+a_{2k}\right)^q$, $x=2^qA^q$, $v=\sum_{k\in b_T}L_k\left[\left(a_{1k}\right)^q+\left(a_{2k}\right)^q\right]$ and $y=2A^q$.

$$\frac{1}{1-2^{q-1}}\left\{1-2^{q-1}\frac{v}{u}\right\}\geq\frac{1}{1-2^{1-q}}\left\{1-2^{1-q}\frac{v+y}{u+x}\right\}\Leftrightarrow\begin{cases}1-2^{q-1}\frac{v}{u}\geq 1-2^{q-1}\frac{v+y}{u+x} & q<1\\ 1-2^{q-1}\frac{v}{u}\leq 1-2^{q-1}\frac{v+y}{u+x} & q>1\end{cases}$$

$$\Leftrightarrow\begin{cases}\frac{v}{u}\leq\frac{v+y}{u+x} & q<1\\ \frac{v}{u}\geq\frac{v+y}{u+x} & q>1\end{cases}\Leftrightarrow\begin{cases}\frac{u}{v}\geq\frac{u+x}{v+y} & q<1\\ \frac{u}{v}\leq\frac{u+x}{v+y} & q>1\end{cases}$$

Same demonstration as above.

When $q\rightarrow 1$, $1-\bar{U}_{12,abs}=1-\bar{C}_{12,abs}={}^{1}evoD_{Rényi,abs}$ (see above).

$$1-\bar{U}_{q2,rel}(1,2\,|\,T)=\frac{1}{1-2^{q-1}}\left\{1-2^{q-1}\frac{\sum_{k\in b_T}L_k\left[\left(\frac{a_{1k}}{\sum_{k\in b_T}L_k a_{1k}}\right)^q+\left(\frac{a_{2k}}{\sum_{k\in b_T}L_k a_{2k}}\right)^q\right]}{\sum_{k\in b_T}L_k\left(\frac{a_{1k}}{\sum_{k\in b_T}L_k a_{1k}}+\frac{a_{2k}}{\sum_{k\in b_T}L_k a_{2k}}\right)^q}\right\}$$

Considering $b_{jk}=a_{jk}/\sum_{k\in b_T}L_k a_{jk}$ for any site *j* and branch *k*,

$$1-\bar{U}_{q2,rel}(1,2\,|\,T)=\frac{1}{1-2^{q-1}}\left\{1-2^{q-1}\frac{\sum_{k\in b_T}L_k\left[(b_{1k})^q+(b_{2k})^q\right]}{\sum_{k\in b_T}L_k\left(b_{1k}+b_{2k}\right)^q}\right\}$$

Adding the new evolutionary unit,

$$1-\bar{U}_{q2,rel}\left(1,2\,|\,Te\right)=\frac{1}{1-2^{q-1}}\left\{1-2^{q-1}\frac{\sum_{k\in b_T}L_k\left[\left(\frac{b_{1k}}{1+A}\right)^q+\left(\frac{b_{2k}}{1+A}\right)^q\right]+2\left(\frac{A}{1+A}\right)^q}{\sum_{k\in b_T}L_k\left(\frac{b_{1k}}{1+A}+\frac{b_{2k}}{1+A}\right)^q+\left(\frac{2A}{1+A}\right)^q}\right\},\ q\geq 0,\ q\neq 1$$

$$=\frac{1}{1-2^{q-1}}\left\{1-2^{q-1}\frac{\sum_{k\in b_T}L_k\left[(b_{1k})^q+(b_{2k})^q\right]+2A^q}{\sum_{k\in b_T}L_k\left(b_{1k}+b_{2k}\right)^q+2^qA^q}\right\}$$

The demonstration can thus follow the same reasoning as for $1-\bar{U}_{q2,abs}$.

When $q\rightarrow 1$, $1-\bar{U}_{12,rel}=1-\bar{C}_{12,rel}={}^1evoD_{Rényi,rel}$ (see above).

$1-\bar{C}_{q2,abs}$, $1-\bar{C}_{q2,rel}$, $1-\bar{U}_{q2,abs}$, $1-\bar{U}_{q2,rel}$ thus satisfy P4.

● Property P5 (sites without species in common have the largest dissimilarity) was satisfied by most of the indices (all except $evoD_{Minkowski}$ (that includes $evoD_{Manhattan}$ and $evoD_{Euclidean}$), $evoD_{Profile}$, and $evoD_{\chi^2}$).

● Legendre and De Cáceres provided two versions of Property P6

**I start with the strong version:**

Strong P6 considers that the dissimilarity should not decrease in series of nested species assemblages. In the PD-dissimilarity context, P6 thus considers that the dissimilarity should not decrease in series of nested assemblages of evolutionary units: the dissimilarity should be the same or increase with the number of unique evolutionary units in a site.

**For indices that depend on a parameter *q*, I consider that a parametric index does not satisfy strong P6 as soon as it does not satisfy it for a value of *q*.**

Indices that use Nipperess et al. (2010) components *a*, *b*, *c* or *A*, *B*, C as presented in the main text satisfy P6. Indeed, when an evolutionary unit is added with a positive abundance in, say, site2. Nipperess et al. components *a*, *b*, *A*, and *B* are unchanged, whereas *c*, and *C* are increased.

To test for P6, I have added a new evolutionary unit in one site with an abundance value equal to *A*. Then I demonstrated that the coefficient of PD-dissimilarity is a non-decreasing function of *A*. If a coefficient of PD-dissimilarity is a non-decreasing function of *A*, then adding an evolutionary unit with any positive abundance increases the value of the coefficient or leaves it unchanged.

For $evoD_{Chord}$, adding the new evolutionary unit to site 2 can be viewed as adding a positive value *A* so that

$$evoD_{Chord}\left(1,2\,|\,Te\right)=\sqrt{\sum_{k\in b_T}L_k\left(b_{1k}-\frac{b_{2k}}{\sqrt{\sum_{k\in b_T}L_k b_{2k}^2+A^2}}\right)^2+\left(0-\frac{A}{\sqrt{\sum_{k\in b_T}L_k b_{2k}^2+A^2}}\right)^2}$$

$$=\sqrt{\sum_{k\in b_T}L_k\left(b_{1k}-\frac{b_{2k}}{\sqrt{1+A^2}}\right)^2+\left(\frac{A}{\sqrt{1+A^2}}\right)^2}$$

Here $b_{jk}=a_{jk}/\sqrt{\sum_{k\in b_T}L_k a_{jk}^2}$ . Consider the function *h* defined as follows:

$$h(A)=\sum_{k\in b_T}L_k\left(b_{1k}-\frac{b_{2k}}{\sqrt{1+A^2}}\right)^2+\left(\frac{A}{\sqrt{1+A^2}}\right)^2$$

$$=\frac{1}{1+A^2}\left[\sum_{k\in b_T}L_k\left(b_{1k}\sqrt{1+A^2}-b_{2k}\right)^2+A^2\right]$$

$$\frac{dh}{A}=-\frac{2A}{\left(1+A^2\right)^2}\left[\sum_{k\in b_T}L_k\left(b_{1k}\sqrt{1+A^2}-b_{2k}\right)^2+A^2\right]+\frac{1}{1+A^2}\left[\sum_{k\in b_T}L_k\frac{\left(b_{1k}\sqrt{1+A^2}-b_{2k}\right)b_{1k}2A}{\sqrt{1+A^2}}+2A\right]$$

$$=\frac{2A}{\left(1+A^2\right)^2}\left[\sum_{k\in b_T}L_k b_{1k}\sqrt{1+A^2}\left(b_{1k}\sqrt{1+A^2}-b_{2k}\right)+1+A^2-\sum_{k\in b_T}L_k\left(b_{1k}\sqrt{1+A^2}-b_{2k}\right)^2-A^2\right]$$

$$=\frac{2A}{\left(1+A^2\right)^2}\left[1+\sum_{k\in b_T}L_k b_{1k}\sqrt{1+A^2}\left(b_{1k}\sqrt{1+A^2}-b_{2k}\right)-\sum_{k\in b_T}L_k\left(b_{1k}\sqrt{1+A^2}-b_{2k}\right)^2\right]$$

Given that $2A/(1+A^2)^2\geq 0$ and that

$$1+\sum_{k\in b_T} L_k b_{1k}\sqrt{1+A^2}\left(b_{1k}\sqrt{1+A^2}-b_{2k}\right)-\sum_{k\in b_T} L_k\left(b_{1k}\sqrt{1+A^2}-b_{2k}\right)^2$$
$$=1+\sum_{k\in b_T} L_k b_{2k}\left(b_{1k}\sqrt{1+A^2}-b_{2k}\right)=1+\sum_{k\in b_T} L_k b_{1k} b_{2k}\sqrt{1+A^2}-\sum_{k\in b_T} L_k\left(b_{2k}\right)^2$$
$$=\sum_{k\in b_T} L_k b_{1k} b_{2k}\sqrt{1+A^2}\geq 0$$

$evoD_{Chord}$ satisfies P6.

For $evoD_{Hellinger}$, adding the new evolutionary unit can be viewed as adding a positive value $A$ so that

$$evoD_{Hellinger}\left(1,2\,|\,Te\right)=\sqrt{\sum_{k\in b_T} L_k\left(\sqrt{b_{1k}}-\sqrt{\frac{b_{2k}}{\sum_{k\in b_T} L_k b_{2k}+A}}\right)^2+\left(0-\sqrt{\frac{A}{\sum_{k\in b_T} L_k b_{2k}+A}}\right)^2}$$
$$=\sqrt{\sum_{k\in b_T} L_k\left(\sqrt{b_{1k}}-\sqrt{\frac{b_{2k}}{1+A}}\right)^2+\frac{A}{1+A}}$$

Here $b_{jk}=a_{jk}/\sum_{k\in b_T} L_k a_{jk}$. Consider the function $h$ defined as follows:

$$h(A)=\sum_{k\in b_T} L_k\left(\sqrt{b_{1k}}-\sqrt{\frac{b_{2k}}{1+A}}\right)^2+\frac{A}{1+A}$$

$$h(A)=\frac{1}{(1+A)}\left[\sum_{k\in b_T} L_k\left(\sqrt{b_{1k}}\sqrt{1+A}-\sqrt{b_{2k}}\right)^2+A\right]$$

$$\frac{dh}{dA}=-\frac{1}{(1+A)^2}\left[\sum_{k\in b_T} L_k\left(\sqrt{b_{1k}}\sqrt{1+A}-\sqrt{b_{2k}}\right)^2+A\right]+\frac{1}{1+A}\left[\sum_{k\in b_T} L_k\frac{\sqrt{b_{1k}}}{\sqrt{1+A}}\left(\sqrt{b_{1k}}\sqrt{1+A}-\sqrt{b_{2k}}\right)+1\right]$$
$$=\frac{1}{(1+A)^2}\left[\sum_{k\in b_T} L_k\sqrt{b_{1k}}\sqrt{1+A}\left(\sqrt{b_{1k}}\sqrt{1+A}-\sqrt{b_{2k}}\right)+1+A-A-\sum_{k\in b_T} L_k\left(\sqrt{b_{1k}}\sqrt{1+A}-\sqrt{b_{2k}}\right)^2\right]$$
$$=\frac{1}{(1+A)^2}\left[1+\sum_{k\in b_T} L_k\sqrt{b_{1k}}\sqrt{1+A}\left(\sqrt{b_{1k}}\sqrt{1+A}-\sqrt{b_{2k}}\right)-\sum_{k\in b_T} L_k\left(\sqrt{b_{1k}}\sqrt{1+A}-\sqrt{b_{2k}}\right)^2\right]$$
$$=\frac{1}{(1+A)^2}\left[1+\sum_{k\in b_T} L_k\sqrt{b_{2k}}\left(\sqrt{b_{1k}}\sqrt{1+A}-\sqrt{b_{2k}}\right)\right]=\frac{1}{(1+A)^2}\left[1+\sum_{k\in b_T} L_k\sqrt{b_{2k}}\left(\sqrt{b_{1k}}\sqrt{1+A}\right)-\sum_{k\in b_T} L_k b_{2k}\right]$$
$$=\frac{1}{(1+A)^2}\left[\sum_{k\in b_T} L_k\sqrt{b_{2k}}\left(\sqrt{b_{1k}}\sqrt{1+A}\right)\right]\geq 0$$

$evoD_{Hellinger}$ thus satisfies P6.

As observed by Legendre and De Cáceres (2013) for specific compositional data, $evoD_{Profile}$ does not satisfy P6. Consider three communities C1, C2, C3 and four species S1, S2, S3, S4.

Consider that the four species have independent evolution (star phylogeny with unit branch lengths): i.e. each species contributes an independent evolutionary unit. The distributions of species abundance in the communities are

C1 : S1=9, S2=1, S3=0, S4=0
C2 : S1=0, S2=5, S3=5, S4=0
C3 : S1=9, S2=1, S3=0, S4=9

Take C2 as the reference community. Compared to C1, C3 contains a new species (S4) and thus a new evolutionary unit.
$evoD_{Profile}$ (C1,C2)≈ 1.10453
$evoD_{Profile}$ (C3,C2)≈ 0.94810
This provides a counter-example for $evoD_{Profile}$.

Index $evoD_{\chi^2}$ and the $\chi^2$ distance do not satisfy P6 as shown by the following counter-example (same data as for $evoD_{Profile}$):

$evoD_{\chi^2}$(C1,C2)≈ 2.14392
$evoD_{\chi^2}$(C3,C2)≈ 2.12685
This provides a counter-example for $evoD_{\chi^2}$.

This result is related to the fact that $evoD_{\chi^2}$ and the $\chi^2$ distance do not satisfy P5 and to the fact that $evoD_{\chi^2}$ and the $\chi^2$ distance are sensitive to the evenness of the abundances within each site.

For the indices derived from the decomposition of Hill numbers, it is sufficient to analyze $1-\bar{V}_{q2}$ as the logarithm function used in ${}^{q}evoD_{Rényi}$, is monotonously increasing.

With absolute abundances, the behavior of $1-\bar{V}_{q2,abs}$ ($q$>0, $q$≠1) when adding the new evolutionary unit, can be studied thanks to the function (see also Chiu et al. 2014, appendix 3)

$$h(A)=\left\{\frac{\sum_{k\in b_T} L_k\left(a_{1k}+a_{2k}\right)^q+A^q}{\sum_{k\in b_T} L_k\left[\left(a_{1k}\right)^q+\left(a_{2k}\right)^q\right]+A^q}\right\}^{\frac{1}{1-q}}$$

Let $u=\sum_{k\in b_T} L_k\left(a_{1k}+a_{2k}\right)^q$ , $v=\sum_{k\in b_T} L_k\left[\left(a_{1k}\right)^q+\left(a_{2k}\right)^q\right]$, $x=A^q$

$$\left\{\frac{u+x}{v+x}\right\}^{\frac{1}{1-q}}\geq\left\{\frac{u}{v}\right\}^{\frac{1}{1-q}}\Leftrightarrow\begin{cases}\dfrac{u+x}{v+x}\geq\dfrac{u}{v} & q<1\\ \dfrac{u+x}{v+x}\leq\dfrac{u}{v} & q>1\end{cases}\Leftrightarrow\begin{cases}\dfrac{u+x}{u}\geq\dfrac{v+x}{v} & q<1\\ \dfrac{u+x}{u}\leq\dfrac{v+x}{v} & q>1\end{cases}\Leftrightarrow\begin{cases}\dfrac{x}{u}\geq\dfrac{x}{v} & q<1\\ \dfrac{x}{u}\leq\dfrac{x}{v} & q>1\end{cases}\Leftrightarrow\begin{cases}1\geq\dfrac{u}{v} & q<1\\ 1\leq\dfrac{u}{v} & q>1\end{cases}$$

$1-\bar{V}_{q2,abs}$ and ${}^{q}evoD_{Rényi,abs}$, $q$>0, $q$≠1, thus satisfy P6.

With absolute abundances, when $q \to 1$, $1-\bar{V}_{q2}$ can be written as

$$1-\bar{V}_{12,abs}\left(1,2\,|\,T\right)=2\exp\left[-\frac{1}{\sum_{k\in b_T} L_k\left(a_{1k}+a_{2k}\right)}\sum_{k\in b_T} L_k a_{1k}\ln\left(1+\frac{a_{2k}}{a_{1k}}\right)-\frac{1}{\sum_{k\in b_T} L_k\left(a_{1k}+a_{2k}\right)}\sum_{k\in b_T} L_k a_{2k}\ln\left(1+\frac{a_{1k}}{a_{2k}}\right)\right]-1$$

Consider

$$1-\bar{V}_{12,abs}\left(1,2\,|\,Te\right)=2\exp\left[\begin{array}{l}-\dfrac{1}{\sum_{k\in b_T} L_k\left(a_{1k}+a_{2k}\right)+A}\sum_{k\in b_T} L_k a_{1k}\ln\left(1+\dfrac{a_{2k}}{a_{1k}}\right)\\ -\dfrac{1}{\sum_{k\in b_T} L_k\left(a_{1k}+a_{2k}\right)+A}\sum_{k\in b_T} L_k a_{2k}\ln\left(1+\dfrac{a_{1k}}{a_{2k}}\right)\end{array}\right]-1$$

It can be directly concluded that $1-\bar{V}_{12}(1,2|Te) \geq 1-\bar{V}_{12}(1,2|T)$ . $1-\bar{V}_{12}$ and ${}^{1}evoD_{Rényi,abs}$ thus satisfy P6.

$$1-\bar{C}_{q2,abs}\left(1,2\,|\,T\right)=\frac{1}{1-2^{1-q}}\left\{1-2^{1-q}\frac{\sum_{k\in b_T} L_k\left(a_{1k}+a_{2k}\right)^q}{\sum_{k\in b_T} L_k\left[\left(a_{1k}\right)^q+\left(a_{2k}\right)^q\right]}\right\}$$

Adding the new evolutionary unit,

$$1-\bar{C}_{q2,abs}\left(1,2\,|\,Te\right)=\frac{1}{1-2^{1-q}}\left\{1-2^{1-q}\frac{\sum_{k\in b_T} L_k\left(a_{1k}+a_{2k}\right)^q+A^q}{\sum_{k\in b_T} L_k\left[\left(a_{1k}\right)^q+\left(a_{2k}\right)^q\right]+A^q}\right\}$$

Consider $u=L_k\left(a_{1k}+a_{2k}\right)^q$, $x=A^q$, $v=\sum_{k\in b_T} L_k\left[\left(a_{1k}\right)^q+\left(a_{2k}\right)^q\right]$.

$$\frac{1}{1-2^{1-q}}\left\{1-2^{1-q}\frac{u}{v}\right\}\geq\frac{1}{1-2^{1-q}}\left\{1-2^{1-q}\frac{u+x}{v+x}\right\}\Leftrightarrow\begin{cases}1-2^{1-q}\dfrac{u}{v}\leq 1-2^{1-q}\dfrac{u+x}{v+x} & q<1\\ 1-2^{1-q}\dfrac{u}{v}\geq 1-2^{1-q}\dfrac{u+x}{v+x} & q>1\end{cases}$$

$$\Leftrightarrow\begin{cases}2^{1-q}\dfrac{u}{v}\geq 2^{1-q}\dfrac{u+x}{v+x} & q<1\\ 2^{1-q}\dfrac{u}{v}\leq 2^{1-q}\dfrac{u+x}{v+x} & q>1\end{cases}\Leftrightarrow\begin{cases}\dfrac{u}{v}\geq\dfrac{u+x}{v+x} & q<1\\ \dfrac{u}{v}\leq\dfrac{u+x}{v+x} & q>1\end{cases}$$

The demonstration can then follow that of $1-\bar{V}_{q2,abs}$ or

$$\Leftrightarrow\begin{cases}\dfrac{v+x}{v}\geq\dfrac{u+x}{u} & q<1\\ \dfrac{v+x}{v}\leq\dfrac{u+x}{u} & q>1\end{cases}\Leftrightarrow\begin{cases}\dfrac{x}{v}\geq\dfrac{x}{u} & q<1\\ \dfrac{x}{v}\leq\dfrac{x}{u} & q>1\end{cases}\Leftrightarrow\begin{cases}\dfrac{u}{v}\geq 1 & q<1\\ \dfrac{u}{v}\leq 1 & q>1\end{cases}$$

When $q \to 1$,

$$1-\bar{C}_{12,abs}\left(1,2\,|\,T\right)=-\frac{1}{\sum_{k\in b_T} L_k\left(a_{1k}+a_{2k}\right)}\left[\sum_{k\in b_T} L_k a_{1k}\log_2\left(\frac{a_{1k}}{a_{1k}+a_{2k}}\right)+\sum_{k\in b_T} L_k a_{2k}\log_2\left(\frac{a_{2k}}{a_{1k}+a_{2k}}\right)\right]$$

Adding the new evolutionary unit,

$$1-\bar{C}_{12,abs}(1,2|Te)=1+\frac{1}{\sum_{k\in b_T}L_k(a_{1k}+a_{2k})+A}\left[\sum_{k\in b_T}L_k a_{1k}\log_2\left(\frac{a_{1k}}{a_{1k}+a_{2k}}\right)+\sum_{k\in b_T}L_k a_{2k}\log_2\left(\frac{a_{2k}}{a_{1k}+a_{2k}}\right)+A\log_2\left(\frac{A}{A}\right)\right]$$

$$=1+\frac{1}{\sum_{k\in b_T}L_k(a_{1k}+a_{2k})+A}\left[\sum_{k\in b_T}L_k a_{1k}\log_2\left(\frac{a_{1k}}{a_{1k}+a_{2k}}\right)+\sum_{k\in b_T}L_k a_{2k}\log_2\left(\frac{a_{2k}}{a_{1k}+a_{2k}}\right)\right]$$

Because $\sum_{k\in b_T}L_k a_{1k}\log_2\left(\frac{a_{1k}}{a_{1k}+a_{2k}}\right)+\sum_{k\in b_T}L_k a_{2k}\log_2\left(\frac{a_{2k}}{a_{1k}+a_{2k}}\right)$ is negative and A is positive $1-\bar{C}_{12,abs}(1,2|Te)\geq 1-\bar{C}_{12,abs}(1,2|T)$

$$1-\bar{U}_{q2,abs}(1,2|T)=\frac{1}{1-2^{q-1}}\left\{1-2^{q-1}\frac{\sum_{k\in b_T}L_k\left[(a_{1k})^q+(a_{2k})^q\right]}{\sum_{k\in b_T}L_k(a_{1k}+a_{2k})^q}\right\}$$

Adding the new evolutionary unit,

$$1-\bar{U}_{q2,abs}(1,2|Te)=\frac{1}{1-2^{q-1}}\left\{1-2^{q-1}\frac{\sum_{k\in b_T}L_k\left[(a_{1k})^q+(a_{2k})^q\right]+A^q}{\sum_{k\in b_T}L_k(a_{1k}+a_{2k})^q+A^q}\right\}$$

Consider $u=L_k(a_{1k}+a_{2k})^q$, $x=A^q$, $v=\sum_{k\in b_T}L_k\left[(a_{1k})^q+(a_{2k})^q\right]$.

$$\frac{1}{1-2^{q-1}}\left\{1-2^{q-1}\frac{v}{u}\right\}\geq\frac{1}{1-2^{1-q}}\left\{1-2^{1-q}\frac{v+x}{u+x}\right\}\Leftrightarrow\begin{cases}1-2^{q-1}\frac{v}{u}\geq 1-2^{q-1}\frac{v+x}{u+x} & q<1\\ 1-2^{q-1}\frac{v}{u}\leq 1-2^{q-1}\frac{v+x}{u+x} & q>1\end{cases}$$

$$\Leftrightarrow\begin{cases}\frac{v}{u}\leq\frac{v+x}{u+x} & q<1\\ \frac{v}{u}\geq\frac{v+x}{u+x} & q>1\end{cases}\Leftrightarrow\begin{cases}\frac{u}{v}\geq\frac{u+x}{v+x} & q<1\\ \frac{u}{v}\leq\frac{u+x}{v+x} & q>1\end{cases}$$

Same demonstration as above.

$1-\bar{C}_{q2,abs}$ and $1-\bar{U}_{q2,abs}$ satisfy P6.

Consider three communities C1, C2, C3 and four species S1, S2, S3, S4. Consider that the four species have independent evolution (star phylogeny with unit branch lengths): i.e. each species contributes an independent evolutionary unit. The distributions of species abundance in the communities are

C1 : S1=9, S2=1, S3=0, S4=0
C2 : S1=0, S2=5, S3=5, S4=0
C3 : S1=9, S2=1, S3=0, S4=1

Take C2 as the reference community. Compared to C1, C3 contains a new species and thus a new evolutionary unit.

$1-\bar{U}_{22,rel}=1-\bar{V}_{22,rel}$ (C1,C2)≈0.85916

$1-\bar{U}_{22,rel}=1-\bar{V}_{22,rel}$ (C3,C2)≈0.85761

This provides a counter-example for $1-\bar{V}_{q2,rel}$ and $1-\bar{U}_{q2,rel}$. In addition,

$1-\bar{C}_{22,rel}$ (C1,C2)≈ 0.92424
$1-\bar{C}_{22,rel}$ (C3,C2)≈ 0.92334

It thus also provides a counter-example for $1-\bar{C}_{q2,rel}$.

Note that even if these indices do not satisfy the stronger version of P6 for all values of *q*. They do satisfy them for some values of *q*. For example, $1-\bar{U}_{02,rel}=1-\bar{U}_{02,abs}$, $1-\bar{C}_{02,rel}=1-\bar{C}_{02,abs}$, $1-\bar{V}_{02,rel}=1-\bar{V}_{02,abs}$, ${}^{0}evoD_{Rényi,rel}={}^{0}evoD_{Rényi,abs}$, all satisfy the stronger version of P6.

Also, $1-\bar{U}_{12,rel}$, $1-\bar{C}_{12,rel}$, $1-\bar{V}_{12,rel}$, ${}^{1}evoD_{Rényi,rel}$, all satisfy the stronger version of P6.

<u>Proof:</u>
Indeed, $1-\bar{U}_{12,rel}=1-\bar{C}_{12,rel}={}^{1}evoD_{Rényi,rel}$, ${}^{1}evoD_{Rényi,rel}=\log(2-\bar{V}_{12,rel})$, and

$$1-\bar{V}_{12,rel}(1,2|T)=2\exp\left[-\frac{1}{2}\sum_{k\in b_T}L_k\left(\frac{a_{1k}}{\sum_{k\in b_T}L_k a_{1k}}\right)\ln\left(1+\frac{\frac{a_{2k}}{\sum_{k\in b_T}L_k a_{2k}}}{\frac{a_{1k}}{\sum_{k\in b_T}L_k a_{1k}}}\right)-\frac{1}{2}\sum_{k\in b_T}L_k\left(\frac{a_{2k}}{\sum_{k\in b_T}L_k a_{2k}}\right)\ln\left(1+\frac{\frac{a_{1k}}{\sum_{k\in b_T}L_k a_{1k}}}{\frac{a_{2k}}{\sum_{k\in b_T}L_k a_{2k}}}\right)\right]-1$$

Let $b_{jk}=a_{jk}\Big/\sum_{k\in b_T}L_k a_{jk}$.

$$1-\bar{V}_{12,rel}(1,2|T)=2\exp\left[-\frac{1}{2}\sum_{k\in b_T}L_k b_{1k}\log\left(1+\frac{b_{2k}}{b_{1k}}\right)-\frac{1}{2}\sum_{k\in b_T}L_k b_{2k}\log\left(1+\frac{b_{1k}}{b_{2k}}\right)\right]-1$$

Adding the new evolutionary unit can be viewed as adding a positive value *A* so that

$$1-\bar{V}_{12,rel}(1,2|Te)=2\exp\left[-\frac{1}{2}\sum_{k\in b_T}L_k b_{1k}\ln\left(1+\frac{b_{2k}}{(1+A)b_{1k}}\right)-\frac{1}{2}\sum_{k\in b_T}L_k\frac{b_{2k}}{1+A}\ln\left(1+\frac{(1+A)b_{1k}}{b_{2k}}\right)-\frac{1}{2}\frac{A}{1+A}\ln(1)\right]-1$$

$$=2\exp\left\{-\frac{1}{2}\sum_{k\in b_T}L_k b_{1k}\ln\left(\frac{(1+A)b_{1k}+b_{2k}}{(1+A)b_{1k}}\right)-\frac{1}{2}\sum_{k\in b_T}L_k\frac{b_{2k}}{1+A}\ln\left(\frac{(1+A)b_{1k}+b_{2k}}{b_{2k}}\right)\right\}-1$$

$$f(A)=-\frac{1}{2}\sum_{k\in b_T}L_k b_{1k}\ln\left(\frac{(1+A)b_{1k}+b_{2k}}{(1+A)b_{1k}}\right)-\frac{1}{2}\sum_{k\in b_T}L_k\frac{b_{2k}}{1+A}\ln\left(\frac{(1+A)b_{1k}+b_{2k}}{b_{2k}}\right)$$

$$\frac{\partial f}{\partial A}=+\frac{1}{2}\sum_{k\in b_T}L_k b_{1k}\frac{(1+A)b_{1k}}{(1+A)b_{1k}+b_{2k}}\frac{b_{2k}}{(1+A)^2 b_{1k}}-\frac{1}{2}\sum_{k\in b_T}L_k\frac{b_{2k}}{1+A}\frac{b_{2k}}{(1+A)b_{1k}+b_{2k}}b_{1k}$$

$$\frac{\partial f}{\partial A}=+\frac{1}{2}\sum_{k\in b_T}L_k\frac{b_{1k}b_{2k}}{(1+A)b_{1k}+b_{2k}}\frac{1}{1+A}-\frac{1}{2}\sum_{k\in b_T}L_k\frac{b_{1k}b_{2k}}{(1+A)b_{1k}+b_{2k}}\frac{b_{2k}}{1+A}$$

$$\frac{\partial f}{\partial A}=\frac{1}{2}\sum_{k\in b_T}L_k\frac{b_{1k}b_{2k}}{(1+A)b_{1k}+b_{2k}}\frac{1}{1+A}(1-b_{2k})\geq 0$$

□

For those indices that do not satisfy strong P6 I check the weaker version of P6 proposed by Legendre and De Cáceres using an idea by Jost et al. (2011).
**Weak P6:** adding a species (here an evolutionary unit) with abundance A in a site and another different species (here another evolutionary unit) with the same abundance A in the other site should either leave the dissimilarity unchanged or increase it.

*evoD*$_{Profile}$ does not satisfy the weaker version of P6 :

Consider four communities C1, C2, C3, C4 and five species S1, S2, S3, S4, S5. Consider that the five species have independent evolution (star phylogeny with unit branch lengths): i.e. each species contributes an independent evolutionary unit. The distributions of species abundance in the communities are

C1 : S1=2, S2=0, S3=2, S4=0, S5=0

C2 : S1=0, S2=2, S3=2, S4=0, S5=0

C3 : S1=2, S2=0, S3=2, S4=1, S5=0

C4 : S1=0, S2=2, S3=2, S4=0, S5=1

Compared to C1 and C2, C3 and C4 are more different as they differ by species 4 in C3 and species 5 in C4. However,

*evoD*$_{Profile}$(C1,C2)≈ 0.707

*evoD*$_{Profile}$(C3,C4)≈ 0.632

Consider three communities C1, C2, C3 and five species S1, S2, S3, S4, S5. Consider that the four species have independent evolution (star phylogeny with unit branch lengths): i.e. each species contributes an independent evolutionary unit. The distributions of species abundance in the communities are (Case 1)

C1 : S1=1, S2=1, S3=0, S4=0, S5=0
C2 : S1=0, S2=0, S3=1, S4=0, S5=0
C3 : S1=1, S2=1, S3=1, S4=1, S5=1

*evoD*$_{\chi^2}$ (C1,C2)≈2.44949

Now consider the following compositions for communities C1, C2 (C3 is unchanged) (Case 2)
C1 : S1=1, S2=1, S3=0, S4=2, S5=0
C2 : S1=0, S2=0, S3=1, S4=0, S5=1
C3 : S1=1, S2=1, S3=1, S4=1, S5=1

*evoD*$_{\chi^2}$ (C1,C2)≈2.08666

Considering a set of communities and adding evolutionary units one in C1 and another in C2 with similar relative abundances can thus decrease the dissimilarity among C1 and C2. I thus considered that *evoD*$_{\chi^2}$ does not satisfy the weaker version of P6. However if C1 and C2 are considered independently of any other communities, their dissimilarity might increase with the addition of evolutionary units one in C1 and another in C2 with similar relative abundances. For example, in the example above, removing C3, the dissimilarity between C1 and C2 is higher in case 2 than in case 1.

Another example, where C1 and C2 share species

Consider three communities C1, C2, C3 and five species S1, S2, S3, S4, S5. Consider that the four species have independent evolution (star phylogeny with unit branch lengths): i.e. each species contributes an independent evolutionary unit. The distributions of species abundance in the communities are (Case 1)

C1 : S1=1, S2=0.5, S3=0, S4=0, S5=0
C2 : S1=0, S2=0.5, S3=1, S4=0, S5=0
C3 : S1=1, S2=1, S3=1, S4=1, S5=1

$evoD_{\chi^2}$ (C1,C2)≈1.88562

Now consider the following compositions for communities C1, C2 (C3 is unchanged) (Case 2)
C1 : S1=1, S2=0.5, S3=0, S4=1, S5=0
C2 : S1=0, S2=0.5, S3=1, S4=0, S5=1
C3 : S1=1, S2=1, S3=1, S4=1, S5=1

$evoD_{\chi^2}$ (C1,C2)≈ 1.78885

With relative abundances,

$$1-\bar{V}_{q2,rel}\left(1,2\,|\,T\right)=2\left\{\frac{\sum_{k\in b_T} L_k\left(\frac{a_{1k}}{\sum_{k\in b_T} L_k a_{1k}}+\frac{a_{2k}}{\sum_{k\in b_T} L_k a_{2k}}\right)^q}{\sum_{k\in b_T} L_k\left[\left(\frac{a_{1k}}{\sum_{k\in b_T} L_k a_{1k}}\right)^q+\left(\frac{a_{2k}}{\sum_{k\in b_T} L_k a_{2k}}\right)^q\right]}\right\}^{\frac{1}{1-q}}-1,\ q>0,\ q\neq 1$$

Let $b_{jk}=a_{jk}\Big/\sum_{k\in b_T} L_k a_{jk}$ .

$$1-\bar{V}_{q2,rel}=2\left\{\frac{\sum_{k\in b_T} L_k\left(b_{1k}+b_{2k}\right)^q}{\sum_{k\in b_T} L_k\left[\left(b_{1k}\right)^q+\left(b_{2k}\right)^q\right]}\right\}^{\frac{1}{1-q}}-1,\ q\geq 0,\ q\neq 1$$

Adding the new evolutionary units can be viewed as adding a positive value A such that

$$1-\bar{V}_{q2,rel}\left(1,2\,|\,Te\right)=2\left\{\frac{\sum_{k\in b_T}L_k\left(\frac{b_{1k}}{1+A}+\frac{b_{2k}}{1+A}\right)^q+2\left(\frac{A}{1+A}\right)^q}{\sum_{k\in b_T}L_k\left[\left(\frac{b_{1k}}{1+A}\right)^q+\left(\frac{b_{2k}}{1+A}\right)^q\right]+2\left(\frac{A}{1+A}\right)^q}\right\}^{\frac{1}{1-q}}-1,\ q\geq 0,\ q\neq 1$$

$$=2\left\{\frac{\sum_{k\in b_T}L_k\left(b_{1k}+b_{2k}\right)^q+2A^q}{\sum_{k\in b_T}L_k\left[\left(b_{1k}\right)^q+\left(b_{2k}\right)^q\right]+2A^q}\right\}^{\frac{1}{1-q}}-1$$

The demonstration that $1-\bar{V}_{q2,rel}\left(1,2\,|\,Te\right)\geq 1-\bar{V}_{q2,rel}\left(1,2\,|\,T\right)$ can thus follow the same reasoning as for $1-\bar{V}_{q2,abs}$

With relative abundances and when $q\rightarrow 1$, the limit of $1-\bar{V}_{q2,rel}$ is

$$1-\bar{V}_{12,rel}\left(1,2\,|\,T\right)=2\exp\left[-\frac{1}{2}\sum_{k\in b_T}L_k\left(\frac{a_{1k}}{\sum_{k\in b_T}L_k a_{1k}}\right)\ln\left(1+\frac{\frac{a_{2k}}{\sum_{k\in b_T}L_k a_{2k}}}{\frac{a_{1k}}{\sum_{k\in b_T}L_k a_{1k}}}\right)-\frac{1}{2}\sum_{k\in b_T}L_k\left(\frac{a_{2k}}{\sum_{k\in b_T}L_k a_{2k}}\right)\ln\left(1+\frac{\frac{a_{1k}}{\sum_{k\in b_T}L_k a_{1k}}}{\frac{a_{2k}}{\sum_{k\in b_T}L_k a_{2k}}}\right)\right]-1$$

This can be simplified as

$$1-\bar{V}_{12,rel}\left(1,2\,|\,T\right)=2\exp\left[-\frac{1}{2}\sum_{k\in b_T}L_k b_{1k}\log\left(1+\frac{b_{2k}}{b_{1k}}\right)-\frac{1}{2}\sum_{k\in b_T}L_k b_{2k}\log\left(1+\frac{b_{1k}}{b_{2k}}\right)\right]-1$$

Adding the new evolutionary unit can be viewed as adding a positive value $A$ so that

$$1-\bar{V}_{12,rel}\left(1,2\,|\,Te\right)=2\exp\left[-\frac{1}{2}\sum_{k\in b_T}L_k\frac{b_{1k}}{1+A}\ln\left(1+\frac{b_{2k}}{b_{1k}}\right)-\frac{1}{2}\sum_{k\in b_T}L_k\frac{b_{2k}}{1+A}\ln\left(1+\frac{b_{1k}}{b_{2k}}\right)-\frac{A}{1+A}\ln\left(1\right)\right]-1$$

$$=2\exp\left\{\left(\frac{1}{1+A}\right)\left[-\frac{1}{2}\sum_{k\in b_T}L_k b_{1k}\ln\left(1+\frac{b_{2k}}{b_{1k}}\right)-\frac{1}{2}\sum_{k\in b_T}L_k b_{2k}\ln\left(1+\frac{b_{1k}}{b_{2k}}\right)\right]\right\}-1$$

$$1-\bar{V}_{12,rel}\left(1,2\,|\,Te\right)\geq 1-\bar{V}_{12,rel}\left(1,2\,|\,T\right)$$

$1-\bar{V}_{q2,rel}$ thus satisfies the weaker version of P6 and so do ${}^{q}evoD_{Rényi}$.

With relative abundances,

$$1-\bar{C}_{q2,rel}(1,2\,|\,T)=\frac{1}{1-2^{1-q}}\left\{1-2^{1-q}\frac{\sum_{k\in b_T}L_k\left(\frac{a_{1k}}{\sum_{k\in b_T}L_k a_{1k}}+\frac{a_{2k}}{\sum_{k\in b_T}L_k a_{2k}}\right)^q}{\sum_{k\in b_T}L_k\left[\left(\frac{a_{1k}}{\sum_{k\in b_T}L_k a_{1k}}\right)^q+\left(\frac{a_{2k}}{\sum_{k\in b_T}L_k a_{2k}}\right)^q\right]}\right\}$$

Considering $b_{jk}=a_{jk}/\sum_{k\in b_T}L_k a_{jk}$ for any site *j* and branch *k*

$$1-\bar{C}_{q2,rel}(1,2\,|\,T)=\frac{1}{1-2^{1-q}}\left\{1-2^{1-q}\frac{\sum_{k\in b_T}L_k\left(b_{1k}+b_{2k}\right)^q}{\sum_{k\in b_T}L_k\left[\left(b_{1k}\right)^q+\left(b_{2k}\right)^q\right]}\right\}$$

Adding the new evolutionary unit,

$$1-\bar{C}_{q2,rel}\left(1,2\,|\,Te\right)=\frac{1}{1-2^{1-q}}\left\{1-2^{1-q}\frac{\sum_{k\in b_T}L_k\left(\frac{b_{1k}}{1+A}+\frac{b_{2k}}{1+A}\right)^q+2\left(\frac{A}{1+A}\right)^q}{\sum_{k\in b_T}L_k\left[\left(\frac{b_{1k}}{1+A}\right)^q+\left(\frac{b_{2k}}{1+A}\right)^q\right]+2\left(\frac{A}{1+A}\right)^q}\right\}$$

$$=\frac{1}{1-2^{1-q}}\left\{1-2^{1-q}\frac{\sum_{k\in b_T}L_k\left(b_{1k}+b_{2k}\right)^q+2A^q}{\sum_{k\in b_T}L_k\left[\left(b_{1k}\right)^q+\left(b_{2k}\right)^q\right]+2A^q}\right\}$$

The demonstration that $1-\bar{C}_{q2,rel}\left(1,2\,|\,Te\right)\geq 1-\bar{C}_{q2,rel}\left(1,2\,|\,T\right)$ can thus follow the same reasoning as for $1-\bar{C}_{q2,abs}$.

When $q\to 1$, $1-\bar{C}_{12,rel}={}^1evoD_{Rényi,rel}$ (see above).
$1-\bar{C}_{q2,rel}$ thus satisfies the weaker version of P6.

$$1-\bar{U}_{q2,rel}(1,2\,|\,T)=\frac{1}{1-2^{q-1}}\left\{1-2^{q-1}\frac{\sum_{k\in b_T}L_k\left[\left(\frac{a_{1k}}{\sum_{k\in b_T}L_k a_{1k}}\right)^q+\left(\frac{a_{2k}}{\sum_{k\in b_T}L_k a_{2k}}\right)^q\right]}{\sum_{k\in b_T}L_k\left(\frac{a_{1k}}{\sum_{k\in b_T}L_k a_{1k}}+\frac{a_{2k}}{\sum_{k\in b_T}L_k a_{2k}}\right)^q}\right\}$$

Considering $b_{jk}=a_{jk}/\sum_{k\in b_T}L_k a_{jk}$ for any site *j* and branch *k*,

$$1-\bar{U}_{q2,rel}(1,2\,|\,T)=\frac{1}{1-2^{q-1}}\left\{1-2^{q-1}\frac{\sum_{k\in b_T}L_k\left[\left(b_{1k}\right)^q+\left(b_{2k}\right)^q\right]}{\sum_{k\in b_T}L_k\left(b_{1k}+b_{2k}\right)^q}\right\}$$

Adding the new evolutionary units,

$$1-\bar{U}_{q2,rel}\left(1,2\,|\,Te\right)=\frac{1}{1-2^{q-1}}\left\{1-2^{q-1}\frac{\sum_{k\in b_T}L_k\left[\left(\frac{b_{1k}}{1+A}\right)^q+\left(\frac{b_{2k}}{1+A}\right)^q\right]+2\left(\frac{A}{1+A}\right)^q}{\sum_{k\in b_T}L_k\left(\frac{b_{1k}}{1+A}+\frac{b_{2k}}{1+A}\right)^q+2\left(\frac{A}{1+A}\right)^q}\right\},\ q\geq 0,\ q\neq 1$$

$$=\frac{1}{1-2^{q-1}}\left\{1-2^{q-1}\frac{\sum_{k\in b_T}L_k\left[\left(b_{1k}\right)^q+\left(b_{2k}\right)^q\right]+2A^q}{\sum_{k\in b_T}L_k\left(b_{1k}+b_{2k}\right)^q+2A^q}\right\}$$

The demonstration that $1-\bar{U}_{q2,rel}\left(1,2\,|\,Te\right)\geq 1-\bar{U}_{q2,rel}\left(1,2\,|\,T\right)$ can thus follow the same reasoning as for $1-\bar{U}_{q2,abs}$.
When $q\to 1$, $1-\bar{U}_{12,rel}={}^1evoD_{Rényi,rel}$ (see above).

$1-\bar{U}_{q2,rel}$ thus satisfies the weaker version of P6.

● Property P7 (a community composition table with the columns in two or several copies should produce the same dissimilarities among sites as the original data table) was satisfied by most of the indices (all except $evoD_{Minkowski}$ (that includes $evoD_{Manhattan}$ and $evoD_{Euclidean}$) and $evoD_{Profile}$).

● Property P8 (invariance to the measurement units) was satisfied by most of the indices (all except $evoD_{Minkowski}$ (that includes $evoD_{Manhattan}$ and $evoD_{Euclidean}$).

● Property P9 (existence of a fixed upper bound) was satisfied by most of the indices (all except $evoD_{Minkowski}$ (that includes $evoD_{Manhattan}$ and $evoD_{Euclidean}$).

As a conclusion, the properties satisfied by each of the PD-dissimilarity index introduced above are given in Table 2-1.

TABLE 2-1. Legendre and DeCáceres (2013) basic necessary properties[‡] adapted for indices of site-to-site PD-dissimilarity.

| | **P1-3** | **P4** | **P5** | **P6*** | **P7** | **P8** | **P9** |
|---|---|---|---|---|---|---|---|
| $evoD_{Jaccard}$ | 1 | 1 | 1 | 1 | 1 | 1 | 1 |
| $evoD_{Sørensen}$ | 1 | 1 | 1 | 1 | 1 | 1 | 1 |
| $evoD_{Ochiai}$ | 1 | 1 | 1 | 1 | 1 | 1 | 1 |
| $evoD_{TJ}$ | 1 | 1 | 1 | 1 | 1 | 1 | 1 |
| $evoD_{TS}$ | 1 | 1 | 1 | 1 | 1 | 1 | 1 |
| $evoD_{TO}$ | 1 | 1 | 1 | 1 | 1 | 1 | 1 |
| ${}^{q}evoD_{Minkowski}$ | 1 | 0 | 0 | 1 | 0 | 0 | 0 |
| $evoD_{Chord}$ | 1 | 1 | 1 | 1 | 1 | 1 | 1 |
| $evoD_{Hellinger}$ | 1 | 1 | 1 | 1 | 1 | 1 | 1 |
| $evoD_{Profile}$ | 1 | 1 | 0 | 0 | 0 | 1 | 1 |
| $evoD_{\chi^2}$ | 1 | 1 | 0 | 0 | 1 | 1 | 1 |
| ${}^{q}evoD_{Rényi}$, $1-\bar{V}_{q2}$, $1-\bar{C}_{q2}$, $1-\bar{U}_{q2}$ | | | | | | | |
| Absolute abundances | 1 | 1 | 1 | 1 | 1 | 1 | 1 |
| Relative abundances | 1 | 1 | 1 | 1 | 1 | 1 | 1 |
| $evoD_{Bray\text{-}Curtis}$ and $evoD_{Morisita\text{-}Horn}$ | 1 | 1 | 1 | 1 | 1 | 1 | 1 |
| $evoD_{ScaledCanberra}$ and $evoD_{Divergence}$ | 1 | 1 | 1 | 1 | 1 | 1 | 1 |

Notes: 1 indicates that the index fulfils the corresponding property and 0 that it does not. See Appendix 1 for formulas of the indices
‡see above in Appendix 2 for a presentation of the basic necessary properties and for proofs.
*I considered that P6 was satisfied if at least the weaker version was satisfied.

# 5. Developing indices of evodiversity and PD-dissimilarity

They are myriads of indices developed in the literature and all can be adapted to the PD-dissimilarity framework. Many of the developed indices of PD-dissimilarity have simple links. For example, adapting Kulczynski index (Legendre and Legendre 1998 and references therein) to PD-dissimilarity would have yielded to

$$evoD_{Kulczynski} = 1 - \frac{\sum_{k \in b_T} L_k \min_{k \in b_T} \{a_{1k}, a_{2k}\}}{2 / \left(1 / \sum_{k \in b_T} L_k a_{1k} + 1 / \sum_{k \in b_T} L_k a_{2k}\right)}$$

where $2 / \left(1 / \sum_{k \in b_T} L_k a_{1k} + 1 / \sum_{k \in b_T} L_k a_{2k}\right)$ is the harmonic mean between the sum of the abundances of all evolutionary units in site 1 and the same sum in site 2. Changing the harmonic mean for the more widespread arithmetic mean leads to

$$1 - \frac{\sum_{k \in b_T} L_k \min_{k \in b_T} \{a_{1k}, a_{2k}\}}{\frac{1}{2} \sum_{k \in b_T} L_k (a_{1k} + a_{2k})} = \frac{\sum_{k \in b_T} L_k |a_{1k} - a_{2k}|}{\sum_{k \in b_T} L_k (a_{1k} + a_{2k})}$$

that is to say to $evoD_{Bray\text{-}Curtis}$. Changing an arithmetic mean, with a geometric, or harmonic one might have consequences on the values taken by an index and on its mathematical properties, such as, for a dissimilarity index, being Euclidean or not (see Appendix 1).

Apart from these indices, traditional quantitative indices of dissimilarity (e.g. Legendre and Legendre 1998) could also be applied to the abundances of evolutionary units normalized by the total abundance of all species in a site. Although they measure phylogenetic dissimilarity, I prefer to exclude them from the class of PD-dissimilarity indices. An example of such index is the weighted Unifrac index (Lozupone et al. 2007):

$$\text{weighted Unifrac} = \sum_{k \in b_T} L_k \left| \frac{a_{1k}}{\sum_{i \in t_T} A_{1i}} - \frac{a_{2k}}{\sum_{i \in t_T} A_{2i}} \right|$$

where $A_{ji}$ is the abundance of species $i$ in site $j$, and its normalized version

$$\text{normalized weighted Unifrac} = \sum_{k \in b_T} L_k \left| \frac{a_{1k}}{\sum_{i \in t_T} A_{1i}} - \frac{a_{2k}}{\sum_{i \in t_T} A_{2i}} \right| \Bigg/ \sum_{k \in b_T} L_k \left| \frac{a_{1k}}{\sum_{i \in t_T} A_{1i}} + \frac{a_{2k}}{\sum_{i \in t_T} A_{2i}} \right|$$

(see also Chang et al. 2011; Chen et al. 2012). Several indices of PD-dissimilarity would be affected if the abundances of evolutionary units were normalized by the total abundance of all species in a site (e.g. all indices of Appendix 1 except those, $evoD_{Chord}$, $evoD_{Hellinger}$, $evoD_{Profile}$, $evoD_{\chi^2}$, $1-\bar{V}_{q2,rel}$, ${}^{q}evoD_{Rényi,rel}$, where $a_{jk} / \sum_{k \in b_T} L_k a_{jk}$ or $a_{jk} / \sqrt{\sum_{k \in b_T} L_k a_{jk}^2}$ is used as relative abundance of any evolutionary unit supported by any branch $k$ in any site $j$).

Finally, a common problem with species dissimilarity measures is that they often have to be estimated from samples. It is well known that it is hardly possible to estimate all those indices that depend on species richness in areas where many species are rare and thus rarely sampled. Estimation problems will be transferred to many PD-dissimilarity measures. New studies will have to acknowledge that the abundances of the species can potentially vary independently whereas those of the evolutionary units vary dependently of their shared descendant species. In addition considering phylogenetic trees means that those trees also have to be accurately estimated. The estimation of PD-dissimilarity requires much more studies, which is beyond the scope of this paper. Instead I have focussed the discussion on another problem that ecologists are faced with: the enormous amount of available indices.

## References

Chang, Q. et al. 2011. Variance adjusted weighted UniFrac: a powerful beta diversity measure for comparing communities based on phylogeny. – BMC Bioinformatics 12: 118.

Chen, J. et al. 2012. Associating microbiome composition with environmental covariates using generalized UniFrac distances. – Bioinformatics 28: 2106–2113.

Chiu, C.-H. et al. 2014. Phylogenetic beta diversity, similarity, and differentiation measures based on Hill numbers. – Ecol. Monogr. 84: 21–44.

Faith, D. P. et al. 2009. The cladistic basis for the phylogenetic diversity (PD) measure links evolutionary features to environmental gradients and supports broad applications of microbial ecology's "phylogenetic beta diversity" Framework. – International Journal of Molecular Sciences 10: 4723–4741.

Hadju, L. J. 1981. Geographical comparison of resemblance measures in phytosociology. – Vegetatio 48: 47–59.

Jost, L. et al. 2011. Compositional similarity and beta diversity. – In: Magurran, A. and McGill, B. (eds) Biological diversity: frontiers in measurement and assessment. – Oxford University Press, pp. 66–84.

Legendre, P. and De Cáceres, M. 2013. Beta diversity as the variance of community data: dissimilarity coefficients and partitioning. – Ecol. Lett. 16: 951–963.

Legendre, P. and Legendre, L. 1998. Numerical ecology. 2nd English ed. – Elsevier Science BV.

Lozupone, C. A. et al. 2007. Quantitative and qualitative beta diversity measures lead to different insights into factors that structure microbial communities. – Applied and Environmental Microbiology 73: 1576–1585.

Nipperess, D. A. et al. 2010. Resemblance in phylogenetic diversity among ecological assemblages. J. Veg. Sci. 21: 809–820.

Pillar, V. D. et al. 2010. A framework for metacommunity analysis of phylogenetic structure. – Ecol. Lett. 13: 587–596.

Appendix 3. Results of evoCA and evoNSCA applied to Medellín et al. (2000) data set

The results of evoCA and evoNSCA applied to Medellín et al. (2000) data set are given in Figs 3-1 and 3-2. Close positions for branches on the ordination space represent closely related branches in the phylogenetic tree; and close positions for sites represent sites with close phylogenetic compositions. Finally, a site positioned near a branch on the ordination space means descendants of this branch (and/or of sister branches depending on the ordination method) occur in the site with notable abundances.

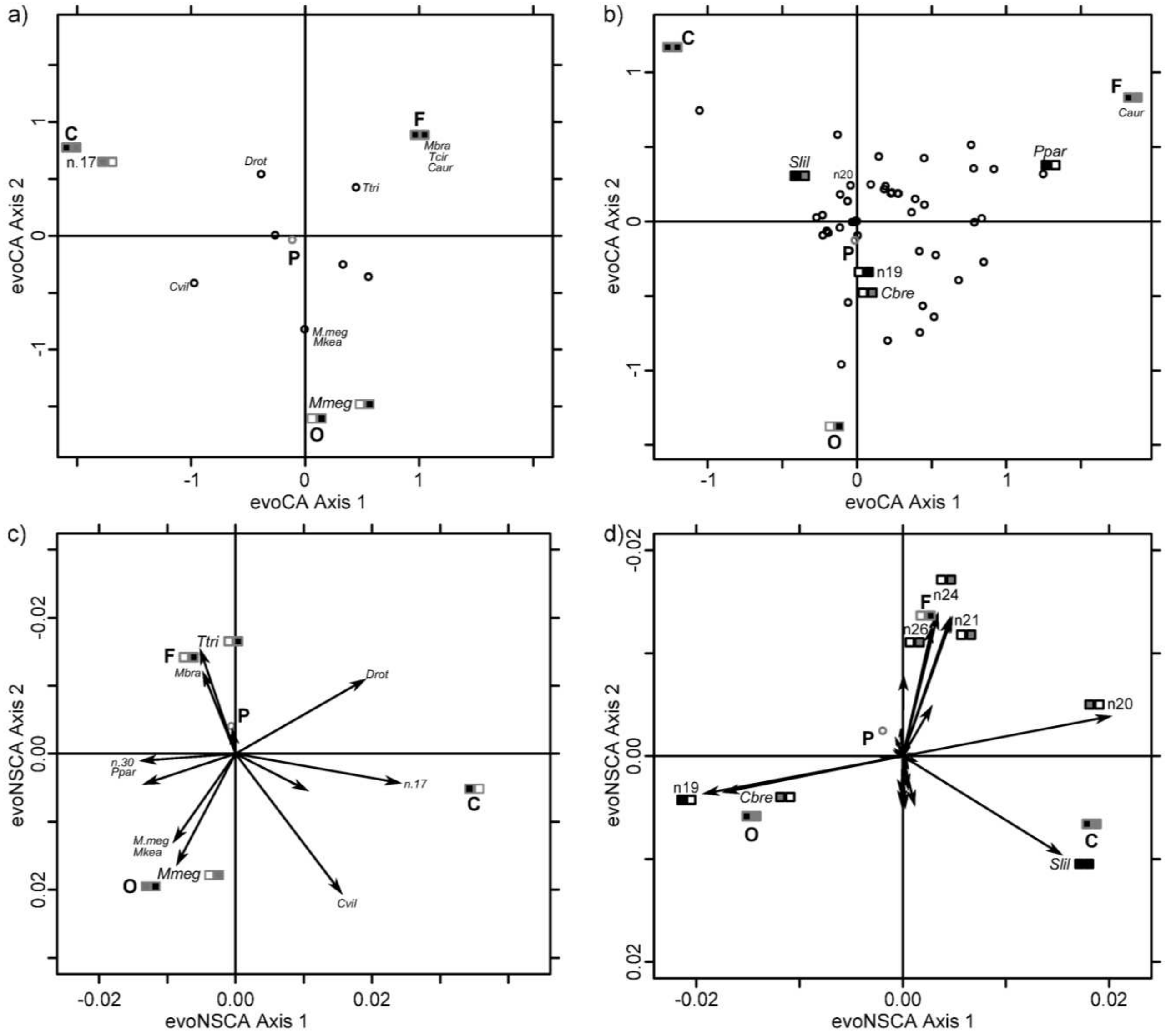

Figure 3-1. Factorial maps obtained with evoCA a)b) and evoNSCA c)d) with the data expressed as a)c) presence/absence and b)d) abundance of the evolutionary units in each habitat. The first two axes of evoCA, expressed 41% and 37%, respectively with presence/absence data and 51%, 41% with abundance data, of the averaged squared phylogenetic distances among communities as measured by index $evoD_{\chi^2}$. The two first axes of evoNSCA expressed 47% and 32%, respectively, with presence/absence data and 71%, 28% with abundance data of the averaged squared phylogenetic distances among communities as measured by index $evoD_{Profile}$. Codes for species are: Caur (*Chrotopterus auritus*), Cbre (*Carollia brevicauda*), Cvil (*Chiroderma villosum*), Drot (*Desmodus rotundus*), Mbra (*Micronycteris brachyotis*) and M.meg (*M. megalotis*), Mkea (*Myotis keaysi*), Mmeg (*Mormoops megalophylla*), Ppar (*Pteronotus parnelii*), Slil (*Sturnira lilium*), Tcir (*Trachops cirrhosus*), Ttri (*Thyroptera tricolor*). Codes for habitats, branches, internal nodes, and for the contributions of habitats and branches to the analyses are given in the legend of Figure 3 in the main text.

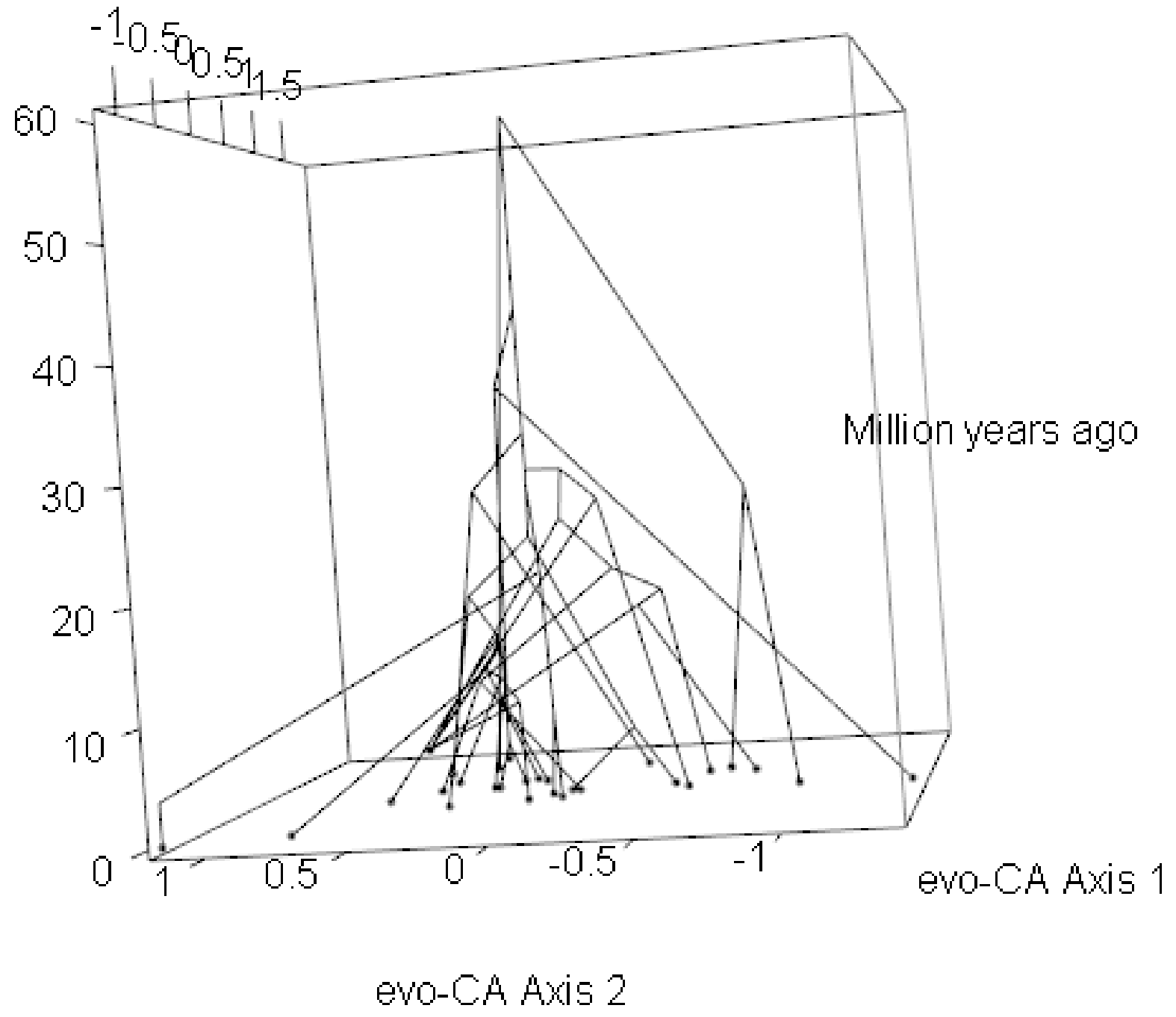

Figure 3-2. To illustrate the particularity of evoCA that positions lineages at the barycenter of their descending species, the phylogenetic tree is displayed on the evoCA map (here with abundance data). It is shown here that the phylogenetic structure is preserved in the sense that a branch is positioned in the evoCA map at the barycenter of its descending species. However, by using a 3D-plot with evoCA axis 1, evoCA axis 2, and the time of evolution (in million years) as the three axes, we can observe that several branches cross others. The fact that some branches cross each other means that some species descending from the same node have different abundance distributions among the habitats. The clearer the tree, the more conserved the associations between species and habitats are.

All these results on the structure of bat communities are contingent on the quality of the phylogenetic tree. For example, if the Vespertilionidae species were actually not as original (i.e. isolated descendants from a long, unshared branch) as they are in Fritz et al. (2009) tree, then their contributions to the PD-dissimilarity among the habitats of the Selva Lacandona of Chiapas would likely decrease and eventually vanish due to the low abundances of these species. Controversy on the evolutionary relationships are nevertheless mostly among bat families due to poor fossil record, whereas most of the phylogenetic structures among habitats, when considering species' abundances, were driven by species estimated to be rather related (descending from node #18 in Fig. 1 from the main text). For example, most of the differences between cornfields and rainforest were driven by the relative abundance of *Sturnina lilium* compared to other Stenodermatinae. Although plants have been the most studied group in phylogenetic community ecology over the last decades, the application of this field of research to other groups, including bats (e.g. Villalobos et al. 2013), is growing.

## References

Fritz, S. A. et al. 2009. Geographic variation in predictors of mammalian extinction risk: big is bad, but only in the tropics. – Ecol. Lett. 12: 538–549.

Medellín, R. et al. 2000. Bat diversity and abundance as indicators of disturbance in Neotropical rainforest. – Conserv. Biol. 14: 1666–1675.

Villalobos, F. et al. 2013. Phylogenetic fields of species: cross-species patterns of phylogenetic structure and geographical coexistence. – Proc. R. Soc. B 280: 20122570.

Appendix 4. Further results on patterns of PD-dissimilarities among habitats in Medellín et al. (2000) data set

I also analysed the components *a*, *b*, *c*, and *A*, *B*, *C* of PD-similarity introduced by Nipperess et al. (2010) using ternary plots (Koleff et al. 2003). In complement, I applied the PD-dissimilarity indices $evoD_{Sørensen}$ and $evoD_{TS}$ to the data and displayed the results using metric multidimensional scaling (mMDS) (see Fig. 4-1).

Ternary plots (Fig. 4-1) provided two main insights. First the PD-similarities among habitat types are higher when considering abundance data than presence/absence data (higher relative values of *A* compared to *a* in Fig. 4-1). Second, with presence/absence data, the cornfields constitute the habitat with the lowest PD-similarity with the other habitat types (lowest values of *a*), whereas, with abundance data the rainforest is the most distinct habitat type (lowest values of *A*). Indices of PD-dissimilarity relying on these parameters reflect these two main insights. As an example, $evoD_{Sørensen}$, which uses presence/absence data, distinguished cornfields from rainforest and oldfields, with cacao plantations having its highest similarity with the rainforest (Fig. 4-1c). In contrast, with $evoD_{TS}$, which uses abundance data, the rainforest is opposed first to cacao plantations and oldfields and second to cornfields (Fig. 4-1d).

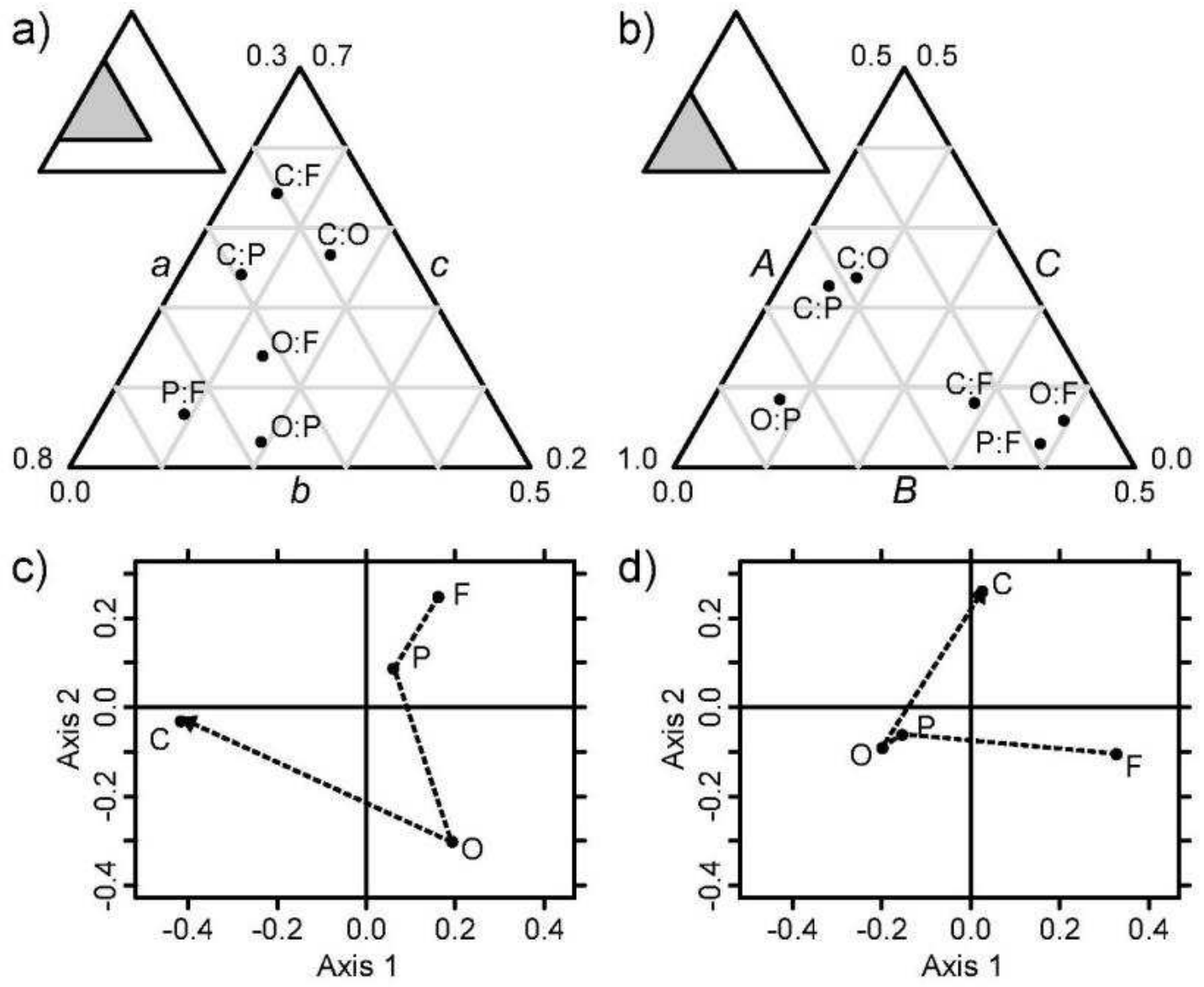

Figure 4-1. Evo(dis)similarity framework introduced by Nipperess et al. (2010). The first two panels provide the ternary plots for the components of PD-similarity introduced by Nipperess et al. (2010) and here applied to the bat PD-dissimilarity among the four habitats in the Selva Lacandona of Chiapas: a) presence/absence data; b) abundance data. A ternary plot represents proportions that sum to one thanks to three axes. In panel a), the proportions are $a/(a+b+c)$,

$b/(a+b+c)$ and $c/(a+b+c)$; in panel b) they are $A/(A+B+C)$, $B/(A+B+C)$, $C/(A+B+C)$; with *a*, *b*, *c*, the presence/absence-based, and *A*, *B*, *C* the abundance-based components of PD-similarity introduced by Nipperess et al. (2010) (see *Material and Methods* in the main text). In a ternary plot, a zoom has been applied on a sub-triangle where all points aggregate. The axes of the full triangle vary from 0 to 1. The location of the sub-triangle in the full one is indicated by the scale of each axis and by the embedded close, grey triangle at the top-left hand of each panel. The last two panels provide the values of PD-dissimilarities among the habitats as measured by c) $\sqrt{evoD_{Sørensen}}$ and d) $\sqrt{evoD_{TS}}$ (graphs obtained by metric Multidimensional Scaling; see Supplementary material Appendix 1 for index equations). A broken line indicates the gradient of disturbance from rainforest to cornfields. Codes for habitats are: F=rainforest; P=cacao plantations; O=oldfields; C=cornfields.

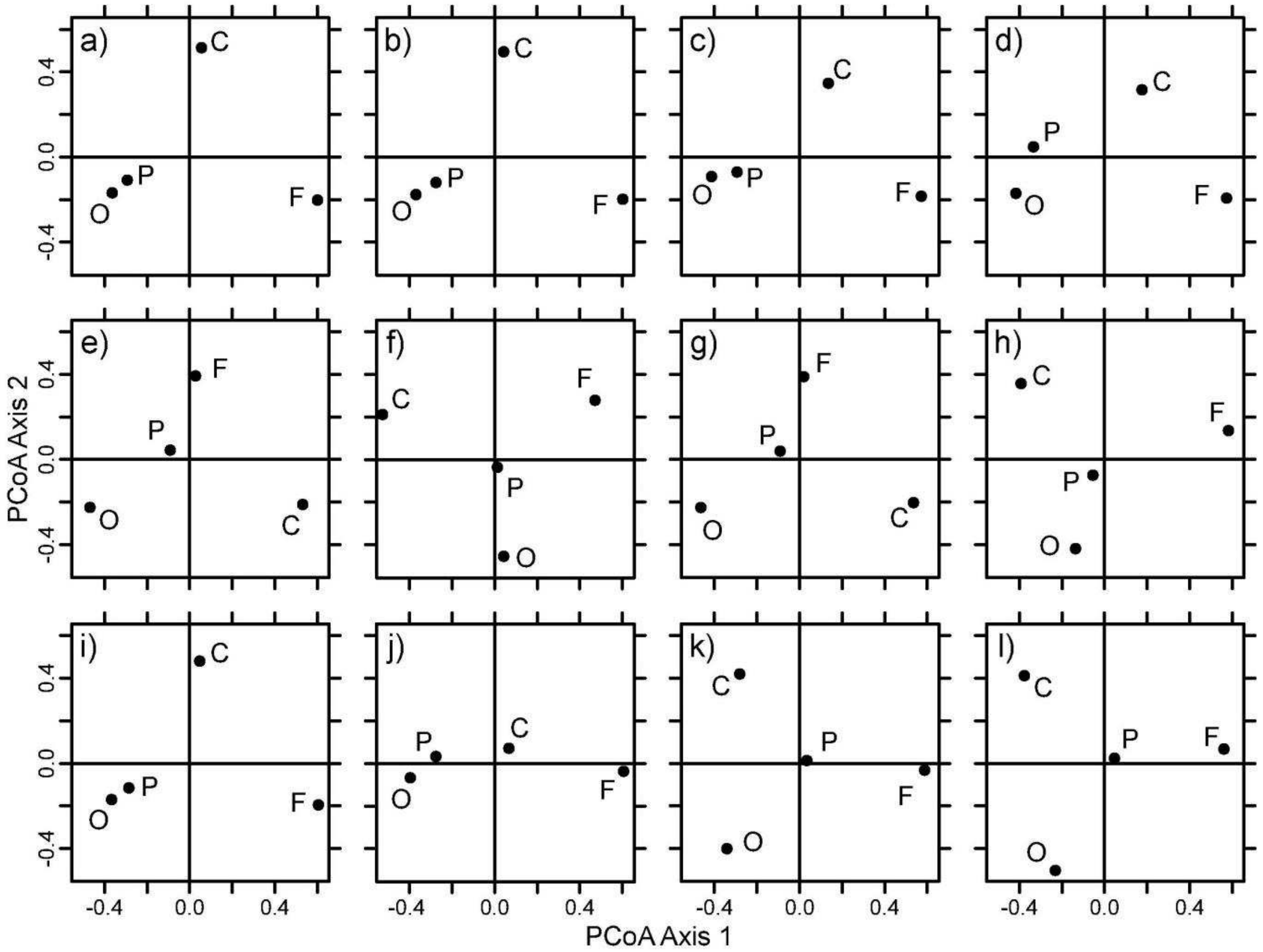

Figure 4-2. Two-dimensional metric multidimensional scaling (=principal coordinate analysis) applied to PD-dissimilarities among habitats as measured by indices (a) $\sqrt{evoD_{TJ}}$, (b) $\sqrt{evoD_{TO}}$, (c) $evoD_{Manhattan}$, (d) $evoD_{Euclidean}$, (e) $evoD_{Chord}$, (f) $evoD_{Hellinger}$, (g) $evoD_{Profile}$, (h) $evoD_{\chi^2}$, (i) $\sqrt{evoD_{Bray\text{-}Curtis}}$, (j) $evoD_{Morisita\text{-}Horn}$, (k) $evoD_{ScaledCanberra}$, and (l) $evoD_{Divergence}$. To ease comparison among measures, each matrix of PD-dissimilarities was divided by the maximum observed PD-dissimilarity between two habitats. Codes for habitats are:

F=rainforest; P=cacao plantations; O=oldfields; C=cornfields. The indices provided different patterns of PD-dissimilarity among habitat. The indices $evoD_{Euclidean}$, $evoD_{Manhattan}$ , $evoD_{TJ}$, $evoD_{TS}$, $evoD_{TO}$, and $evoD_{BC}$ distinguished forest from first cornfields and then both oldfields and cacao plantations considered as the most similar habitats. With $evoD_{ScaledCanberra}$ and $evoD_{Divergence}$, cornfield, oldfield and forested habitats were almost equidistant from each other; and cacao plantations had partial, even similarities with all these habitats. The pattern provided by $evoD_{\chi^2}$, $evoD_{Hellinger}$, and $evoD_{Chord}$ were close to that given by $evoD_{ScaledCanberra}$ and $evoD_{Divergence}$. However, $evoD_{\chi^2}$ showed stronger similarities among cacao plantations and oldfields than among cacao plantations and the other habitats, whereas $evoD_{Hellinger}$ and $evoD_{Chord}$ opposed cornfields to the other habitats. In contrast, $evoD_{Morisita\text{-}Horn}$ showed strong differences between rainforest and both cacao plantations and oldfields with cornfields having an intermediate position and thus an intermediate phylogenetic composition. This figure shows that the choice of an index of PD-dissimilarity influences the results of a study.

## References

Koleff, P. et al. 2003. Measuring beta diversity for presence-absence data. – J. Anim. Ecol. 72: 367–382.

Nipperess, D. A. et al. 2010. Resemblance in phylogenetic diversity among ecological assemblages. – J. Veg. Sci. 21: 809–820.